\documentclass[aps,prd,onecolumn,superscriptaddress,showpacs]{revtex4}

\usepackage{graphicx}
\usepackage{dcolumn}
\usepackage{bm}
\usepackage{pstricks}

\newcommand{\be}{\begin{equation}}
\newcommand{\ee}{\end{equation}}
\newcommand{\bea}{\begin{eqnarray}}
\newcommand{\eea}{\end{eqnarray}}

\newcommand{\kms}{{\rm ~km/s}}

\newcommand{\cm}{{\rm ~cm}}

\newcommand{\kevee}{{\rm ~keVee}}
\newcommand{\kevr}{{\rm ~keVr}}

\newcommand{\GeV}{{\rm ~GeV}}

\begin{document}

\title{Inelastic Dark Matter and DAMA/LIBRA: An \emph{Experimentum Crucis}}

\author{Douglas P. Finkbeiner}
\affiliation{Harvard-Smithsonian Center for Astrophysics, 60 Garden St.,
  Cambridge, MA 02138, USA}

\author{Tongyan Lin}
\affiliation{Physics Department, Harvard University, Cambridge, MA 02138, USA}

\author{Neal Weiner}
\affiliation{Center for Cosmology and Particle Physics, Department of Physics, New York University, 
New York, NY 10003, USA}

\date{\today}

\begin{abstract}

  The DAMA/LIBRA collaboration has detected an annual modulation of
  the recoil rate in NaI crystals with the phase expected for WIMP
  scattering events.  This signal is dramatically inconsistent with
  upper limits from other experiments for elastically scattering
  weak-scale WIMPs.  However, the results are compatible for the case
  of inelastic dark matter (iDM).  The iDM theory, as implemented by
  Tucker-Smith and Weiner, constrains the WIMP to a tight contour in
  $\sigma_n-\delta$ space, where $\delta$ is the mass difference
  between the ground state and excited WIMPs.  An urgent priority in
  direct detection is to test this scenario. The crucial test of the
  iDM explanation of DAMA -- an \emph{experimentum crucis} -- is an
  experiment with directional sensitivity, which can measure the daily
  modulation in \emph{direction}.  Because the contrast can be 100\%,
  it is a sharper test than the much smaller annual modulation in the
  rate.  We estimate the significance of such an experiment as a
  function of the WIMP mass, cross section, background rate, and other
  parameters.  The proposed experiment severely constrains the
  DAMA/iDM scenario even with modest exposure ($\sim 1000$ kg $\cdot$
  day) on gaseous xenon.

\end{abstract}

\pacs{95.35.+d}

\maketitle
\twocolumngrid

\section{Introduction}

The DAMA claim \cite{Bernabei:2000qi} of an annual modulation signal has
long appeared to be in conflict with non-detections in other experiments
\cite{Gaitskell:2004gd}.  Though recent limits by XENON10
\cite{Angle:2007uj,Angle:2008we} and CDMS II \cite{Ahmed:2008eu}
appear to rule out the DAMA region of parameter space by a factor of 100
in cross section, DAMA/LIBRA \cite{Bernabei:2008yi} has recently
confirmed their previous annual modulation result and increased the
significance to $8.2 \sigma$.  This conflict has motivated serious
discussion of models beyond the simplest elastic scattering of
weak-scale WIMPs, with the hope of accommodating DAMA as well as the
other limits.

At least four approaches have been considered: 1. electron scattering \cite{Bernabei:2007gr}; 2. spin
dependent scattering \cite{Ullio:2000bv,Belli:2002yt,Savage:2004fn}; 3. light dark matter \cite{Bottino:2003cz,Gondolo:2005hh}; and 4. inelastic
scattering \cite{Smith:2001hy}.  The first hypothesizes that the signal in DAMA is
scattering of WIMPs off of electrons.  Significant momentum can be
transferred to the electron during the small fraction of the time ($<
0.1$\%) that it finds itself near the nucleus and at moderately
relativistic speeds.  However, this small fraction must be balanced by
an uncomfortably large cross section, which is almost certainly ruled
out by early Universe (CMB) constraints.

The spin-dependent scattering argument attempts to circumvent limits
from CDMS in Si for example by positing that the cross section is
strongly dependent on nuclear spin.  However, recent experiments
\cite{Behnke:2008zza} have significantly tightened constraints on this
scenario, and the allowed regions require a significant drop in the
background in the signal region \cite{Savage:2008er}.  While small
regions of parameter space are still allowed, we do not consider this
here.

Another suggestion is that the DAMA recoil events are not in the
energy range first suspected.  Assuming recoils off of iodine, the
quenching factor of 0.09 implies that the $2-6\kevee$ observed
energy corresponds to a recoil energy of $22-66\kevr$.  It has recently
been suggested that ``channeling'', i.e. alignment of the recoil with
principal directions in the crystal lattice, creates an effective
quenching factor of unity for some fraction of the events \cite{Bernabei:2007hw}.  In this
case, there is a small amount of parameter space available for lighter
WIMPs ($\sim 5 \GeV$) still compatible with other limits
\cite{Bottino:2007qg,Petriello:2008jj,Savage:2008er}.  In general, light WIMPs have difficulty with
constraints from the energy spectrum of the unmodulated DAMA signal
\cite{Chang:2008xa,Fairbairn:2008gz}.  While further exploration of light WIMPs may be
warranted, we do not consider this option here.

\subsection{The DAMA/iDM Scenario}
The inelastic scattering scenario of Tucker-Smith \& Weiner
\cite{Smith:2001hy,Tucker-Smith:2004jv,Chang:2008gd} takes a different
approach: inelastic dark matter (iDM) has an excited state some $\delta \sim$ 100 keV
above the ground state.  The origin of this excited state is unimportant
for the present arguments; see \cite{ArkaniHamed:2008qp} for one
realization of this idea. Elastic scatterings off of nuclei are
suppressed by at least two orders of magnitude with respect to the
inelastic scatterings, leading to a preferred energy threshold with few
events at low energies. The high sensitivity of e.g. XENON10 to
low-energy scatterings (which dominate in the standard elastic
scattering models) means that even a small exposure time (316 kg day) can
place record-beating limits on the elastic cross section.  
Because iDM does not produce such low-energy events, it is plausible
that the much larger combined exposure time of DAMA/LIBRA and DAMA/NaI (300,000 kg day)
could see the higher energy events invisible in the other experiments.

Models of iDM are simple to construct, for instance a fourth-generation (vector-like) neutrino, coupling through the Z-boson \cite{Tucker-Smith:2004jv}, a mixed sneutrino \cite{Smith:2001hy}, KK states in RS theories \cite{Cui:2009xq}, in composite models \cite{Alves:2009nf}, or in theories with light mediators \cite{ArkaniHamed:2008qn}, see also \cite{Pospelov:2008jd,Chun:2008by,Baumgart:2009tn,Cheung:2009qd,Cui:2009xq,Katz:2009qq,Finkbeiner:2009mi,Batell:2009vb,Alves:2009nf,Morrissey:2009ur,Chen:2009ab,Kaplan:2009de}. In fact, off-diagonal couplings are very natural in dark matter theories, with only the small splitting $\delta$ remaining to be explained.

In an annual modulation experiment, iDM enjoys an additional
enhancement relative to elastic models because only WIMPs on the high
velocity tail scatter.  The modulation can be much larger than the
2-3\% expected for elastic scattering, partially compensating
for the fact that the majority of WIMPs are below threshold and do
not scatter.

If the direct detection data from DAMA and others are taken at face
value as nuclear WIMP scattering events, they argue strongly for
further experiments designed to test iDM.  The experiment must make
predictions beyond the already observed annual modulation so that a
positive result would add substantially to the believability of the
result. 
Such a make-or-break experiment is known as a ``critical
experiment,'' or \emph{experimentum crucis} \footnote{The term
\emph{experimentum crucis} was first used by Isaac Newton in a 1672
letter about his Theory of Light and Colors.}.  In the next section we
describe such an experiment and discuss the limits obtained.

\subsection{Advantages of Directional Sensitivity}
The DAMA result is compelling enough to motivate further
experiments involving iodine or other nuclei of similar mass.
Direct detection experiments generally fall into 3 categories, based
on their background rejection strategy.  Some (CDMS II, XENON10, etc.)
reject individual electron scattering events and look for the residual
signal from WIMP scattering.  

Another strategy for dealing with background is to search for the
annual modulation of the signal (DAMA) brought about by the Earth's
velocity around the Sun, added to the velocity of the Sun around the
Galaxy.  The assumption is that the WIMP velocities are nearly
isotropic, and the Sun moves through the WIMPs at roughly 200 km/s.
The Earth moves around the Sun at $v_{orb}\approx 30\kms$ in an orbit
inclined by $i\approx 60^\circ$ with respect to the Sun's velocity,
introducing a modulation of $v_{orb}\cos(i) \approx 15\kms$.  This
method has the virtue of ignoring all steady state instrumental
backgrounds, but is vulnerable to backgrounds that vary with the
seasons.  Though DAMA has placed stringent limits on variations in
temperature, humidity, radon gas, line voltage, and anything else
known to vary by season \cite{Bernabei:2008yi}, this remains a
persistent concern.

A third strategy is to use directional information
\cite{Spergel:1987kx}.  Because the scattering events should
originate, on average, from a specific direction on the sky
($\ell=90^\circ, b=0^\circ$), a daily modulation in \emph{direction}
due to the rotation of the Earth is a sharp test of the WIMP
scattering model.  As with the annual modulation, many other
backgrounds may be expected to vary on a daily timescale.  However, as
the Earth orbits around the Sun, the angle between the Sun direction
and the WIMP signal varies from $60^\circ$ ($\sim$7 March) to
$120^\circ$ ($\sim$9 September).  Also, any Sun-related oscillation (365.25
yr$^{-1}$) is orthogonal to the WIMP signal (366.25 yr$^{-1}$) over
one year.  This separation allows a much sharper test than the annual
modulation alone, even in the limit of low statistics.  Furthermore,
directional detectors have excellent background rejection and can
distinguish between recoils of nuclei and other particles by
correlating the length and energy of recoil tracks.

In the context of iDM, a directional experiment has another
advantage. The minimum velocity $v_{min}$ for a WIMP to scatter with 
a nuclear recoil of energy $E_R$ is:
\begin{eqnarray}
	v_{min} = \sqrt{\frac{1}{2 m_N E_R}} \left( \frac{m_N E_R}{\mu} + \delta \right)
\label{eq:vmin}
\end{eqnarray}
where $\mu$ is the nucleus-WIMP reduced mass $m_\chi m_N/(m_\chi + m_N)$
and $m_\chi$ is the WIMP mass.
Because of the energy threshold, most events result from
WIMPs in the high velocity tail of the WIMP velocity distribution, and therefore
most events happen near threshold.  This is advantageous because events at
threshold have a sharply peaked angular distribution, making the
directional discrimination even more pronounced. 
The energy-dependent maximum recoil angle is
\begin{equation}
	\cos \gamma_{max}(E_R) = \frac{v_{esc} - v_{min}(E_R,\delta)}{v_E}
\end{equation}Here $\gamma$ is the angle between the velocity of the Earth
and the recoil velocity in the Earth frame, and $v_{esc}$ is the Galactic escape velocity
from the Solar neighborhood. For the benchmark models considered here, 
$\gamma$ is constrained to be within $\sim$100 degrees of the Earth's direction.
Furthermore, as with annual modulation, 
the total number of events should vary through the
year in a predictable way.  These advantages allow a decisive test of
the DAMA/iDM scenario with modest experimental effort.

In this article, we evaluate the sensitivities for the
DAMA/iDM scenario as a function of WIMP mass $m_\chi$, $\delta$, and other
parameters. We focus on a set of benchmark models, given in Table~\ref{tab:benchmarks},
that can simultaneously explain DAMA and satisfy constraints from other 
experiments \cite{Chang:2008gd}. Note that the $m_\chi=70$ GeV
benchmark cannot actually explain the DAMA data because of the predicted asymmetry
in the modulation amplitude during summer and winter. 
However, we include the benchmark as a worst-case scenario, as there is 
flexibility in the WIMP parameters due to the uncertainty 
in the halo distribution and astrophysical parameters 
\cite{MarchRussell:2008dy}. These benchmarks
give the general features and sensitivities (within an order of magnitude)
of a directional experiment to the available parameter space of iDM.
We find that in most parts of parameter space, 1000
kg days of exposure is sufficient to confirm or
refute DAMA/iDM at high confidence.

\begin{table}[t]
\begin{center}
\begin{tabular}{ccccc}
$m_{\chi}$ & & $\delta$ & & $\sigma_n$ \\
(GeV) & & (keV) & & ($10^{-40} \text{cm}^2$) \\
\hline
70 & & 119 & & 11.85 \\
150 & & 126 & & 2.92 \\
700 & & 128 & & 4.5 \\
150* & & 130 & & 4
\end{tabular}
\end{center}
\caption{Benchmark models for $v_{esc} = 500$ km/s, $v_0 = 220$ km/s \cite{Chang:2008gd}. In the last row we have listed the benchmark model for $m_\chi = 150$ GeV at $v_{esc} = 600$ km/s.}
\label{tab:benchmarks}
\end{table}

\begin{figure*}[thb]
\begin{center}
a)\includegraphics[width=.45\textwidth]{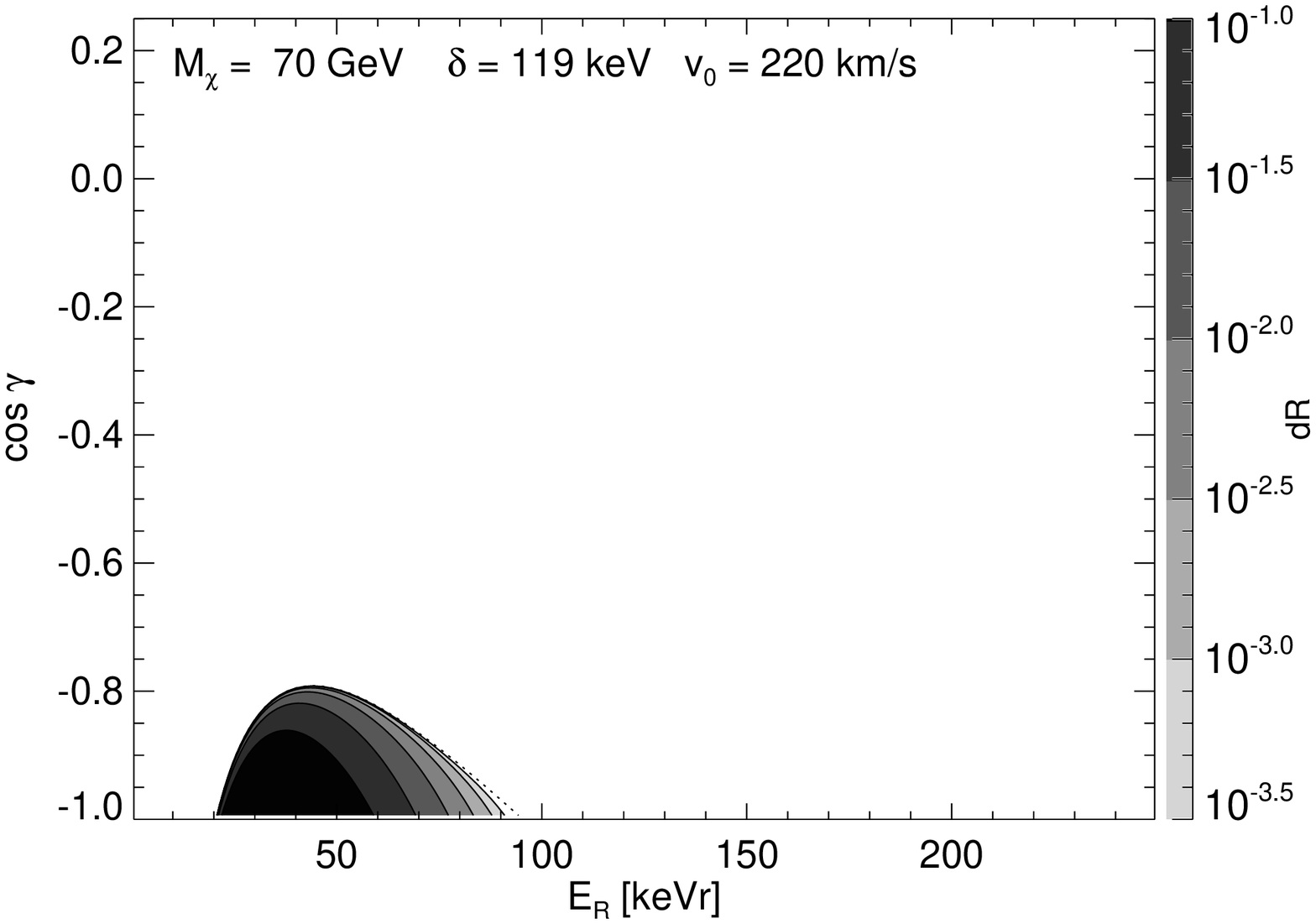}\hskip 0.2in
b)\includegraphics[width=.45\textwidth]{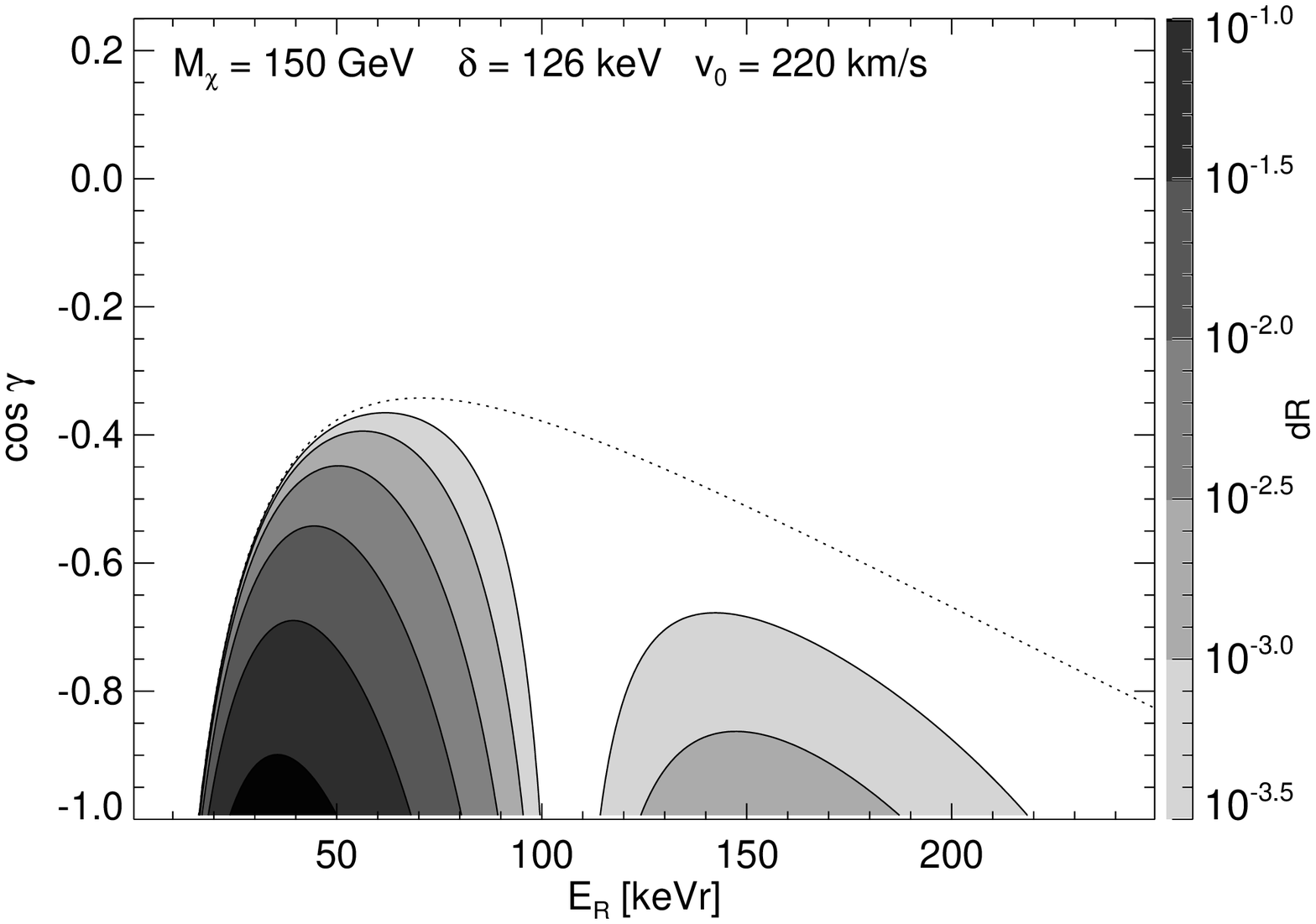} \\
c)\includegraphics[width=.45\textwidth]{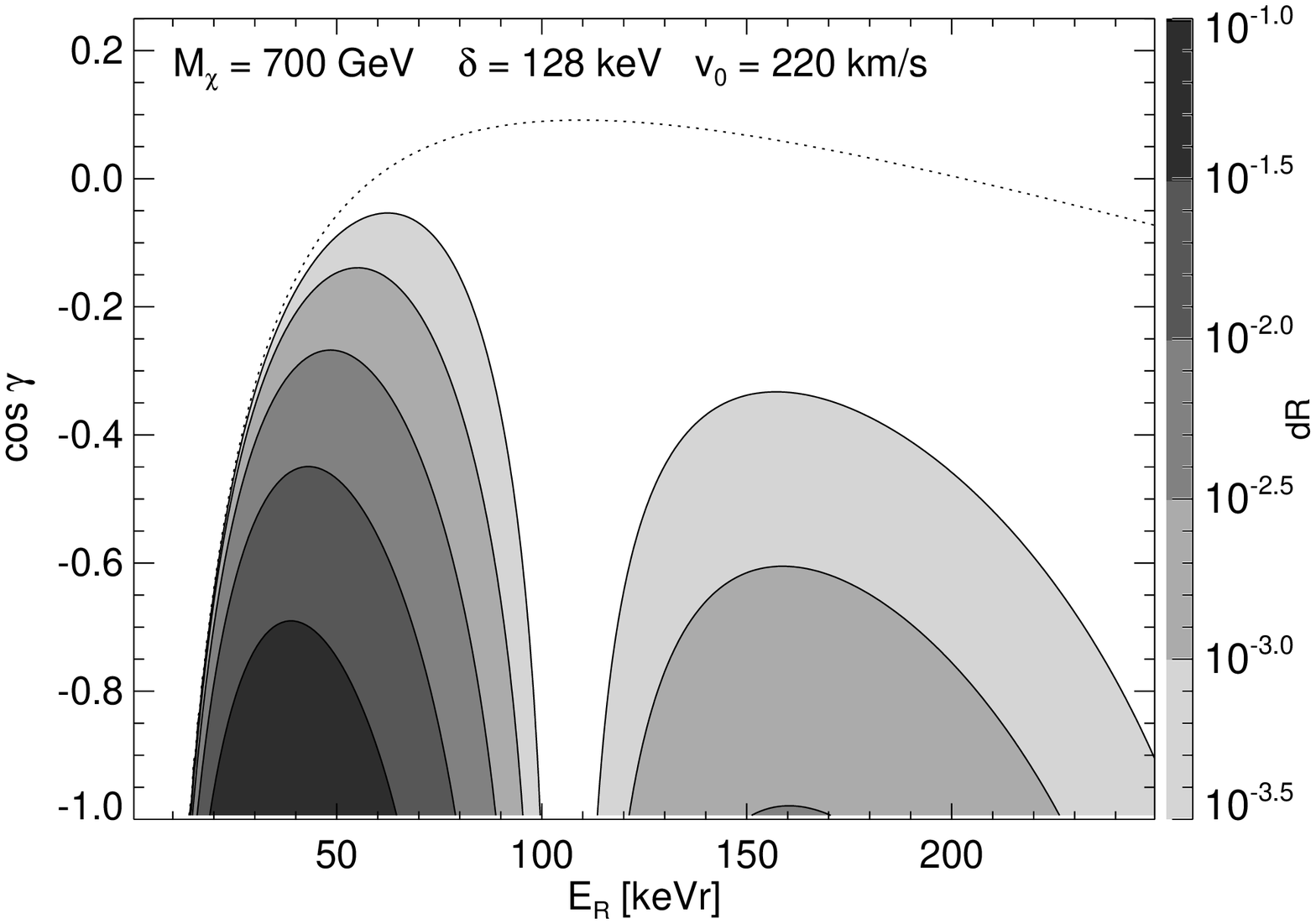}\hskip 0.2in
d)\includegraphics[width=.45\textwidth]{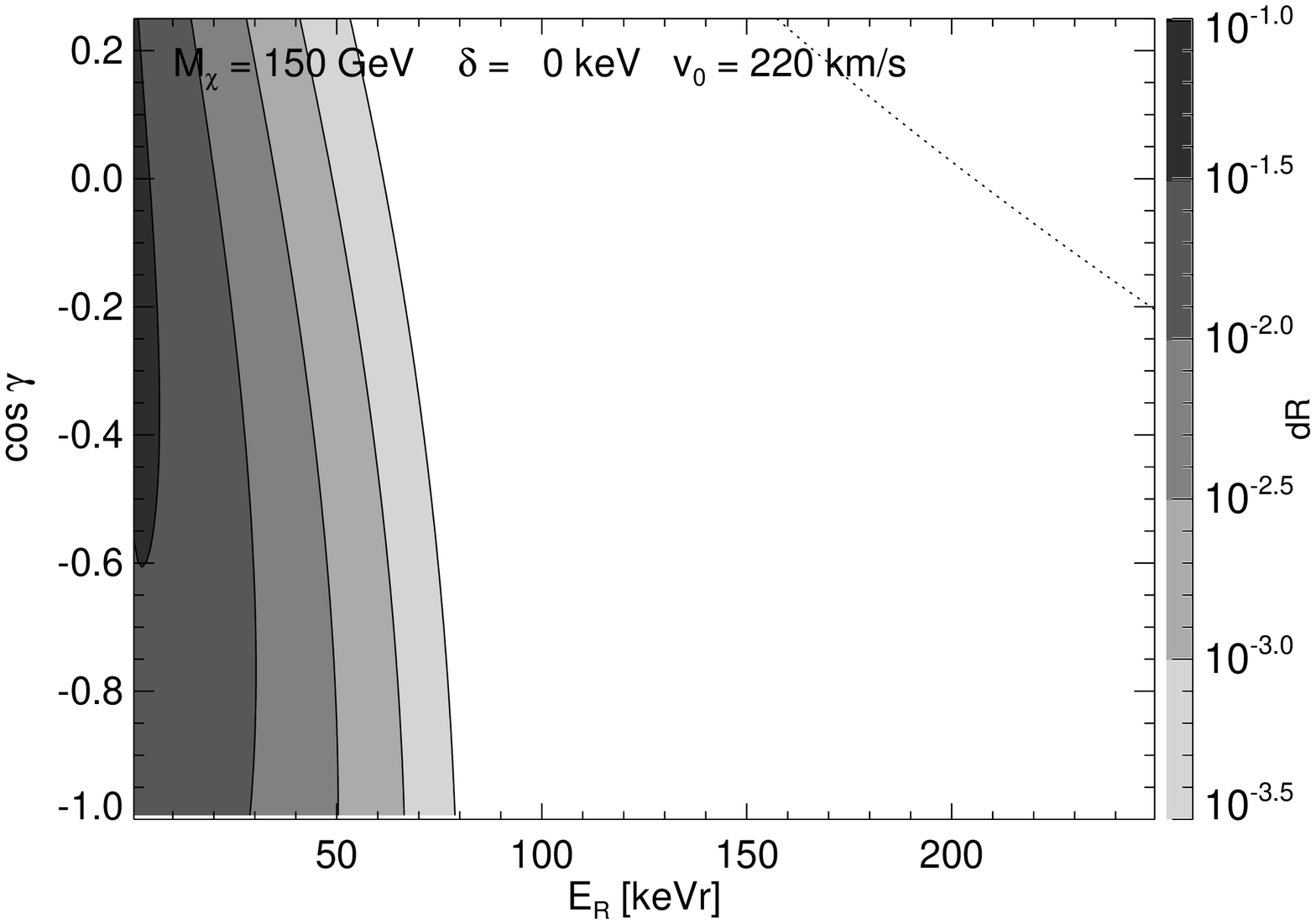}
\end{center}
\caption{Differential rates $dR/(dE_R\,d\cos\gamma)$ for the benchmark
  models given in Table~\ref{tab:benchmarks} for $v_{esc}=500$ km/s, as well as for an elastic WIMP.  
In each case, the differential
  rate is normalized so that the total rate is unity. Outside the region 
  indicated by the dashed line, scattering events are kinematically forbidden.}
\label{fig:rates}
\end{figure*}

\section{Experimental Setup \label{sec:experiment}}
Before discussing the specifics of the experiment, we can address a few
basic questions of exposure and energy range. DAMA/LIBRA reports a cumulative 
modulation in the $2-6 \kevee$ range of $0.052$ counts per day per kg, 
$({\rm cpd/kg})$. 
The quoted energy range is related to the nuclear recoil energy by a 
quenching factor $q = E_{ee}/E_{NR} \simeq 0.09$ for iodine. 
Thus, $2-6 \kevee \approx 22-66 \kevr$.

In the extreme case where the modulation is 100\% (i.e., no scattering at all occurs in the winter), the signal is essentially directional. One would need approximately 400 kg $\cdot$ day in the summer to yield 20 events of signal, roughly the number of events needed for an unambiguous detection at zero background, as we will discuss in Section~\ref{sec:cosgamma}. 
Consistency with other experiments is also possible with $\sim$ 20\% modulation \cite{Chang:2008gd}, with only 40 kg $\cdot$ day needed for a clear discovery.

However, this estimate assumes that the signal occurs in an energy range which is detectable at a directional experiment, and this, we shall see, is very unlikely to be the case. A directional experiment will likely have a higher energy threshold.

The DAMA/LIBRA signal peaks near $E_R \approx 3 \kevee$, after which it
falls significantly. Above 5 \kevee, the total modulation is $0.0034 \pm
0.0024$ cpd/kg, which is consistent with zero. The signal above 4 \kevee\ yields a signal
at DAMA of $0.014 \pm 0.004$ cpd/kg, which requires approximately 
1400 kg $\cdot$ day of exposure for 20 events. Moreover, it is possible 
that the actual signal is at 3.5 \kevee\ and below, and the signal at apparently higher energy is due to the resolution of the DAMA detector \cite{Bernabei:2008yh}.

There is a significant uncertainty in the quenching factor as well. While $q = 0.09$ is a commonly used value, the measurements are uncertain, and values $q = 0.10$ and slightly higher are possible. Since the range of WIMP parameters allowed arises from fitting the DAMA peak, the uncertainty in this factor is hidden from our analyses here. Nonetheless, the presence of a larger quenching factor would result in a lower range of energies for the signal. Thus, it is clear that a {\em robust} test of the DAMA result involves pushing the energy threshold as low as possible. While the models that we consider generally {\em do} have signal above 50 \kevr, this cannot be guaranteed, especially in situations where form factors might suppress the higher energy events \cite{Alves:2009nf}. In the event the experiments as we describe are performed and no signal is seen, we would advocate lowering the threshold, even at the cost of exposure from reduced pressure, to whatever extent possible.

With these important caveats in mind, we can proceed to discuss the details of what such an experiment would look like.

\subsection{Experimental Design}
Gaseous detectors can resolve the nuclear recoil
tracks, which have lengths of several millimeters at sufficiently low
pressures. Several gaseous directional detection experiments are already 
underway, including DMTPC \cite{Sciolla:2009fb}, NEWAGE \cite{Miuchi:2007ga},
DRIFT \cite{Burgos:2008mv}, and MIMAC \cite{Santos:2007ga}, which employ
time-projection chambers to reconstruct tracks.
However, these experiments are typically focused on spin-dependent 
WIMP-nucleus interactions and use the gas CF$_4$ as a detector,
with the exception of DRIFT, which uses CS$_2$. For a review of the various
detector technologies, see \cite{Ahlen:2009ev,Sciolla:2008vp,Sciolla:2009ps}.

We suggest using a gas containing xenon or another heavy element. 
This increases sensitivity to spin-independent interactions
because scattering rates are kinematically highly suppressed for 
lighter nuclei in the iDM scenario, in addition to the overall factor of 
$A^2$ that appears in the cross section. However, heavier
elements have shorter recoil tracks which are more difficult to resolve.
Furthermore, the gas should allow for good electron (or ion) drift and
also have good scintillation properties (at least for DMTPC). 
Choosing a gas will involve some compromise
between these properties. We note that for a splitting of $\delta \sim 120$ keV,
$A$ must be greater than 75 to see \textit{any} signal for the mass range 
$m_\chi \sim 100-1000$ GeV for an earth velocity of $225$ km/s and 
an escape velocity of $500$ km/s.

According to preliminary work of the directional detection experiments
mentioned above, in order to resolve the angles of the tracks, 
the gas chamber must be at a pressure of around 50 torr. 
Furthermore if the recoil energies are too low (below $\sim$ 50 \kevr), 
it is difficult to detect the sense (head-tail discrimination) 
of the track, which reduces sensitivity significantly 
\cite{Morgan:2004ys,Copi:2005ya,Green:2007at}.
The directional resolution of DMTPC is currently estimated to be
around 15 degrees at 100 \kevr\ and 
improves by several degrees at higher energies  \cite{Dujmic:2008ut}.

The dominant irreducible background is neutron recoils arising from
radioactive materials near or in the detector.  Simulations suggest
background rejection is excellent for gamma-rays, electrons, and
$\alpha$'s \cite{SnowdenIfft:1999hz} (see also Fig. 7 of
\cite{Sciolla:2009fb}).  The DRIFT collaboration has reported on
neutron backgrounds; however, they found a radioactive source
($^{222}$Rn) inside the detector \cite{Burgos:2007zz}.  The NEWAGE
experiment at Kamioka estimated their primary background to come from
the fast neutron flux which, when shielded by 50 cm of water, would
contribute only a few events per year \cite{Tanimori:2003xs}. 

\onecolumngrid

\section{Recoil Spectrum}

We derive the differential nuclear recoil spectrum in recoil energy $E_R$ and
$\cos \gamma$, which is defined as $\cos \gamma$ = $\hat v_E \cdot \hat v_R$.
The Earth's motion in the halo rest frame is $\vec v_E$ and the vector
$\vec v_R$ is the nuclear recoil velocity in the Earth's frame. Let $\vec v$
be the incoming WIMP velocity in the Earth's frame.

The single nucleon scattering cross section is:
\begin{equation}
	d\sigma = \frac{\sigma_n m_n}{2 \mu_n^2} \frac{1}{v^2} \ 
		dE_R \ d \cos \gamma \ \delta^{(1)} 
		\left( \hat v \cdot \hat v_R - \frac{v_{min}}{ v} \right)
\end{equation} where $\mu_n$ is the
WIMP-nucleon reduced mass and $\sigma_n$ is a reference cross section that
is assumed to be the same for all nucleons. $m_n$ is nucleon mass. The minimum velocity 
$v_{min}$ for a WIMP to scatter with a nuclear recoil of energy $E_R$
was given in Eq.~\ref{eq:vmin}.

The differential recoil rate for WIMP-nucleus scattering is
\begin{eqnarray}
	\frac{dR}{dE_R d \cos \gamma} &=& N_T \frac{\rho_\chi}{m_\chi} 
		\int d^3 v \ v \ f(\vec v + \vec v_E) 
		\frac{d\sigma}{dE_R d \cos \gamma} 
\end{eqnarray} where $f(\vec v)$, the WIMP distribution in the 
galaxy frame, is boosted to the Earth frame. $N_T$ is the number of target
nuclei per kg and $\rho_\chi$ is the local WIMP energy density. We are now using
the differential scattering cross section $d\sigma$ for the whole nucleus.
Define the constant $\kappa$:
\begin{equation}
	\kappa = N_T \frac{\rho_X}{m_\chi} 
		\frac{\sigma_n m_N}{2 \mu_n^2} \frac{(f_p Z + (A-Z)f_n)^2}{f_n^2}.
\end{equation} 
Changing variables to $\vec v' = \vec v + \vec v_E$ gives:
\begin{equation}
	\frac{dR}{dE_R d \cos \gamma} = \kappa F^2(E_R) 
		\int d^3 v \ f(\vec v) \ \delta^{(1)} \left( \vec v \cdot \hat v_R - 
		\vec v_E \cdot \hat v_R - v_{min}(E_R,\delta) \right) 
\end{equation} and $F^2(E_R)$ is the Helm form factor given in \cite{Lewin:1995rx}.
This formula is discussed in detail (in the context of 
Radon transforms) in \cite{Gondolo:2002np}. 
Thus we can see that at fixed $E_R$, the signal peaks where the delta function
is nonzero over the largest portion of the phase space, or $\cos \gamma = \hat v_E \cdot \hat v_R = -1$. The peak in $E_R$ and fixed $\gamma$ is determined by the competition  between the form factor (which pushes the signal to lower energies) and the inelasticity (whereby the minimum velocity produces a minimum value of $E_R$).

\begin{figure*}[t]
\centering
a) \includegraphics[width=.45\textwidth]{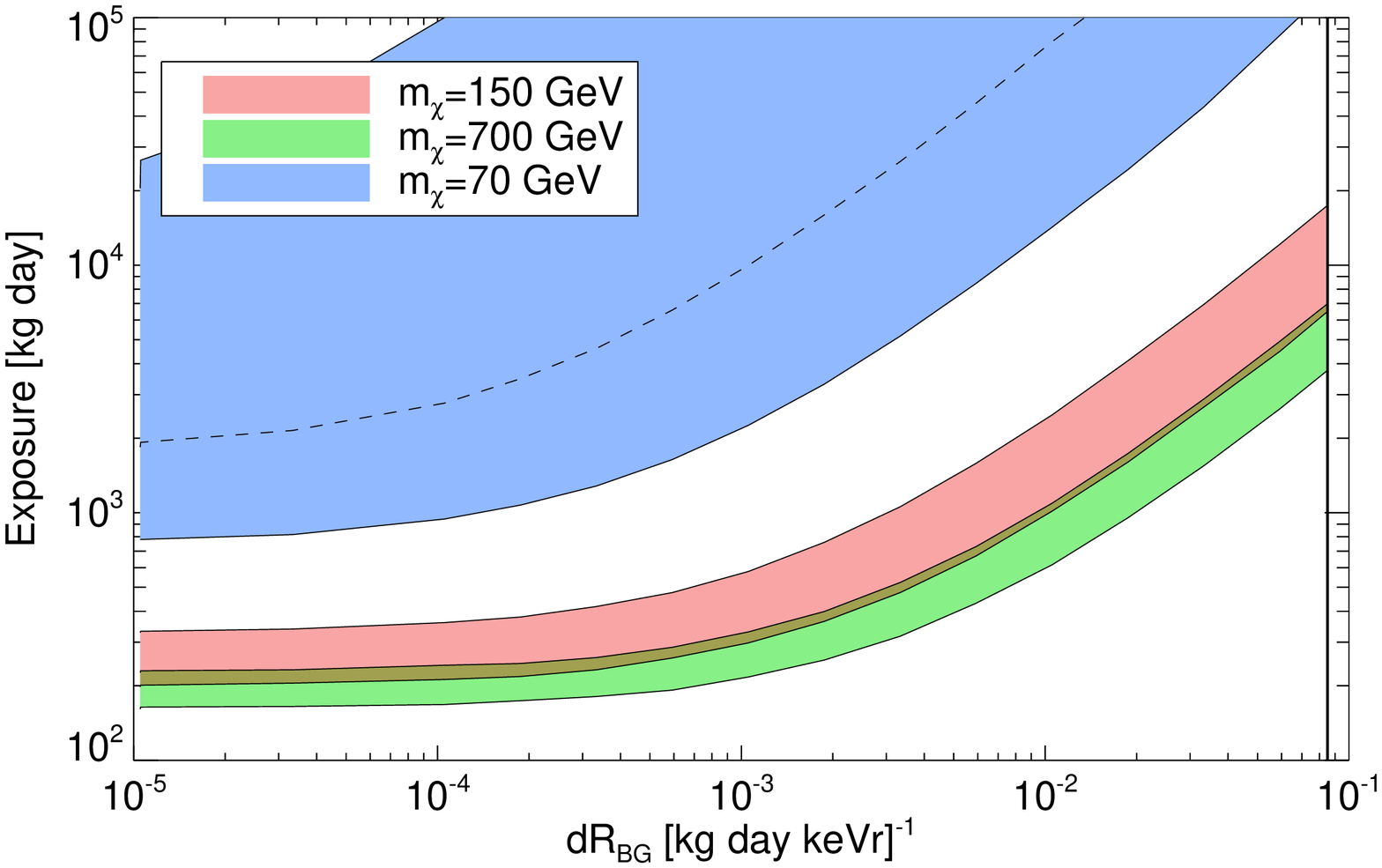} 
b) \includegraphics[width=.45\textwidth]{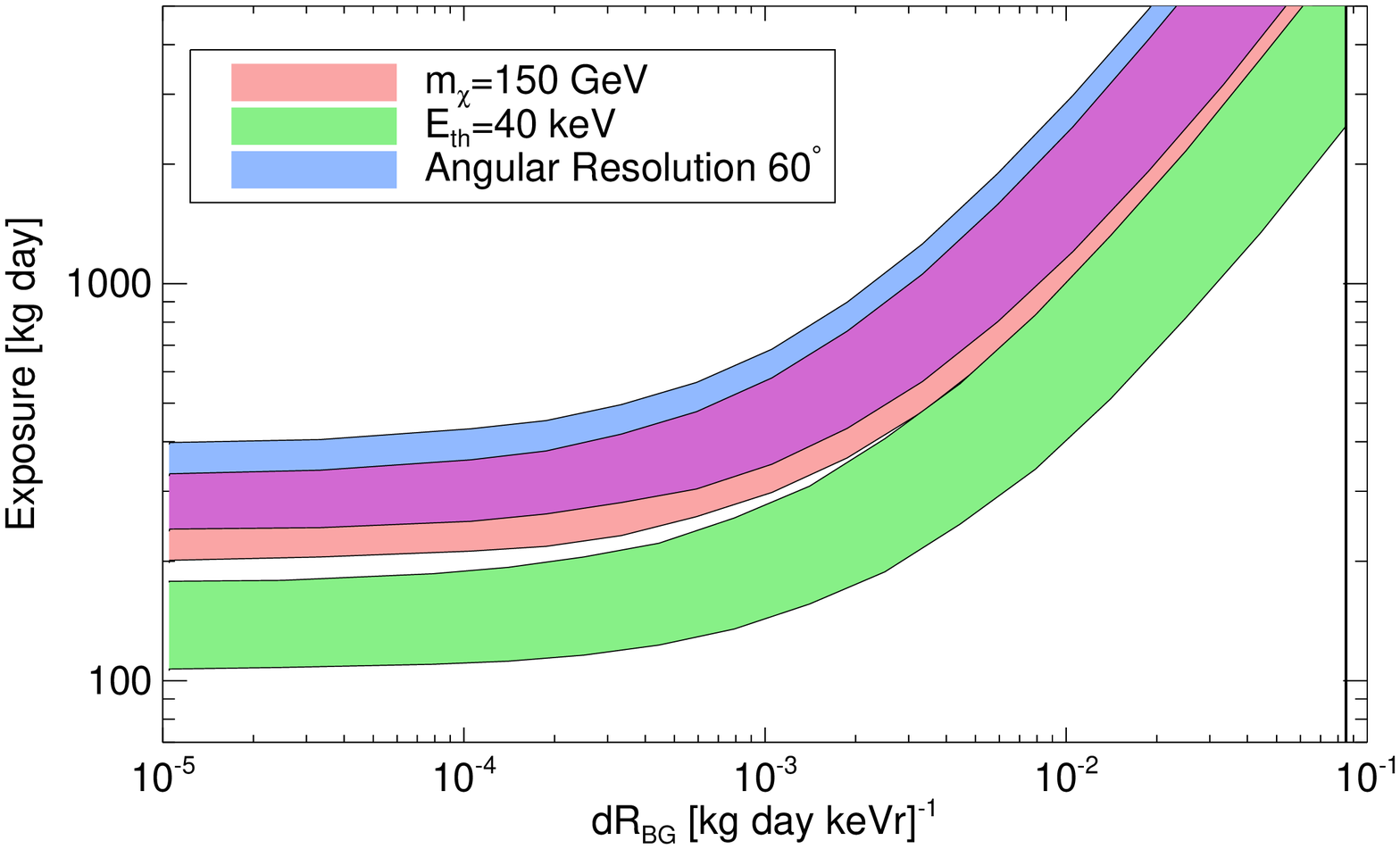}
\centering
\caption{Exposure to obtain a 5$\sigma$ measurement of $\langle \cos \gamma \rangle$ 90$\%$ of the time the experiment is conducted on Earth. The energy range of the experiment is $E_R \in [50,80]$ \kevr. $dR_{BG}$ is the background rate; the DAMA unmodulated background rate is indicated by the solid vertical line at 0.085. The bands shown give the exposures necessary as the rates modulate throughout a year. Since the annual modulation is asymmetric in summer and winter for low mass dark matter, the average exposure for $m_\chi = 70$ GeV is indicated by the dashed line. In (a) we show three mass benchmarks from Table~\ref{tab:benchmarks} and in (b) we show the effect of decreasing the angular resolution of the detector to 60 degrees and of lowering the energy threshold to 40 \kevr. (Darker regions indicate where the bands overlap.)}
\label{fig:expbench}
\end{figure*}

Following \cite{Chang:2008gd}, we use the truncated Maxwell-Boltzmann distribution in the rest of this paper: 
\begin{equation}
  f(\vec v) =  \frac{1}{n(v_0,v_{esc})} \exp \left( -\frac{\vec v^2}{v_0^2} \right) \Theta(v_{esc} - |\vec v|)
\end{equation}
where $n(v_0,v_{esc})$ normalizes $\int d^3v f$ to 1. The resulting spectrum is:
\begin{eqnarray}
	\frac{dR}{dE_R d\cos \gamma} = 
		\frac{\kappa F^2(E_R)}{n(v_0,v_{esc})} \pi v_0^2 \left( 
		\exp \left(-\frac{(\vec v_E \cdot \hat v_R+v_{min}(E_R,\delta))^2}{v_0^2} \right)  
		- \exp \left(-\frac{v_{esc}^2}{v_0^2} \right) \right) 
		\Theta(v_{esc} - |\vec v_E \cdot \hat v_R + v_{min}(E_R,\delta)| )
\end{eqnarray} The values we use for the astrophysical parameters are:
$v_0 = 220$ km/s, $v_E = 225$ km/s, $v_{esc} = 500-600$ km/s \cite{Smith:2006ym}, and
 $\rho_\chi$ = 0.3 GeV/cm$^3$. The normalized rate spectrum of several benchmark
models is shown in Fig.~\ref{fig:rates}.

\vspace{.5cm}

\twocolumngrid
\section{Sensitivity}

A robust detection of a directional modulation is possible with
surprisingly few events, and does not require use of the rate formulas
in the previous section.  In fact, a full likelihood analysis based on
the correct model is only a factor of $\sim2$ better than a simple
technique, and for a convincing detection, simpler is better.  In this
section we assume the detection gas has $A=127$ (for iodine; Xe with
$A=131$ would be similar) and focus on the energy range $E_R \in
[50,80]$ \kevr.

\subsection{Detectability \label{sec:cosgamma}}
For a model-independent statistic we follow \cite{Morgan:2004ys,Green:2006cb} and use the
dipole of the recoil direction, $\langle \cos \gamma \rangle$.  This is
motivated by the fact that the rate should depend only on $\cos \gamma$
and $E_R$, so the directional part can be expanded in spherical
harmonics.

Our detection criterion is a measurement of $\langle \cos \gamma
\rangle$ that is $5\sigma$ relative to the distribution of $\langle \cos
\gamma \rangle$ for the same number of randomly distributed events. For
a fixed exposure, we generate many random sets of model data
(constrained by the DAMA benchmarks in Table~\ref{tab:benchmarks}), and
then demand that 90$\%$ of the time the result is $5\sigma$
from the null hypothesis.  The background is modeled as uniform in
recoil energy and angle.  We assume the detector has an angular resolution of 15
degrees.

In Fig.~\ref{fig:expbench}(a) we show the exposures necessary for such
conditions, as a function of the background rate, for a few benchmark
models. At zero background, roughly 18 events are needed for all
benchmark models, on average.  Fig.~\ref{fig:expbench}(b) shows the effect
of decreasing the angular resolution to 60 degrees and lowering the
energy threshold of the experiment to $E_R = 40$ \kevr. Because of the
sharp angular profile of the recoil spectrum, a poor angular resolution
does not significantly reduce the possibility of a detection. However,
achieving an energy threshold of 30-40 \kevr\ dramatically lowers the
necessary exposures because the peak of the recoil spectrum occurs at
30-40 \kevr\ and falls off exponentially.

\subsection{Parameter Estimation}

We also perform a likelihood analysis as a measure of sensitivity of the experiment to
the parameters of the model, assuming perfect energy and angular resolution. From our analysis in the previous section, we expect this assumption does not affect the results significantly. (See also \cite{Copi:2005ya}, which shows the sensitivity dependence on angular resolution.)
The parameters we consider here are 
$m_\chi, \delta, $ and $\sigma_n$, which we denote together simply by $p$. 
Define
\begin{equation}
  \mu(x; p) \equiv \frac{dR}{dE_R d\cos \gamma}(x; p) + dR_{BG}/2,
\end{equation} which is the rate (cpd/kg/keVr per cos$\gamma$) at a given recoil energy and angle (denoted together by $x$)
for parameters $p$. We assume the background rate, $dR_{BG}$, in units of 
cpd$/$\kevr$/$kg, is known. 

The likelihood is the probability of parameters $p$ given the events $\{x_i\}$.
Given events $\{x_i\}$, bin the events such that in each bin there is only
0 or 1 event and label the bins with one count by $\{X_\alpha\}$ and 
the empty bins by $\{X_\beta\}$. The expected number of counts in a bin is
\begin{equation}
	E(X;p) = {\cal E} \mu(x;p) \Delta x
\end{equation} where ${\cal E}$ is the exposure.  Then the (log) likelihood is
\begin{equation}
\ln{\cal L}_{tot}(p) = \sum_\alpha  \ln  \left(e^{-E(X_\alpha;p)}E(X_\alpha;p)\right)
						+ \sum_\beta \ln  e^{-E(X_\beta;p)}
\label{eq:lnldiscrete}
\end{equation} which is the log of the Poisson probability of obtaining 0
or 1 event in each bin. 
To find the expected average $\ln{\cal L}_{tot}$ for a given exposure ${\cal E}$ 
and true parameters $p_0$, we compute
\begin{equation}
\ln{\cal L}_{tot}(p) ={\cal E} \int dx {\bigg(} \mu(x;p_0) \ln \mu(x;p) - \mu(x;p) {\bigg)} 
\end{equation} which is the continuum, noiseless limit of Eq.~\ref{eq:lnldiscrete}.
Since we can only compare differences in log likelihood, in this equation
we have subtracted an arbitrary constant in $p$ which takes care of the units
in $\ln \mu(x,p)$.

\begin{figure*}[htpb]
\centering
\includegraphics[width=.3\textwidth]{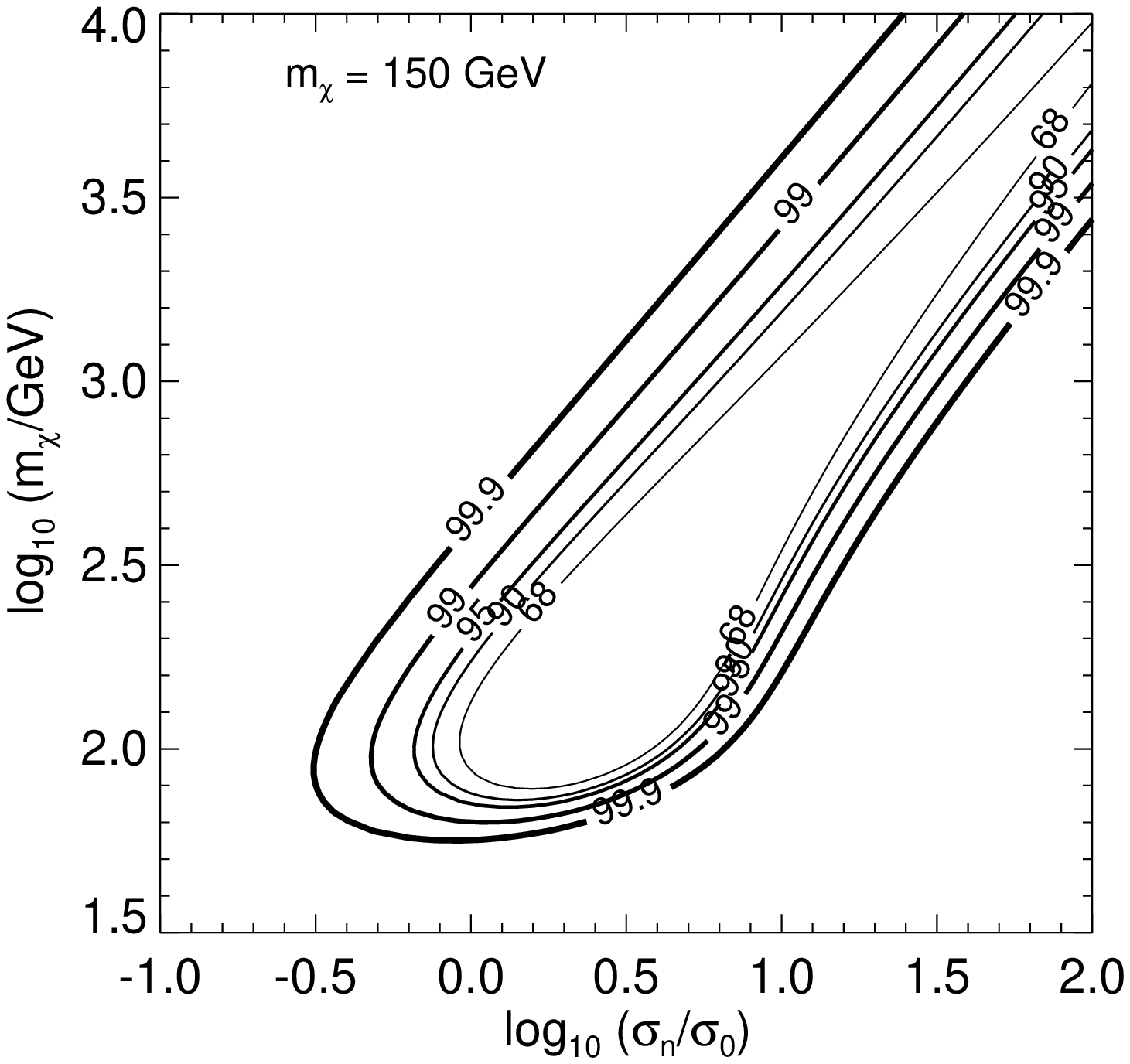}\hskip 0.2in
  \includegraphics[width=.3\textwidth]{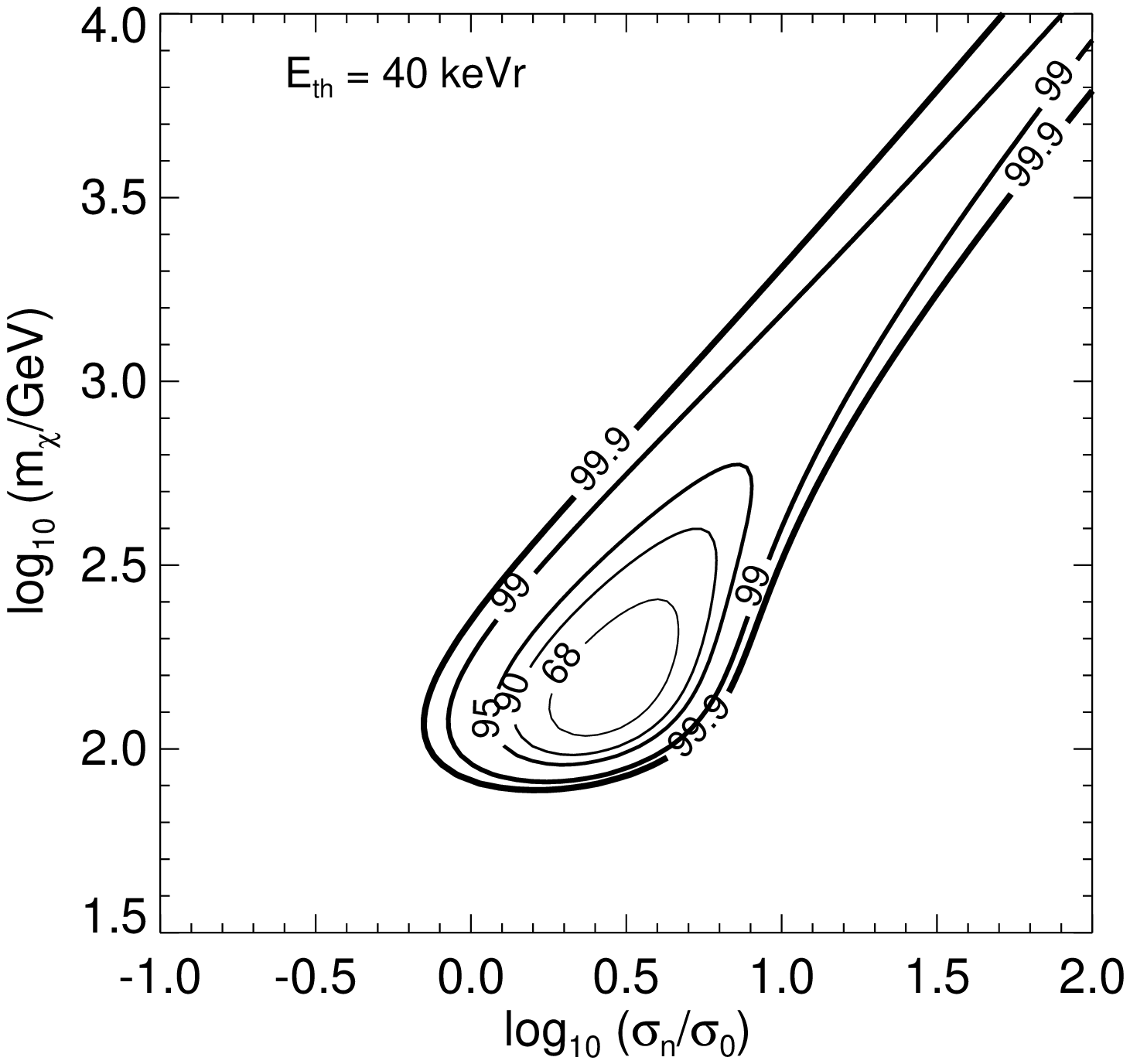}\hskip 0.2in
  \includegraphics[width=.3\textwidth]{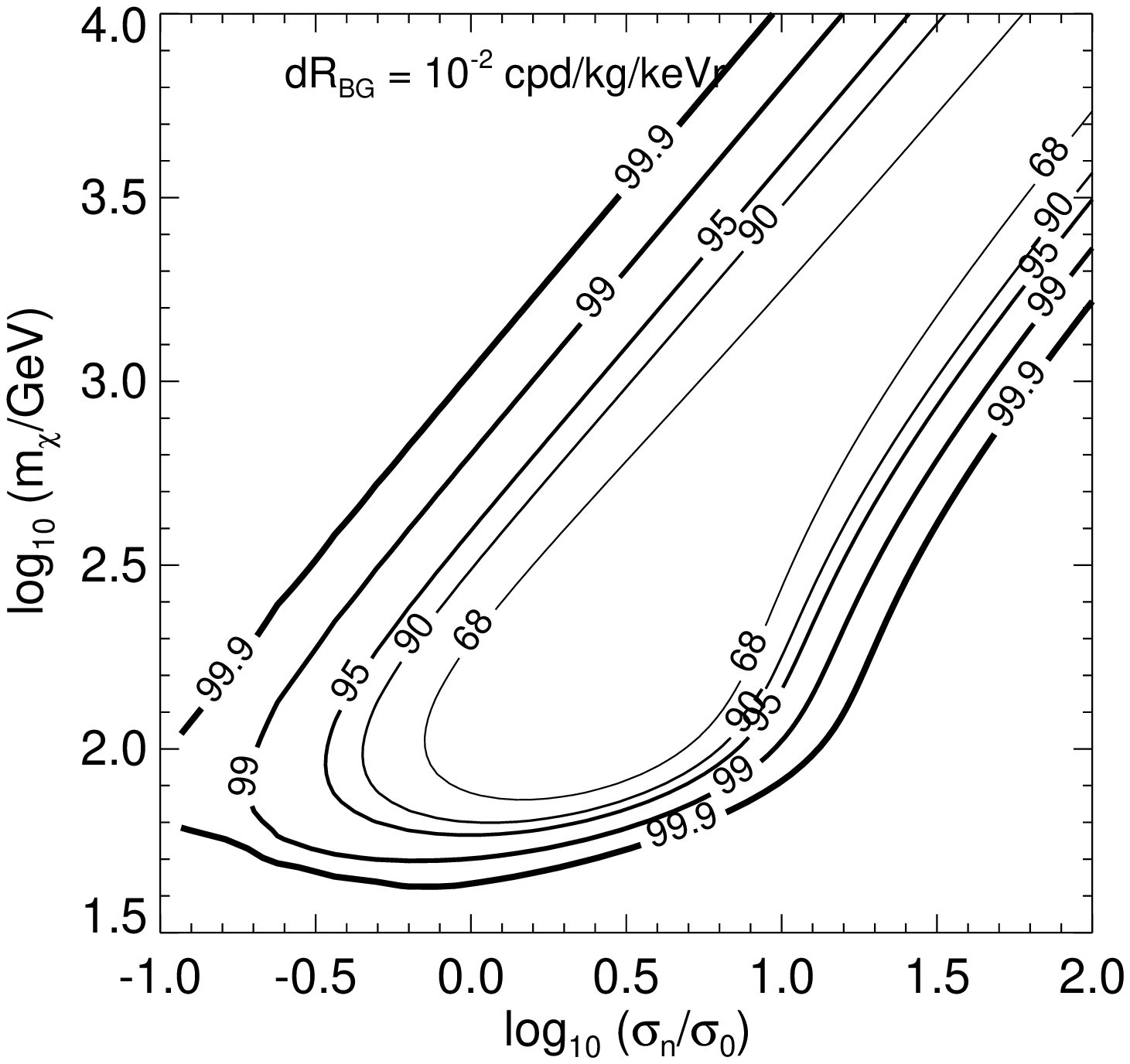}\\
\includegraphics[width=.3\textwidth]{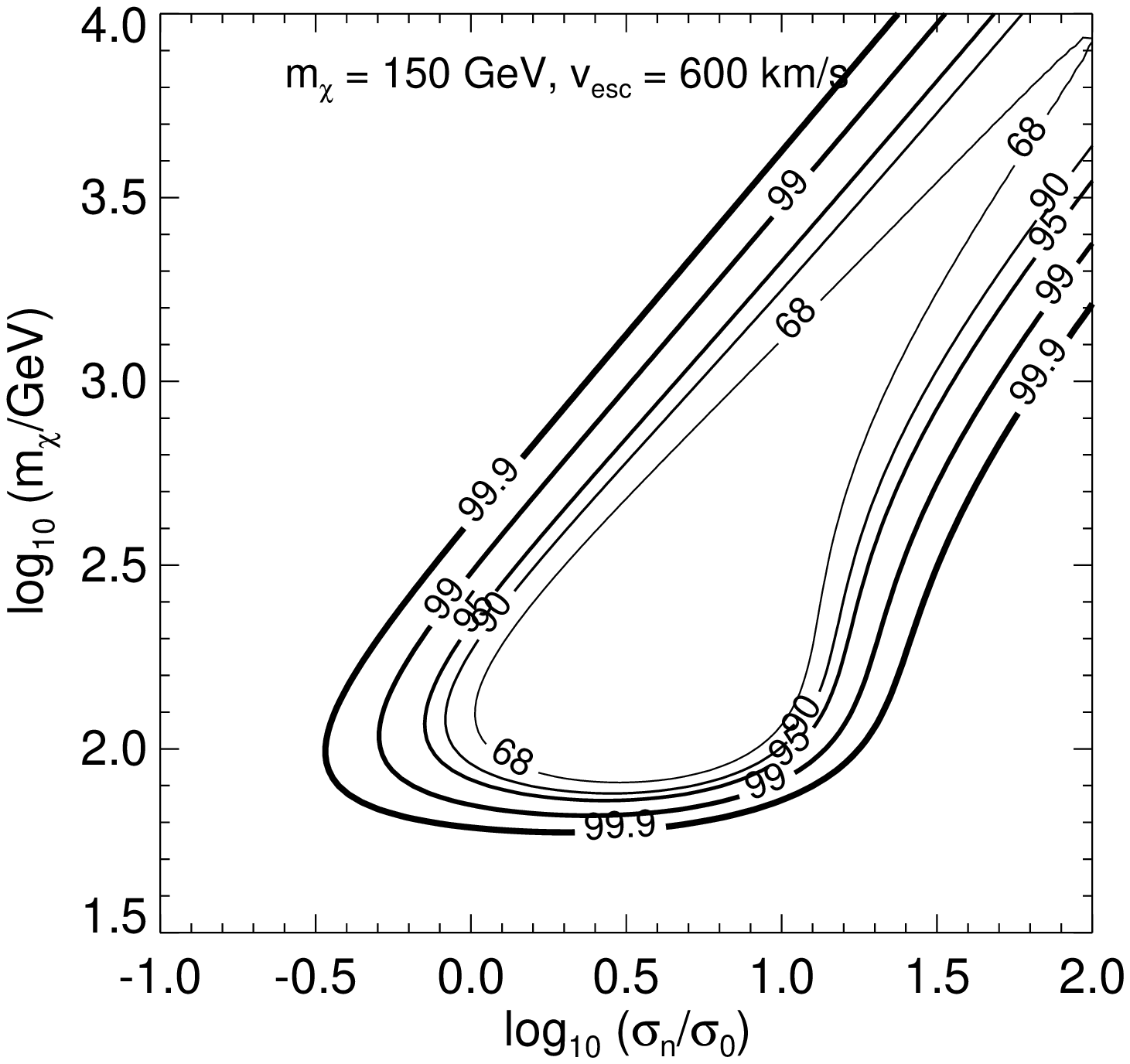}\hskip 0.2in
  \includegraphics[width=.3\textwidth]{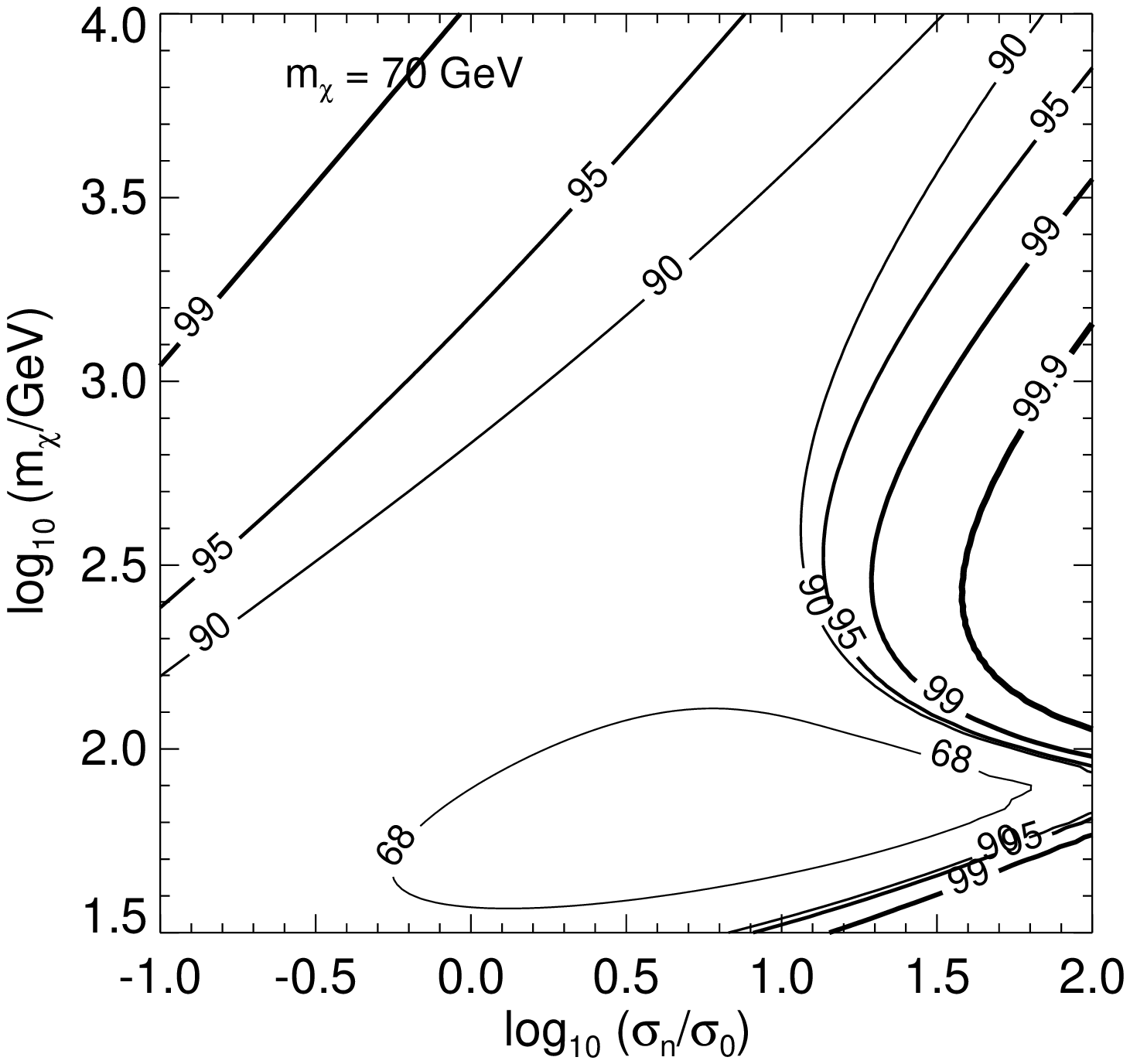}\hskip 0.2in
  \includegraphics[width=.3\textwidth]{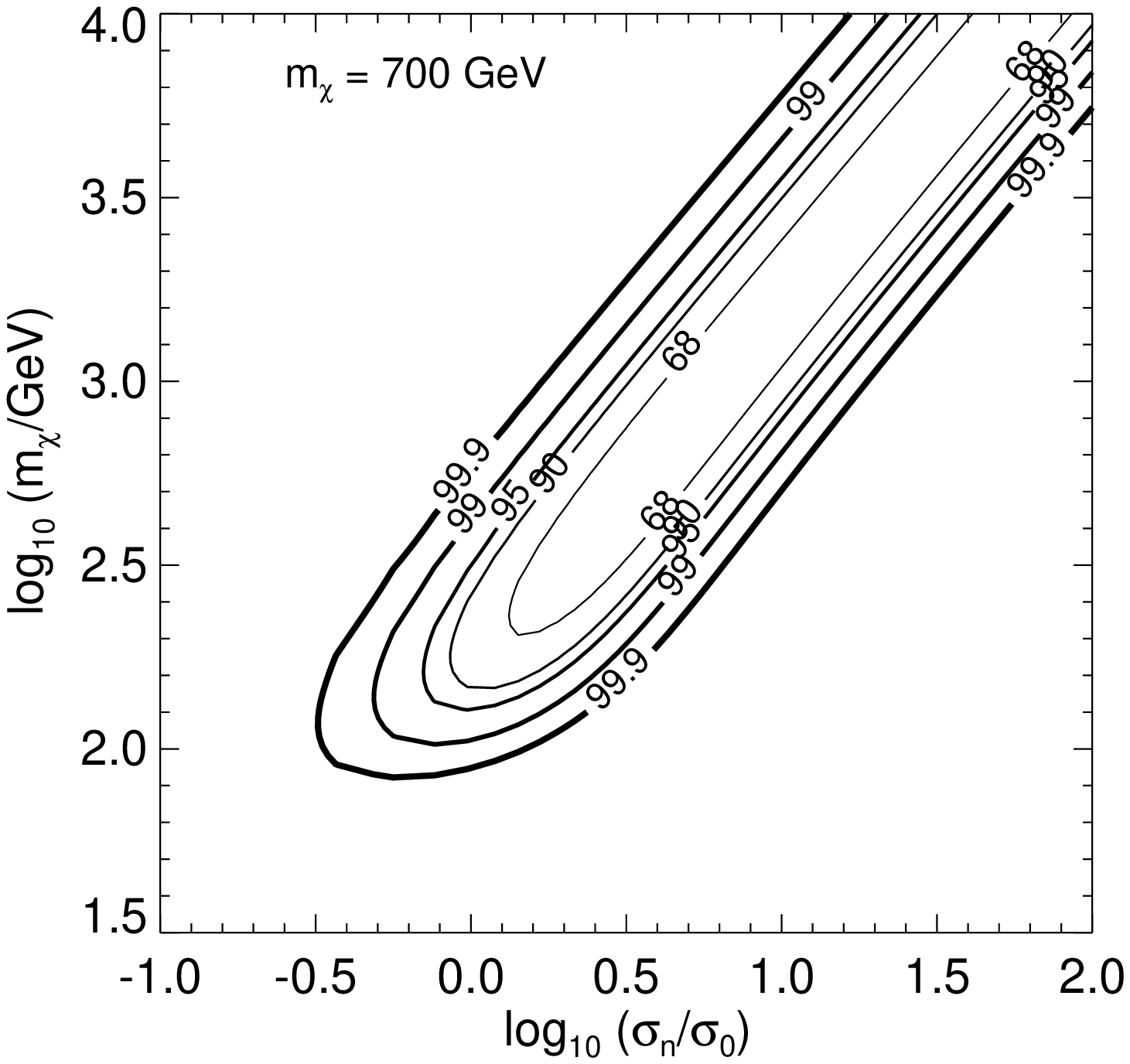}\\
\centering
\caption{Confidence levels for determining $m_\chi$ and $\sigma_n$, where $\delta$ is unknown, with an exposure of 1000 kg $\cdot$ day. $\sigma_0 = 10^{-40} cm^2$.}
\label{fig:lnl_msigma}
\end{figure*}

\begin{figure*}[htpb]
\centering
\includegraphics[width=.3\textwidth]{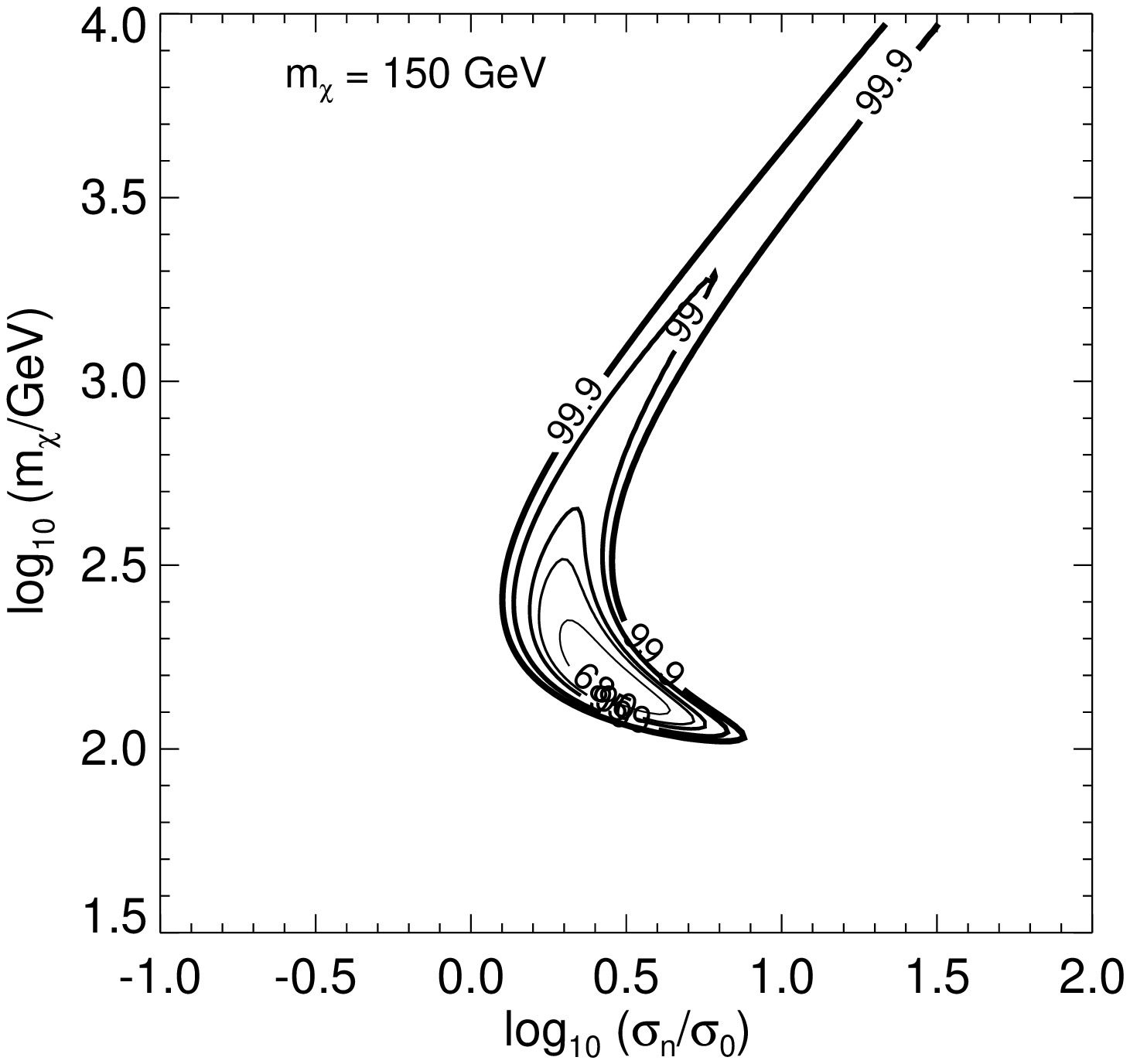}\hskip 0.2in
  \includegraphics[width=.3\textwidth]{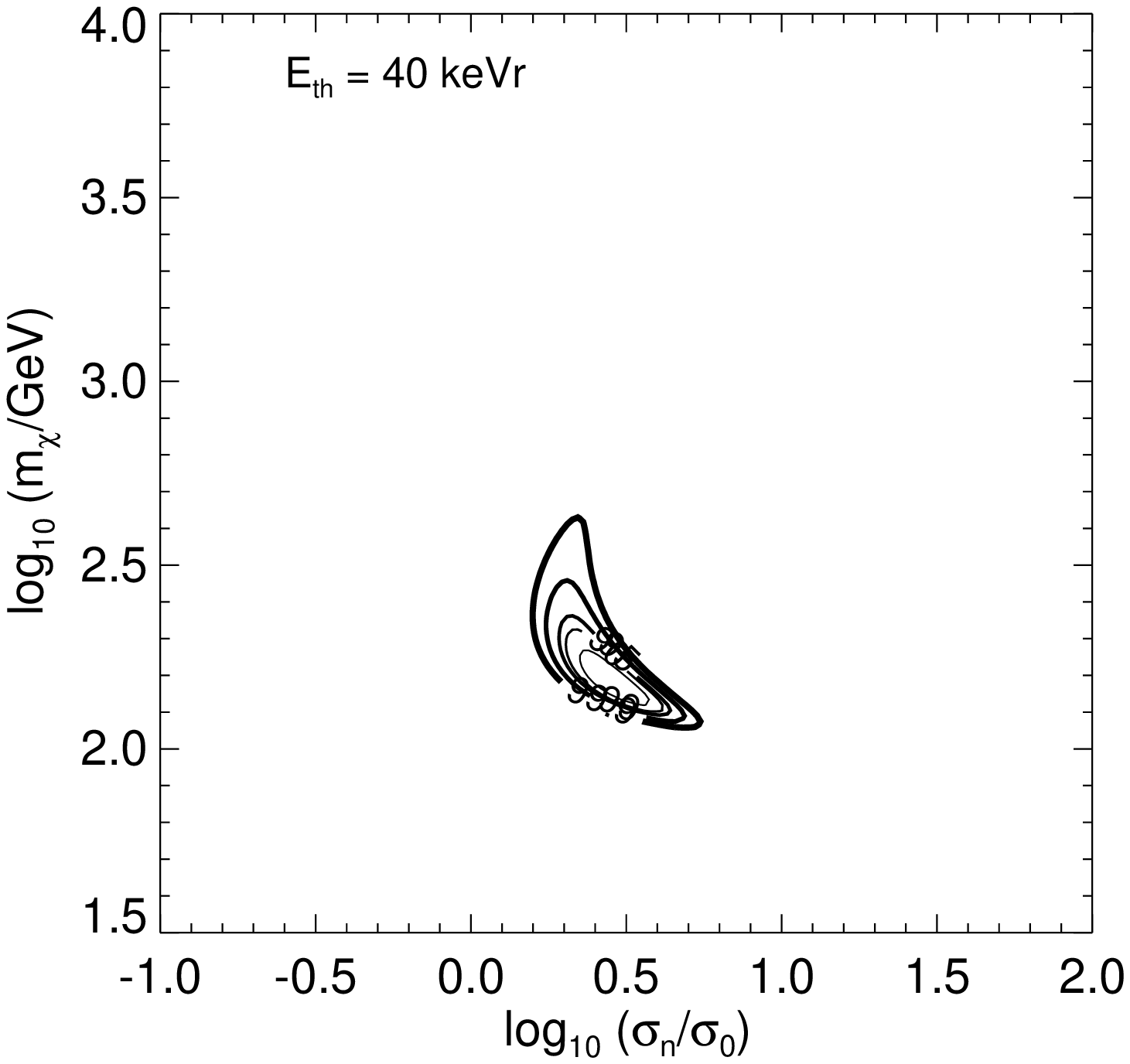}\hskip 0.2in
  \includegraphics[width=.3\textwidth]{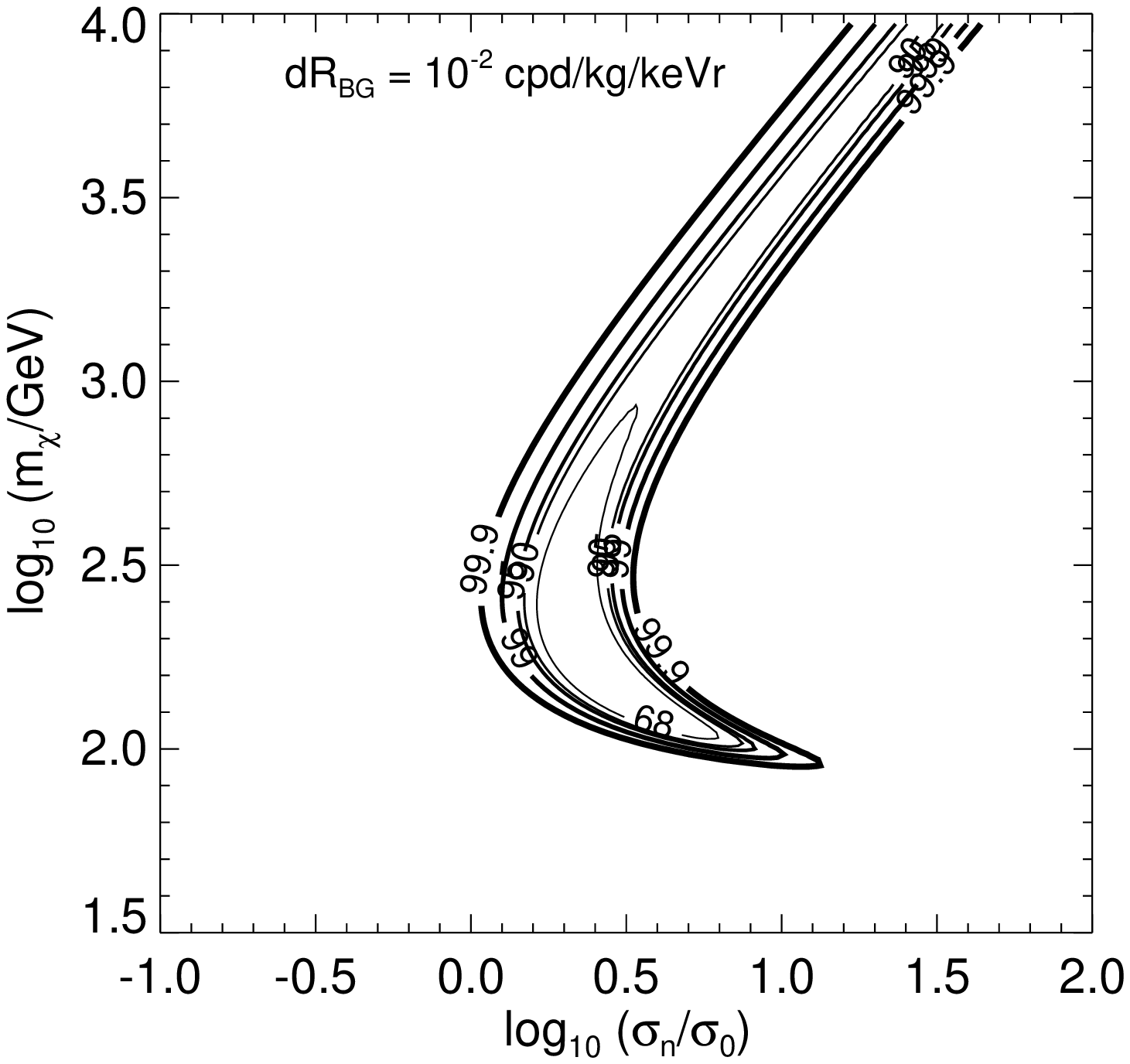}\\
\includegraphics[width=.3\textwidth]{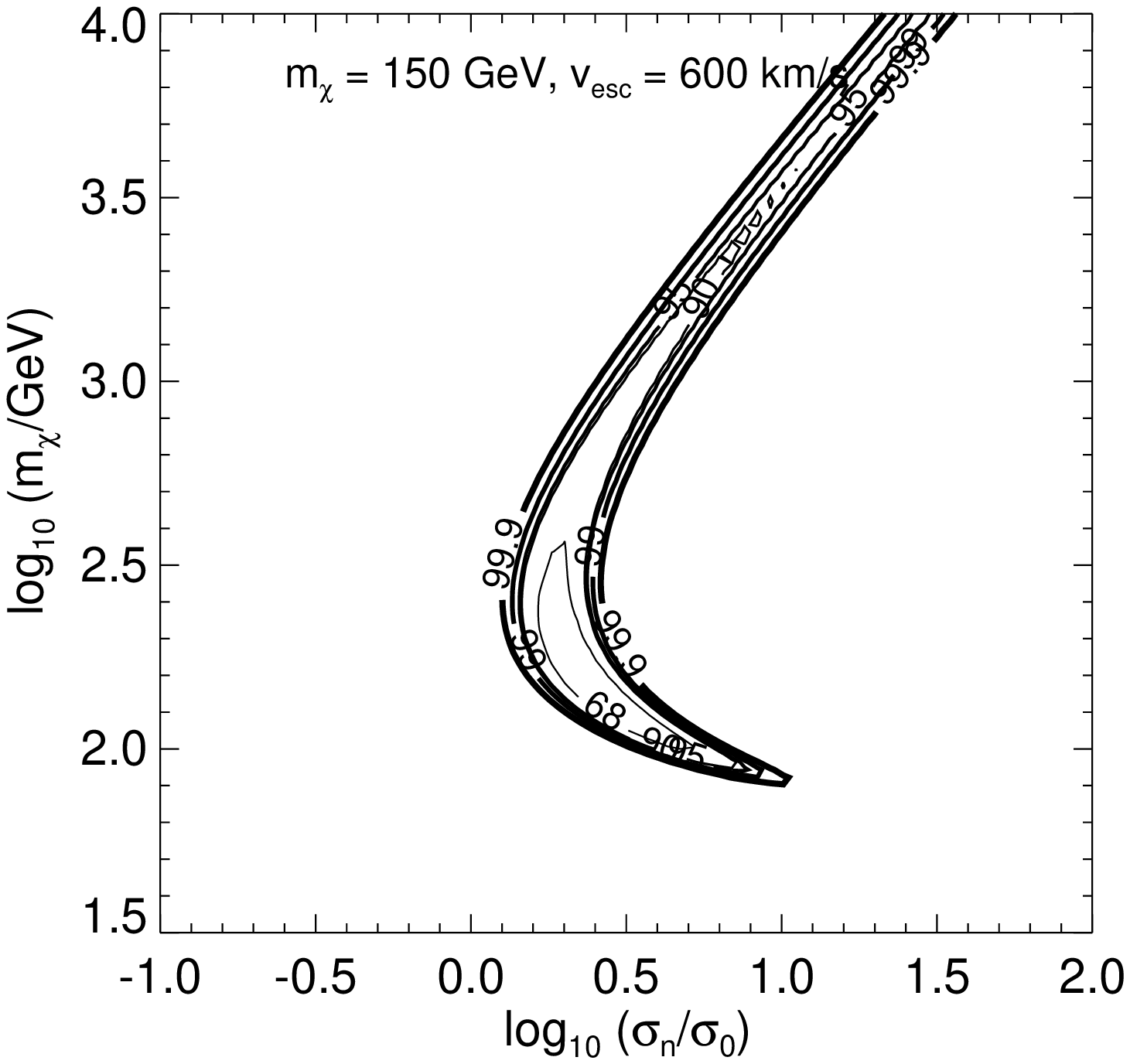}\hskip 0.2in
  \includegraphics[width=.3\textwidth]{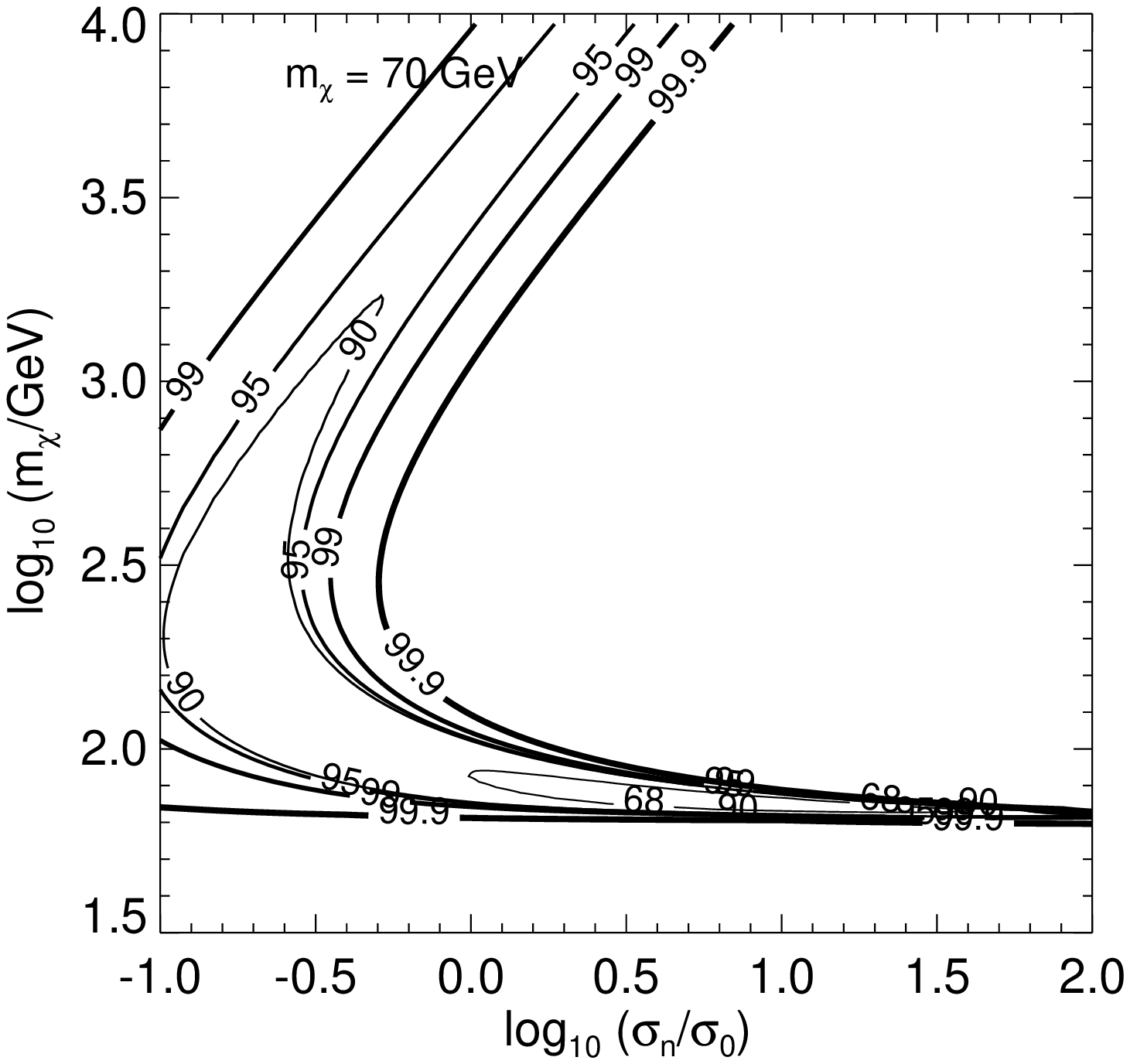}\hskip 0.2in
  \includegraphics[width=.3\textwidth]{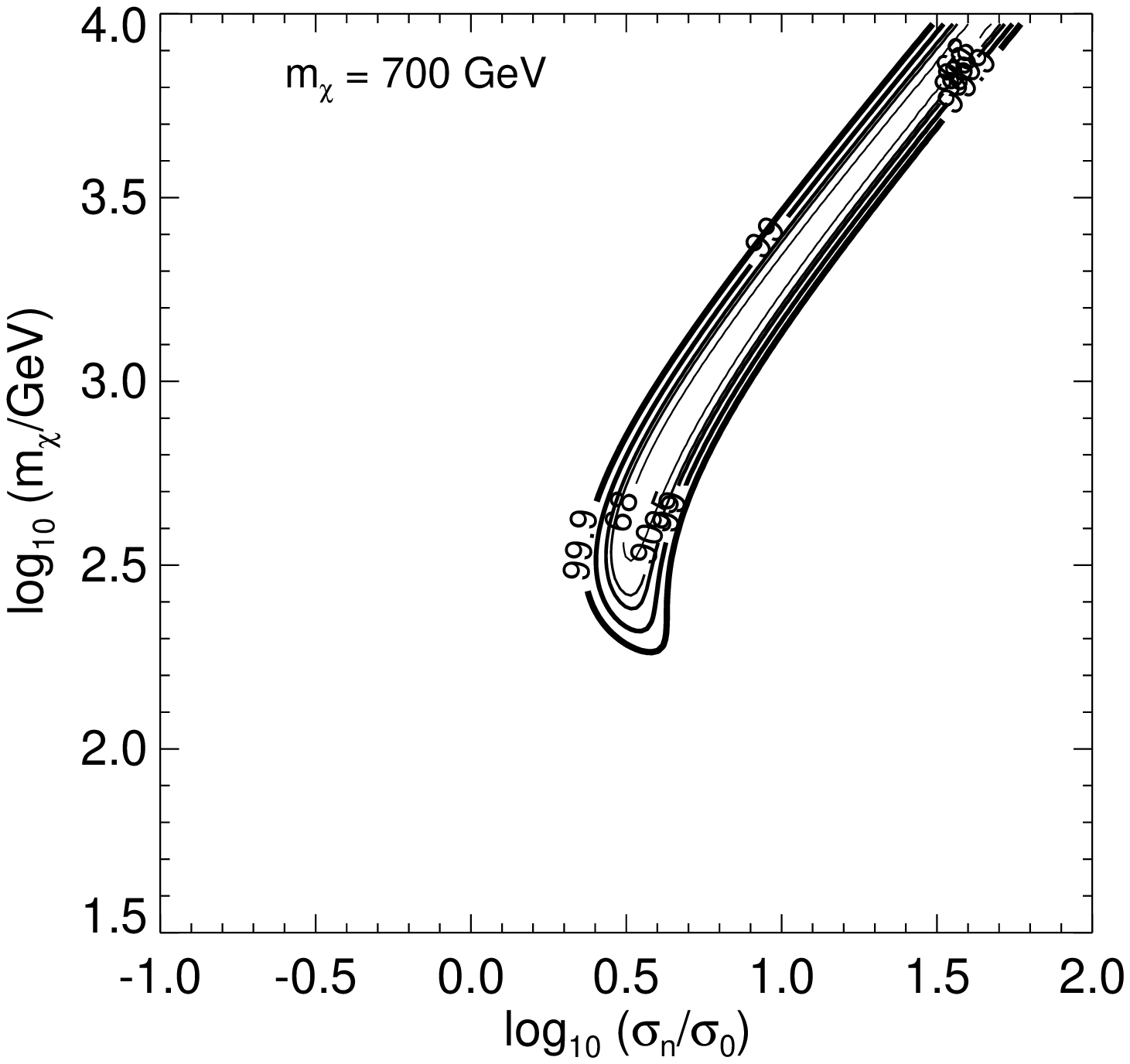}\\
\centering
\caption{Confidence levels for determining $m_\chi$ and $\sigma_n$, where $\delta$ is {\it known} with an exposure of 1000 kg $\cdot$ day. $\sigma_0 = 10^{-40} cm^2$.}
\label{fig:lnl_fixedDelta}
\end{figure*}

\begin{figure*}[htpb]
\centering
\includegraphics[width=.3\textwidth]{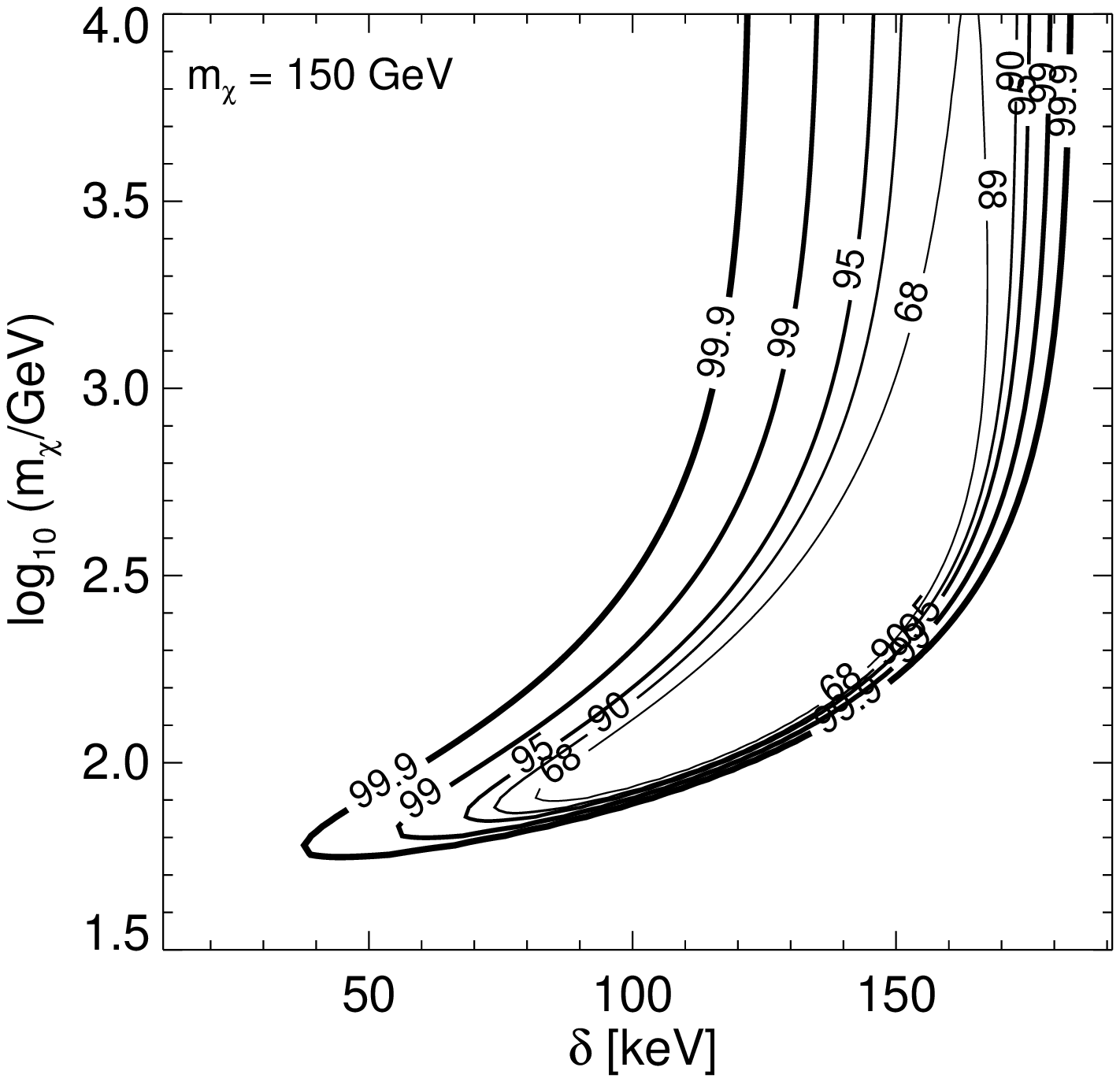}\hskip 0.2in
  \includegraphics[width=.3\textwidth]{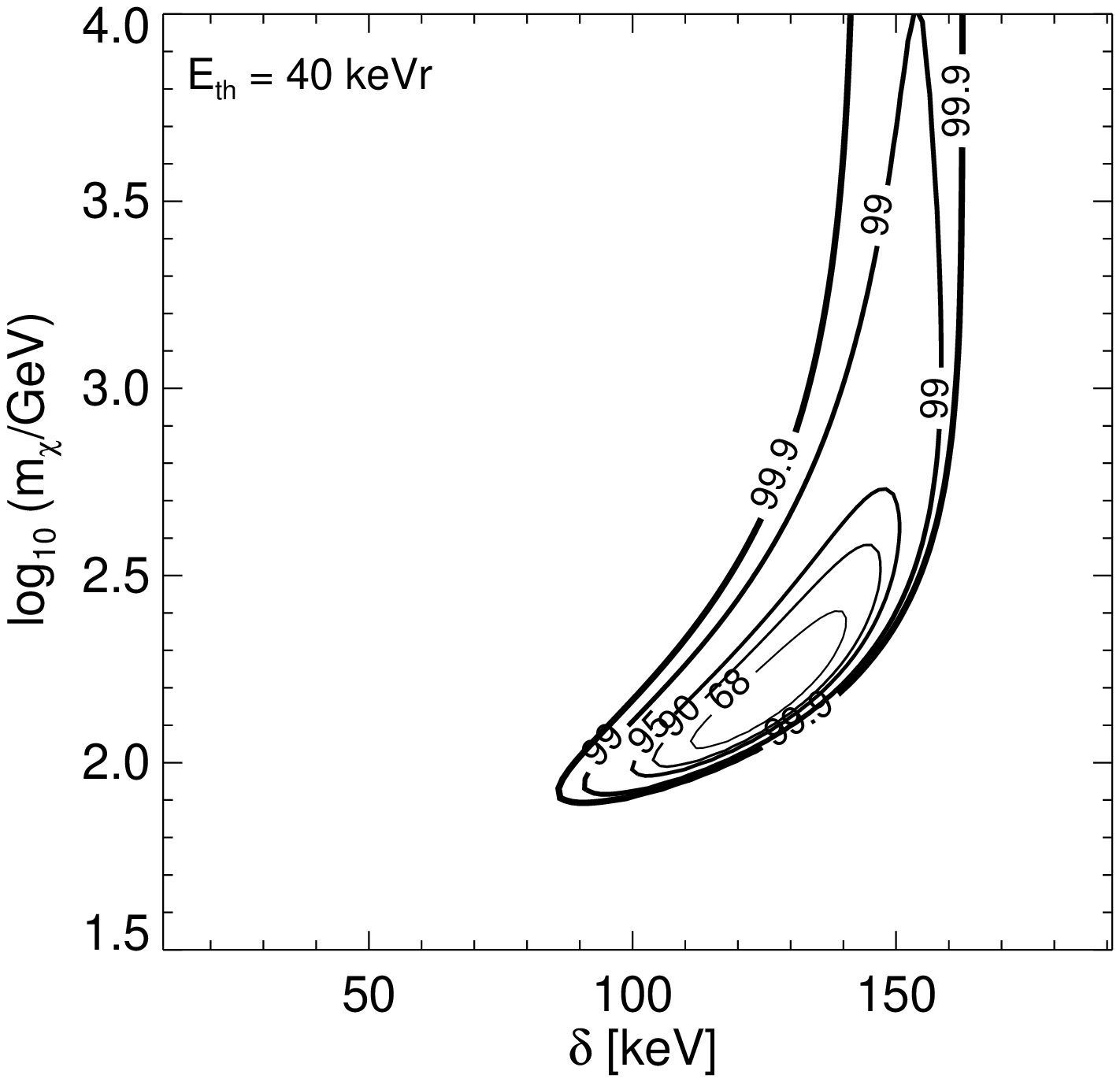}\hskip 0.2in
  \includegraphics[width=.3\textwidth]{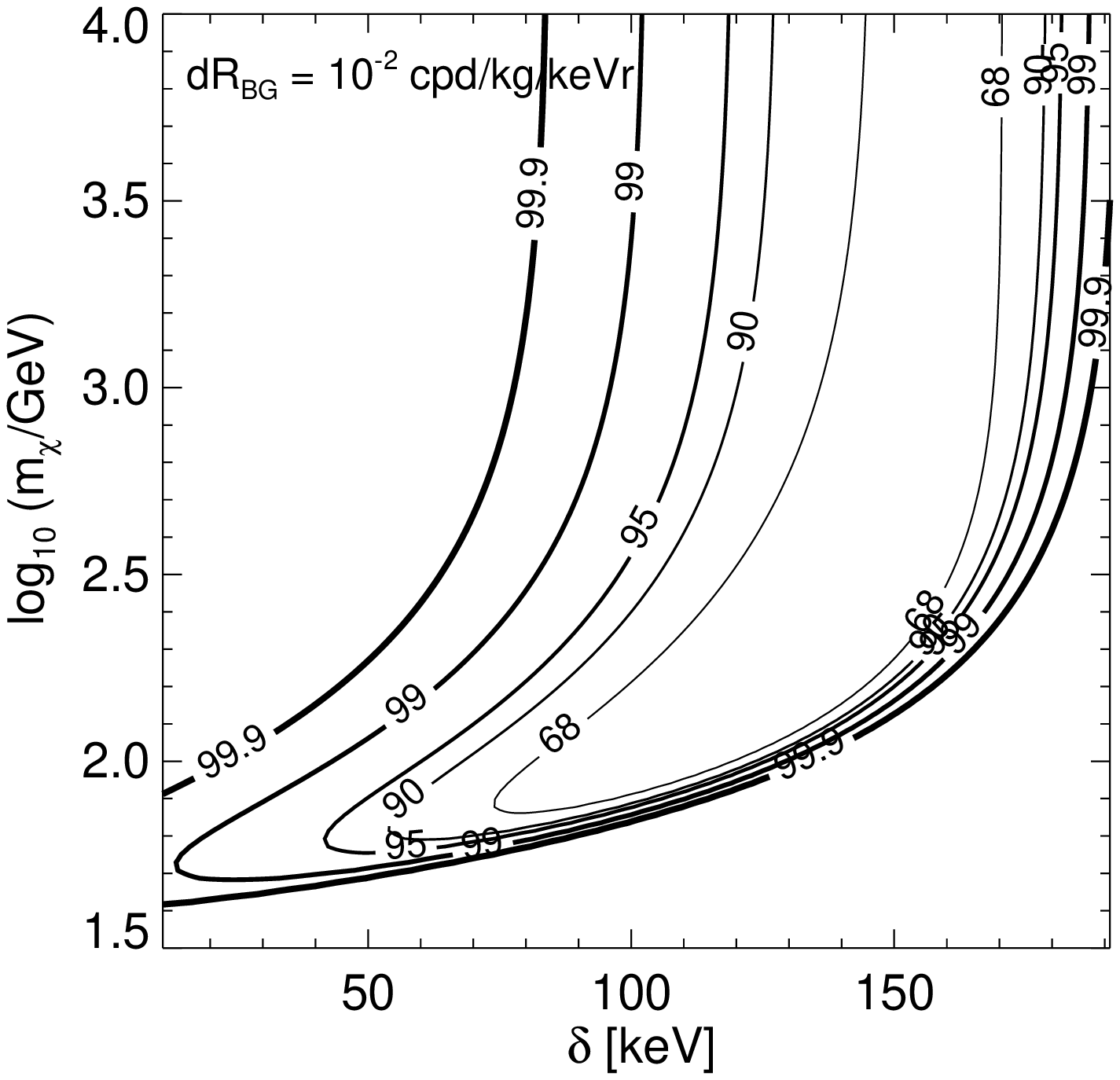}\\
\includegraphics[width=.3\textwidth]{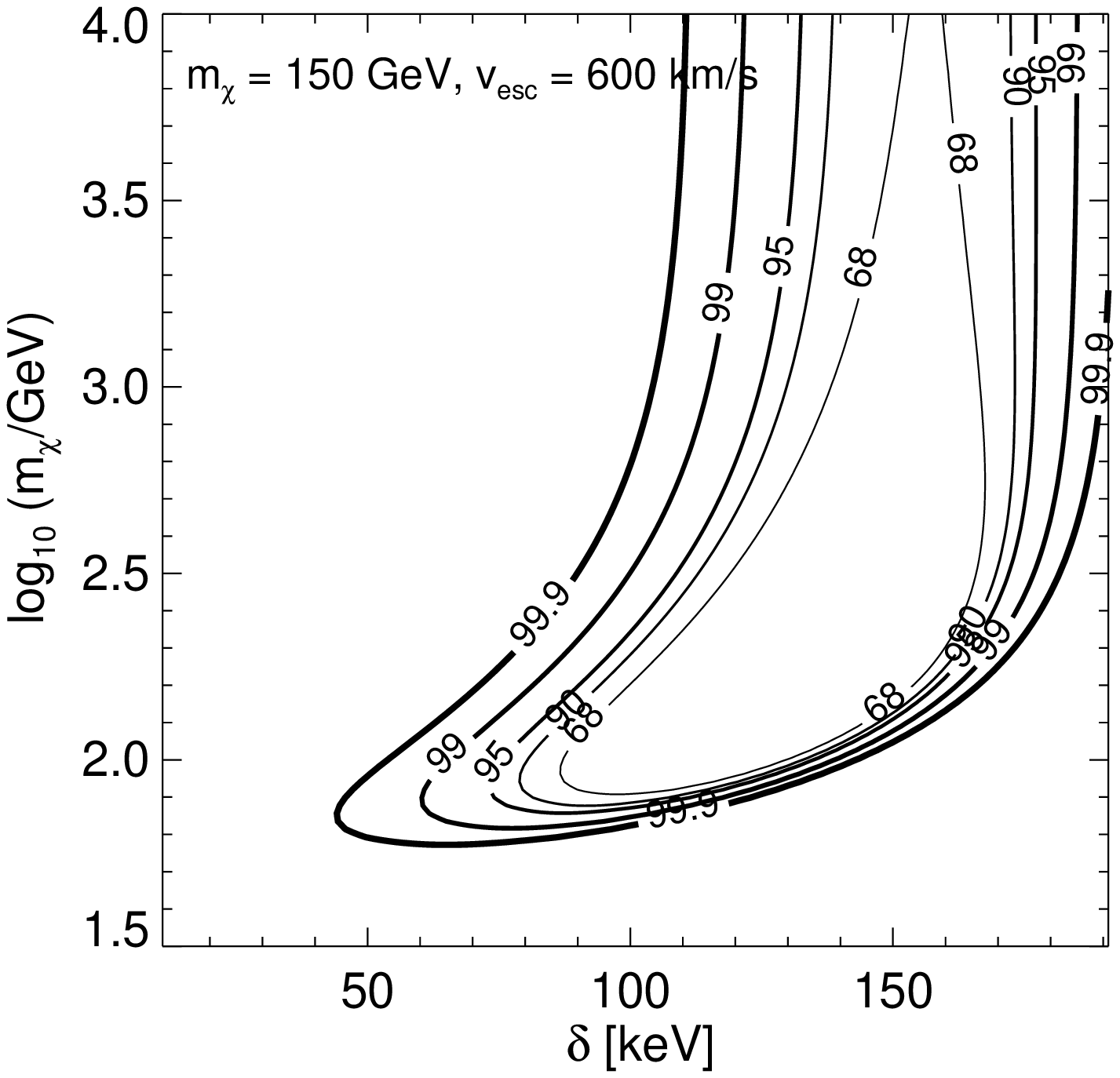}\hskip 0.2in
  \includegraphics[width=.3\textwidth]{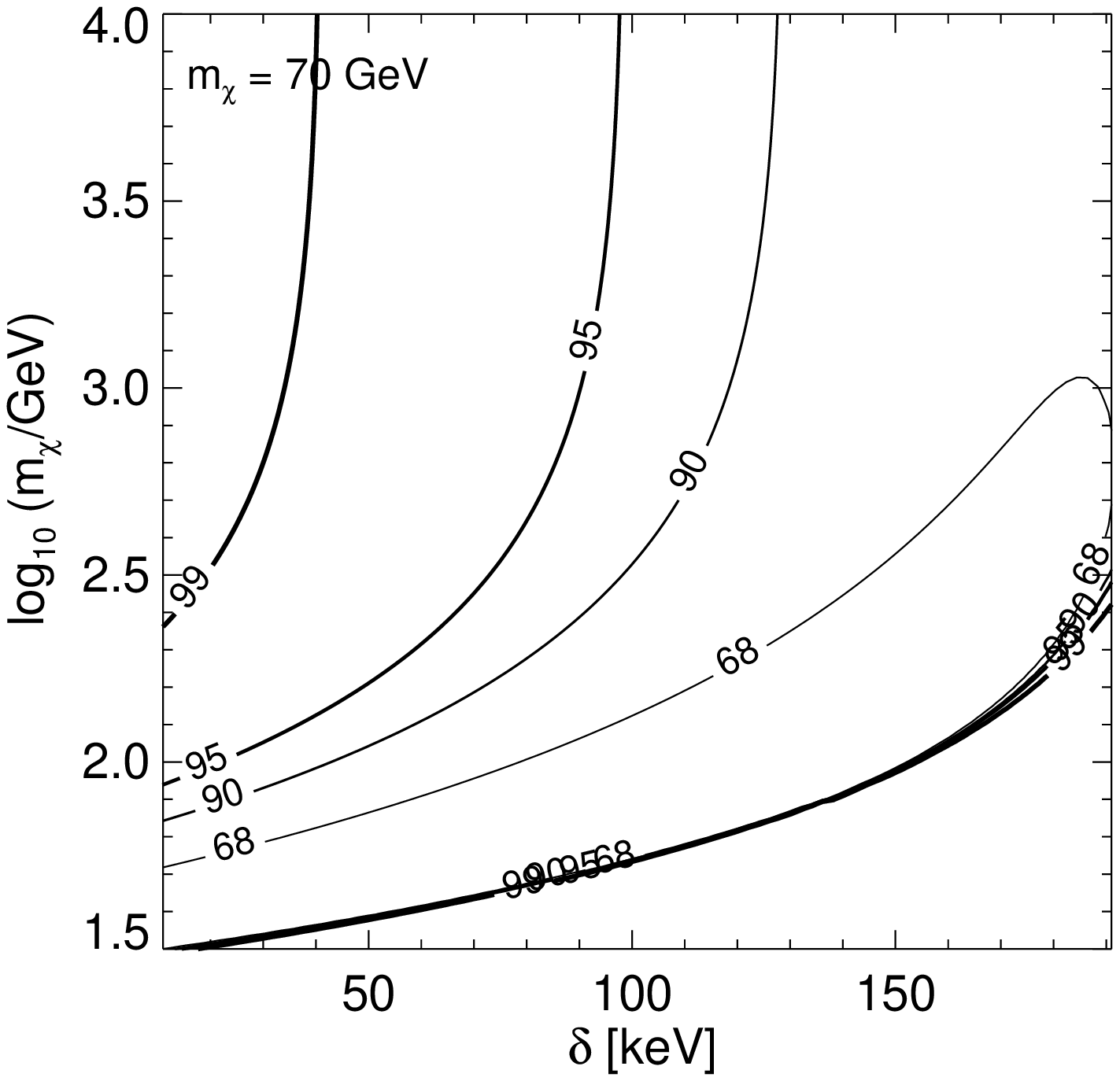}\hskip 0.2in
  \includegraphics[width=.3\textwidth]{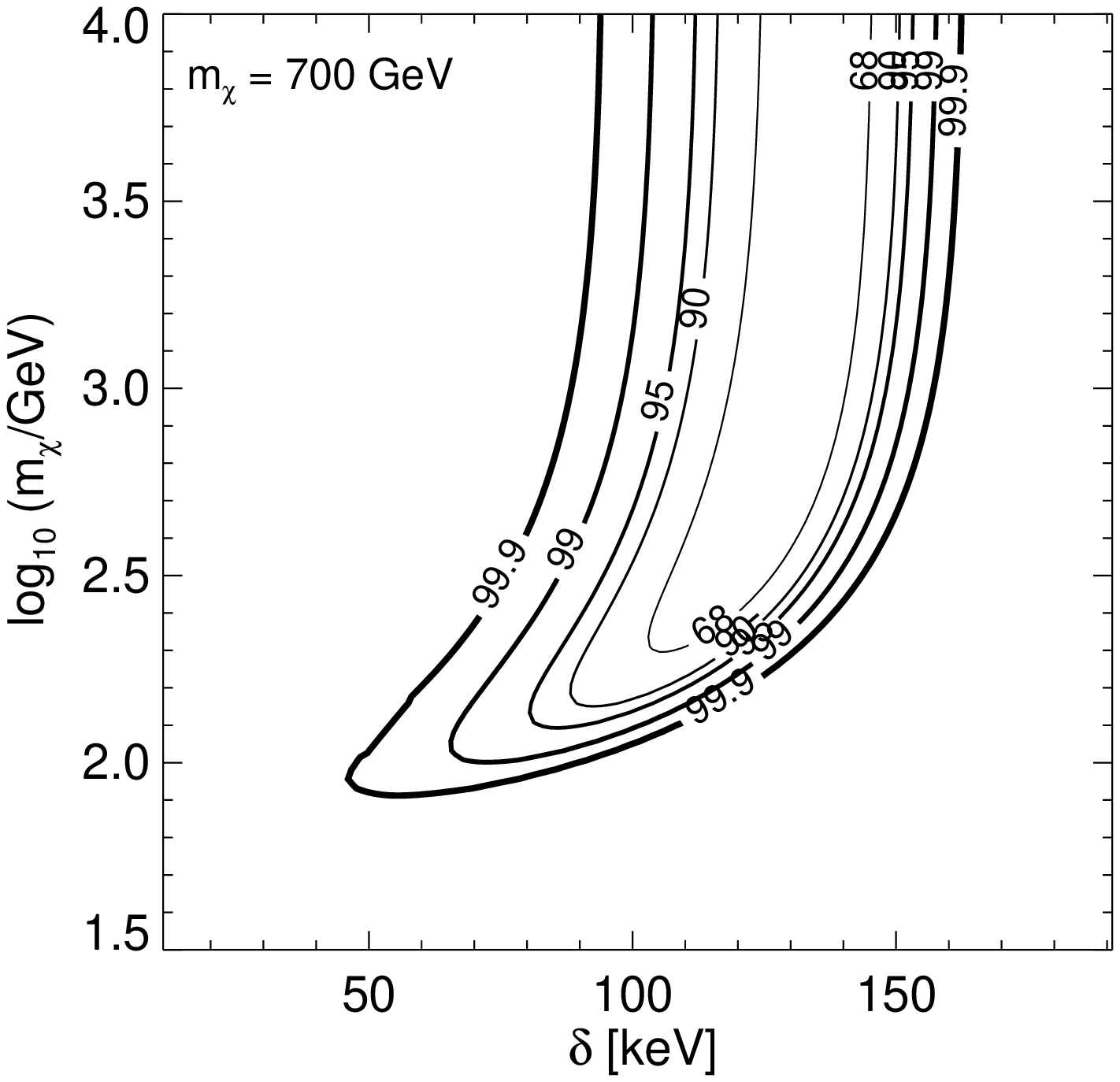}\\
\centering
\caption{Confidence levels for determining $m_\chi$ and $\delta$, where $\sigma_n$ is unknown, with an exposure of 1000 kg $\cdot$ day.}
\label{fig:lnl_mdelta}
\end{figure*}

\begin{figure*}[htpb]
\centering
\includegraphics[width=.3\textwidth]{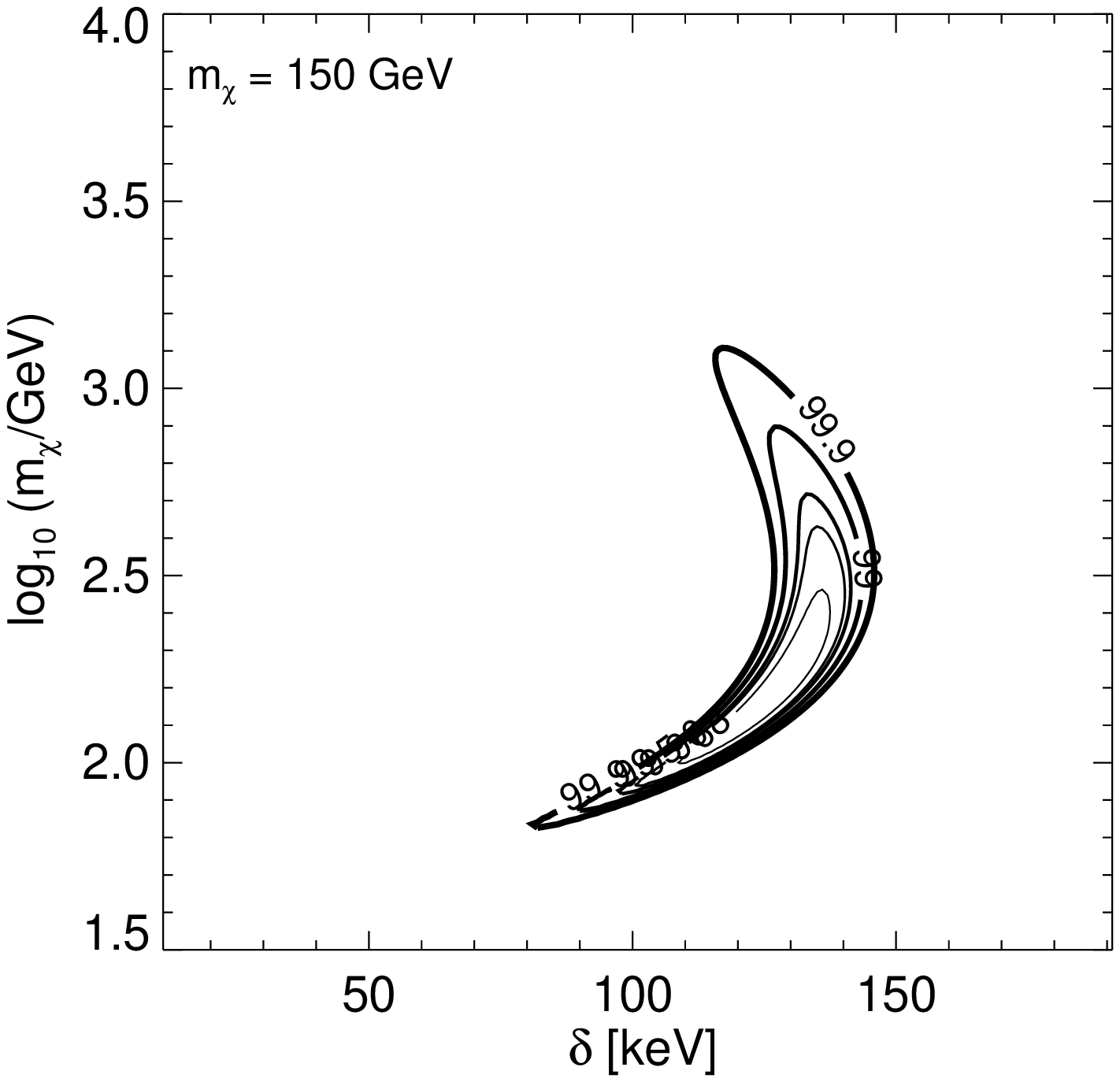}\hskip 0.2in
  \includegraphics[width=.3\textwidth]{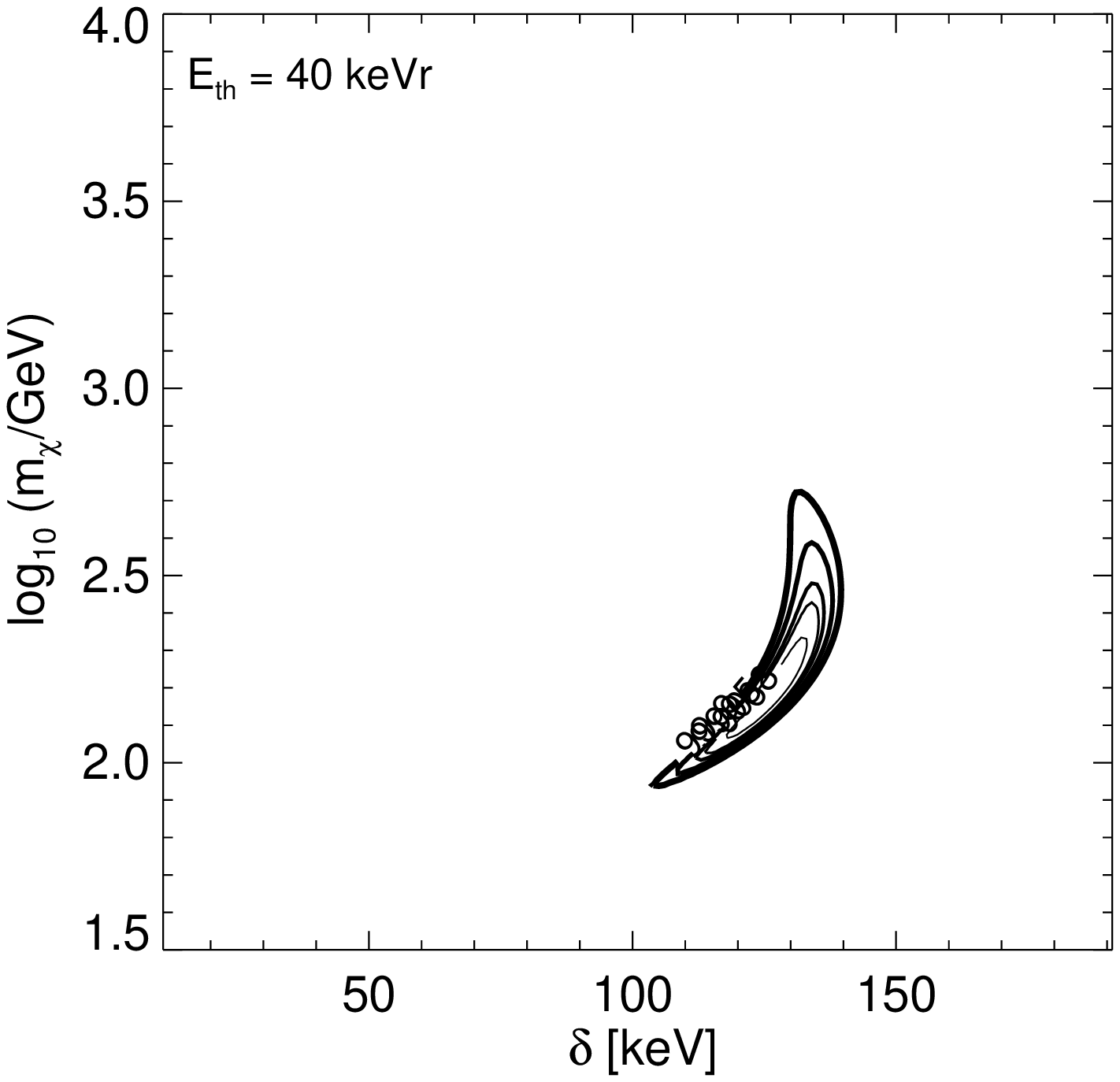}\hskip 0.2in
  \includegraphics[width=.3\textwidth]{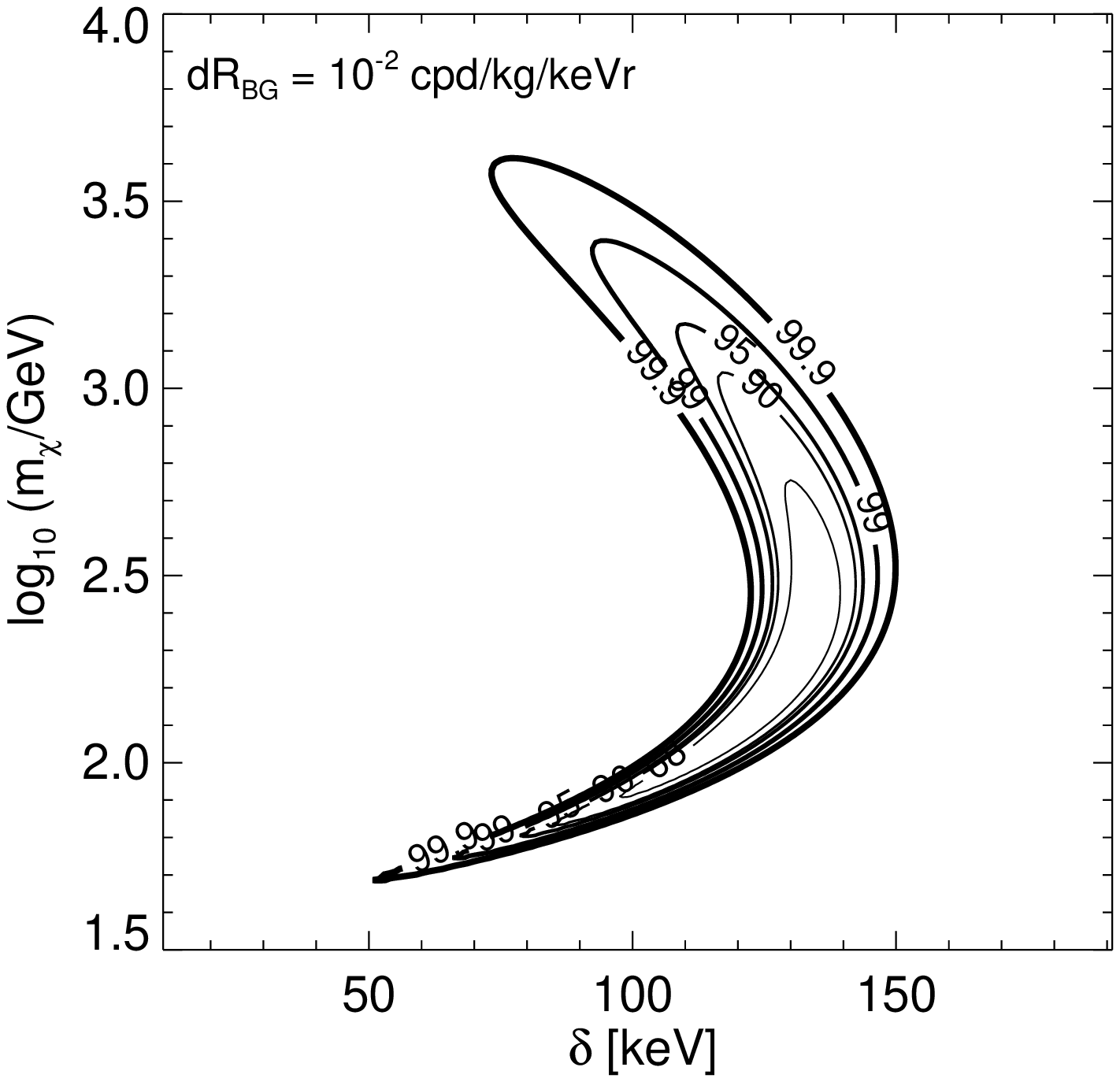}\\
\includegraphics[width=.3\textwidth]{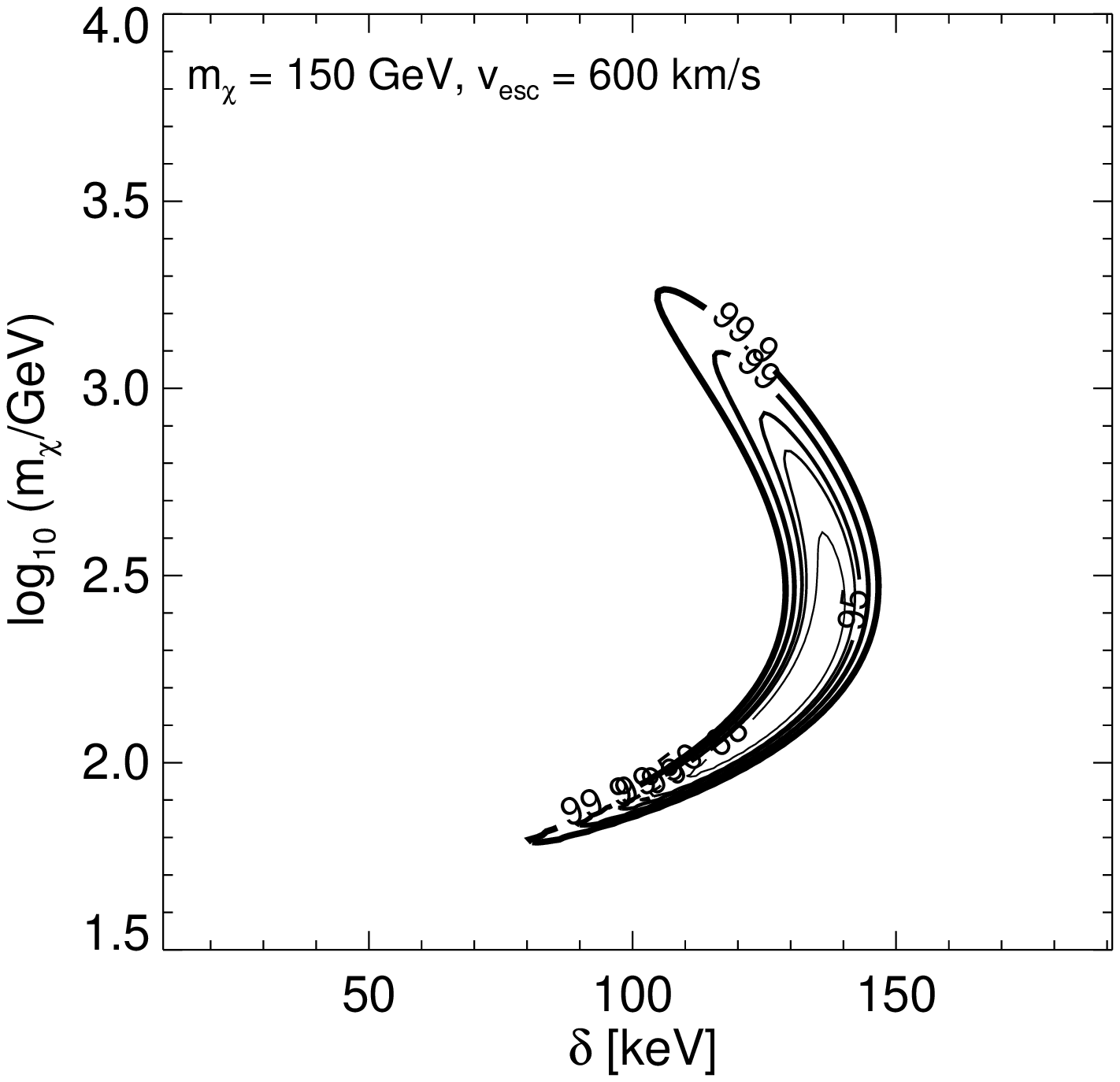}\hskip 0.2in
  \includegraphics[width=.3\textwidth]{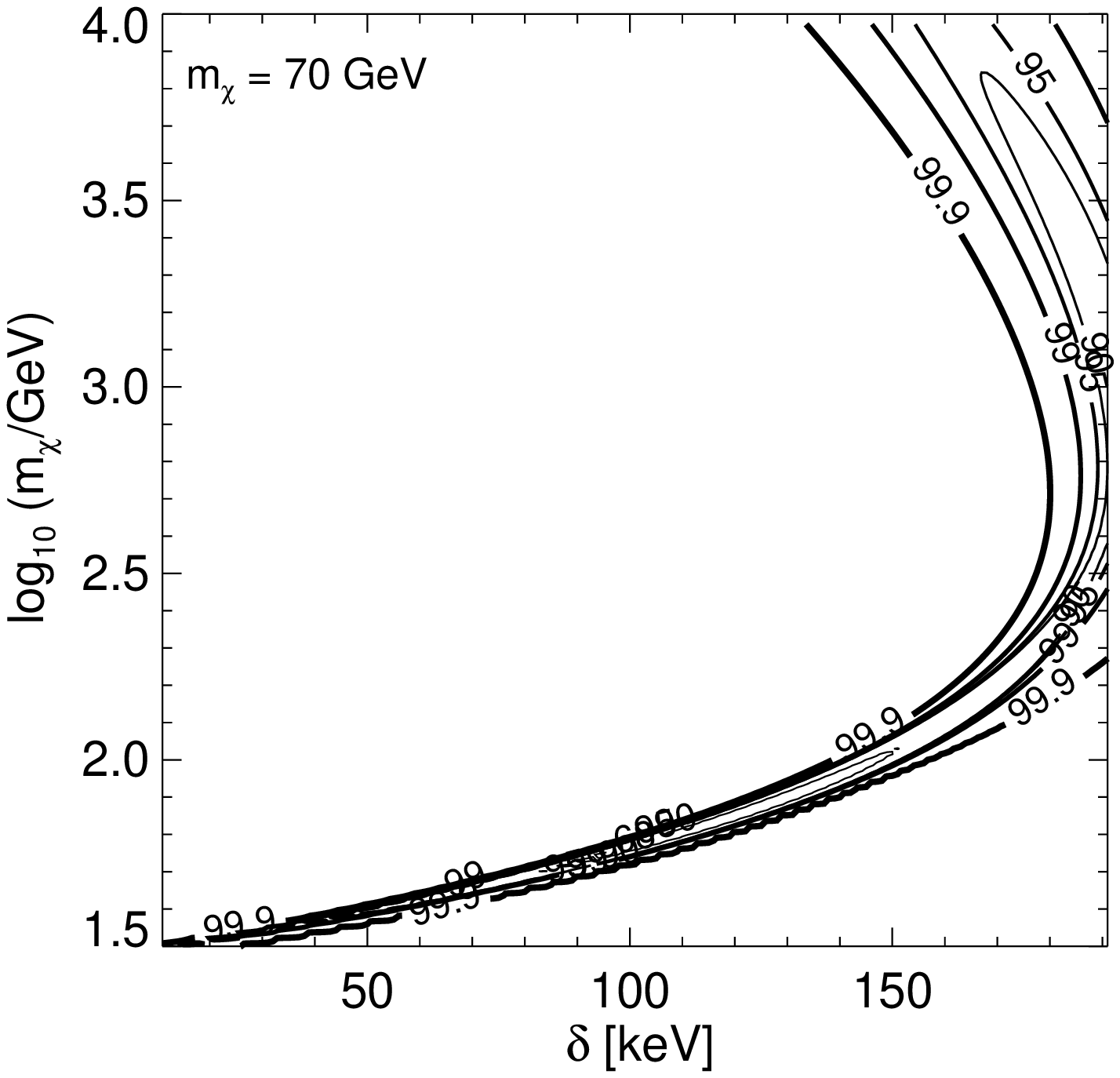}\hskip 0.2in
  \includegraphics[width=.3\textwidth]{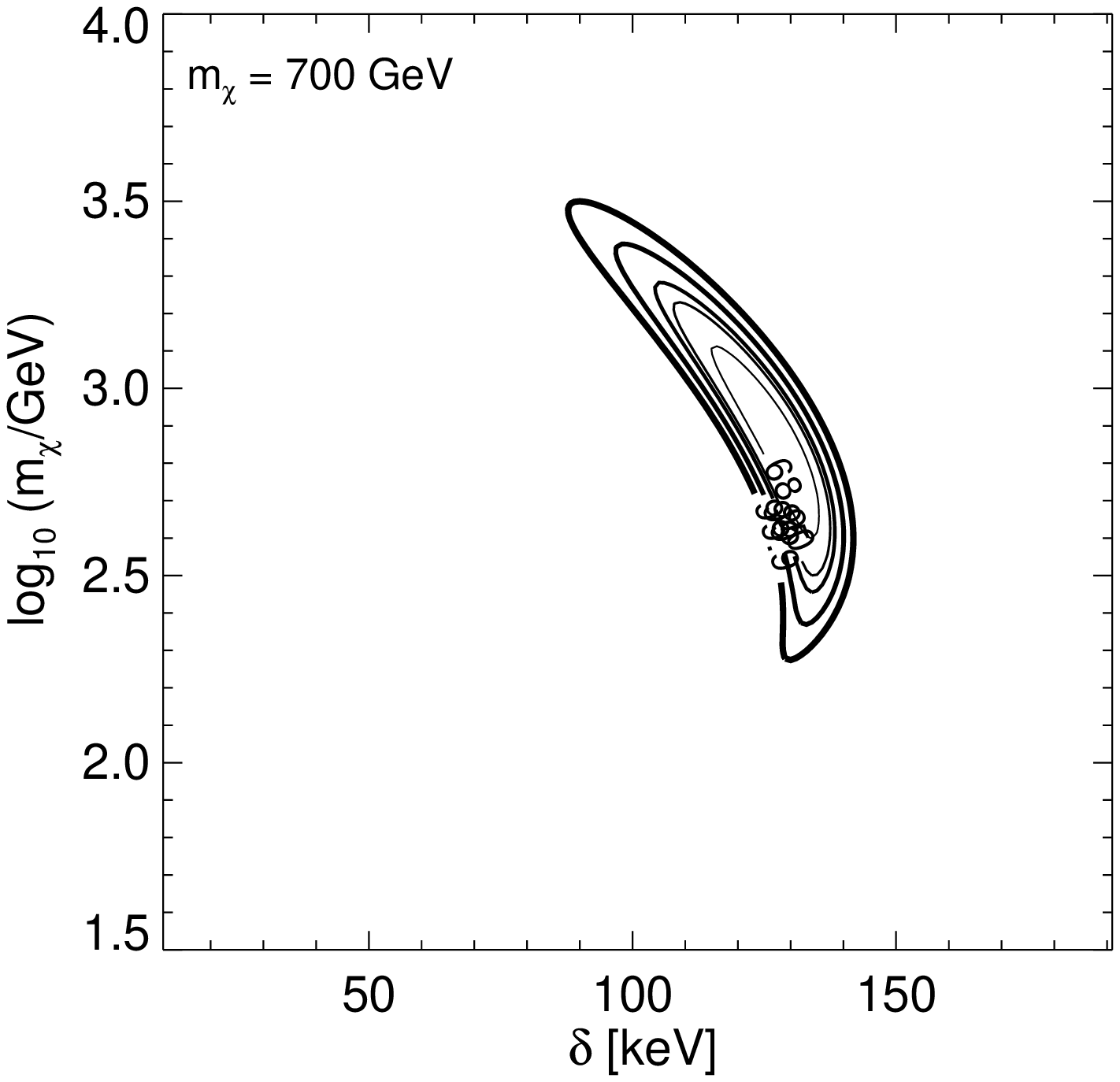}\\
\centering
\caption{Confidence levels for determining $m_\chi$ and $\delta$, where $\sigma_n$ is {\it known}, with an exposure of 1000 kg $\cdot$ day.}
\label{fig:fixedSigma}
\end{figure*}

\begin{figure*}[htpb]
\centering
\includegraphics[width=.3\textwidth]{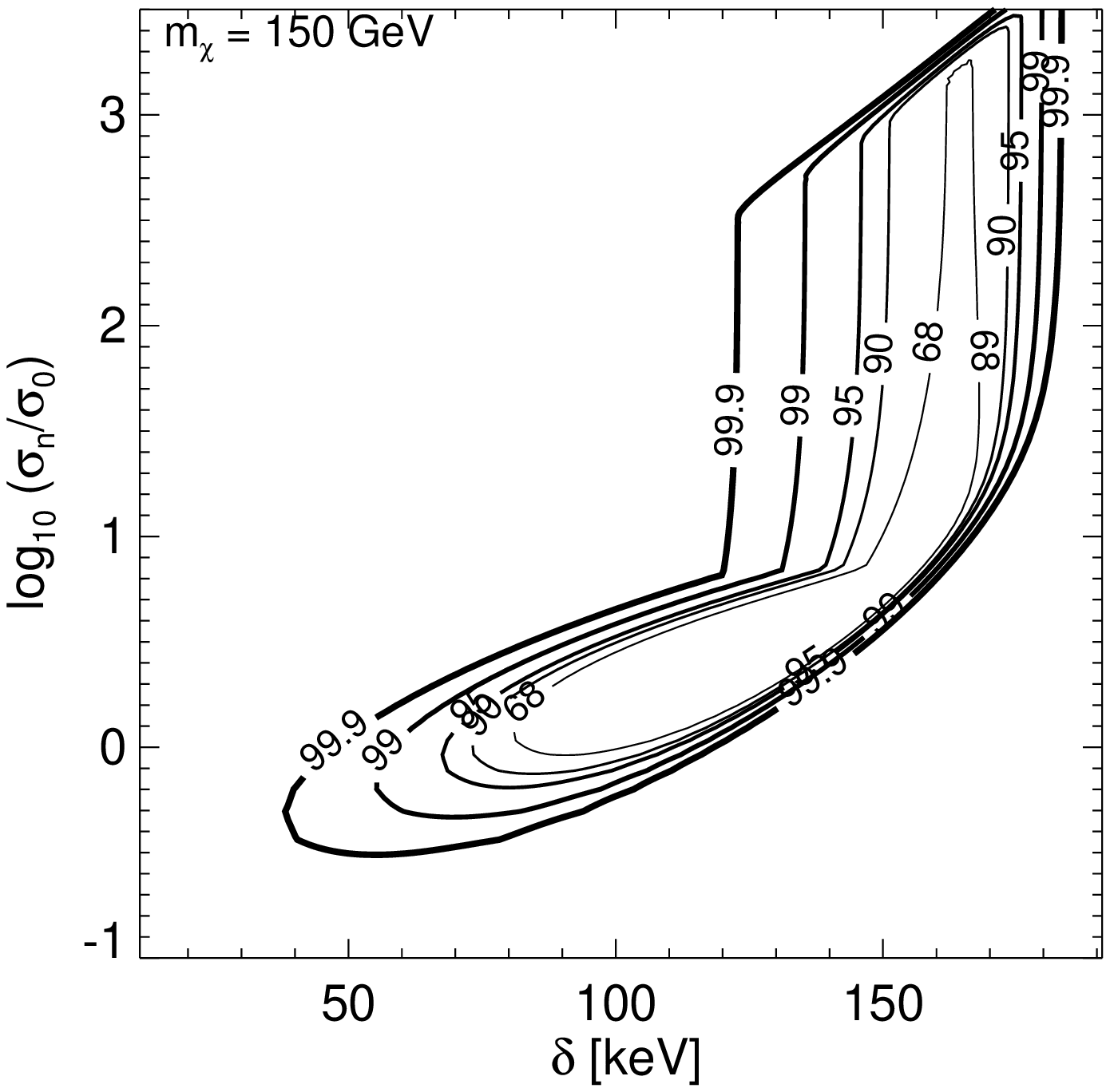}\hskip 0.2in
  \includegraphics[width=.3\textwidth]{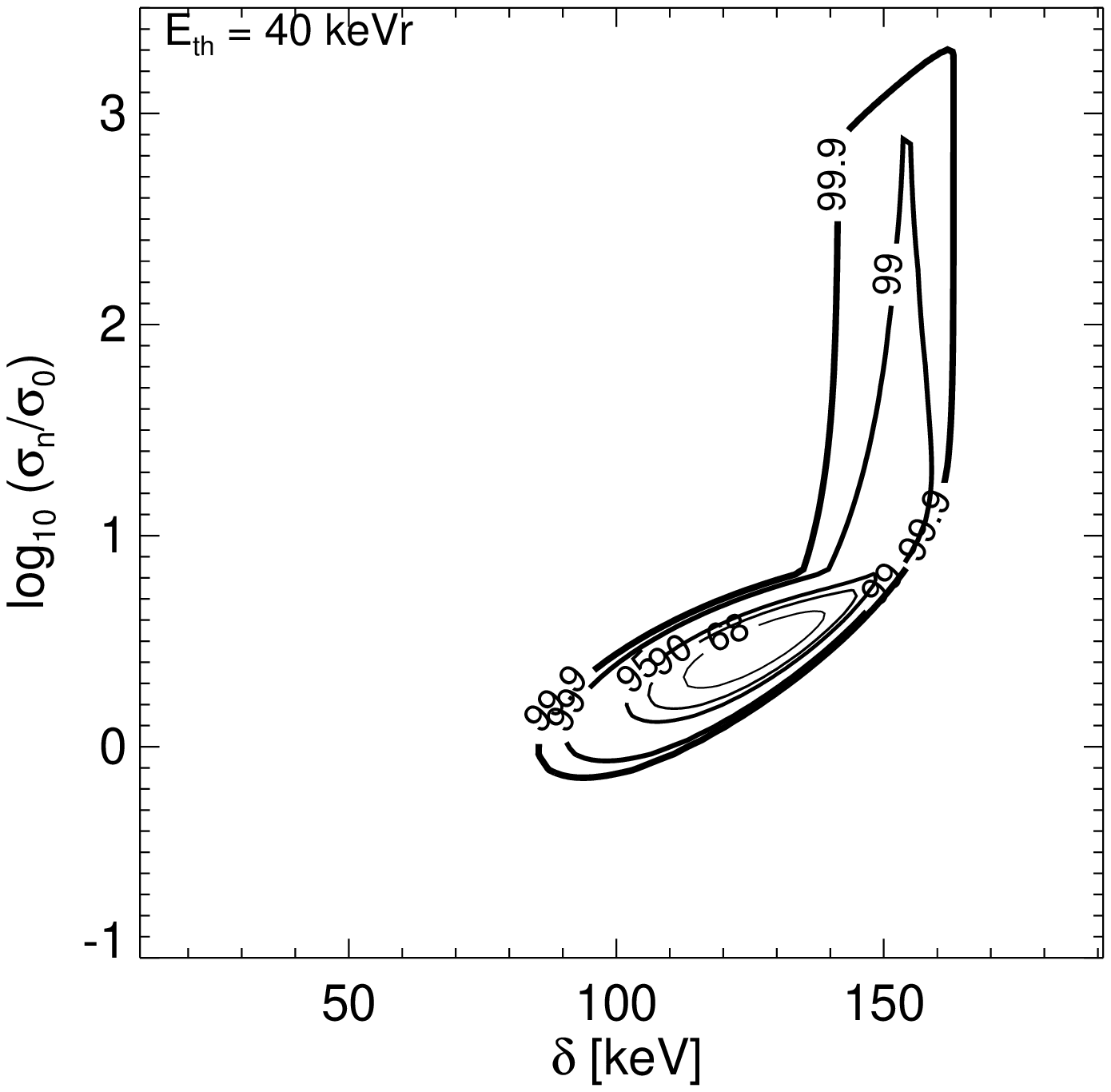}\hskip 0.2in
  \includegraphics[width=.3\textwidth]{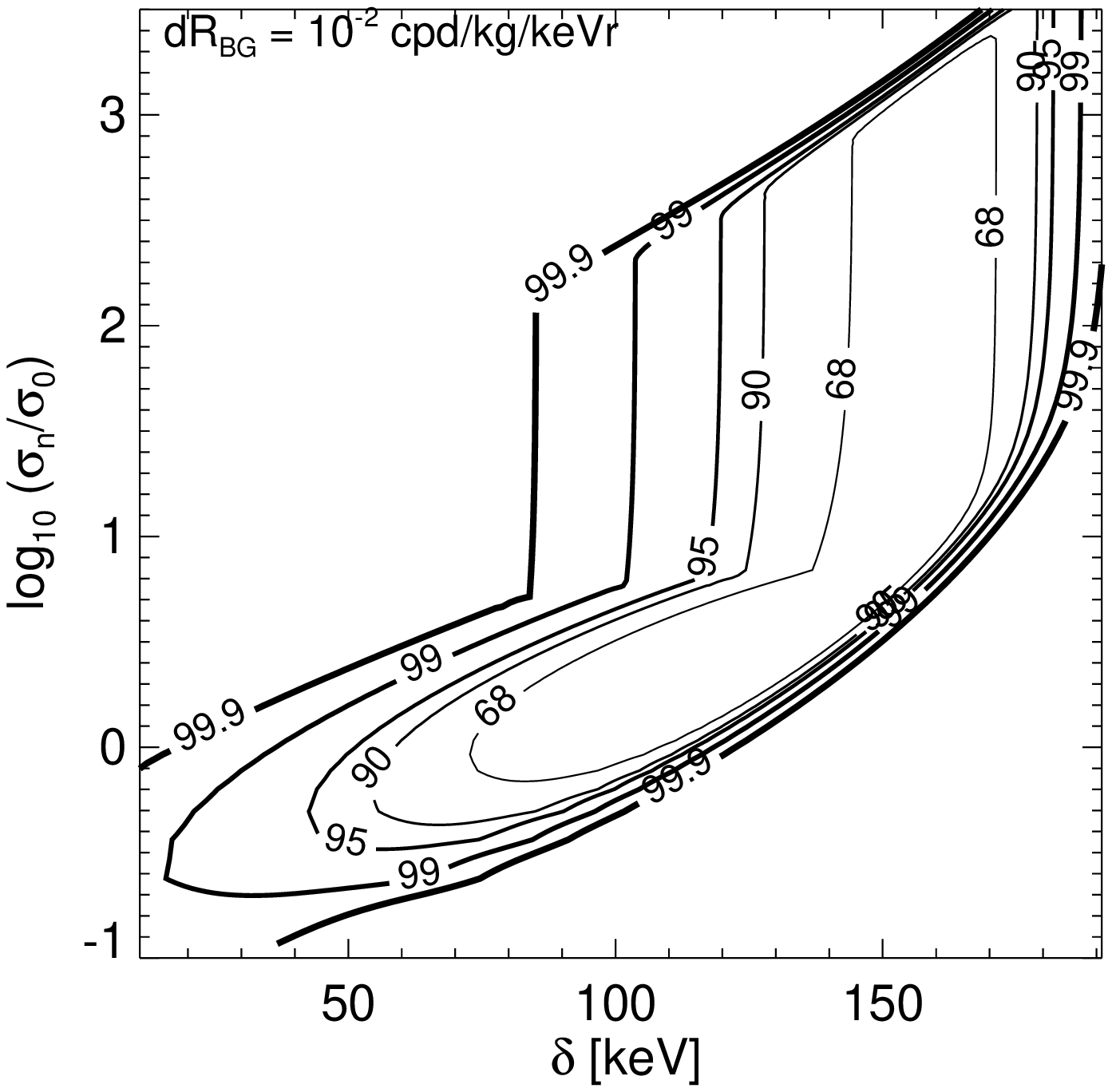}\\
\includegraphics[width=.3\textwidth]{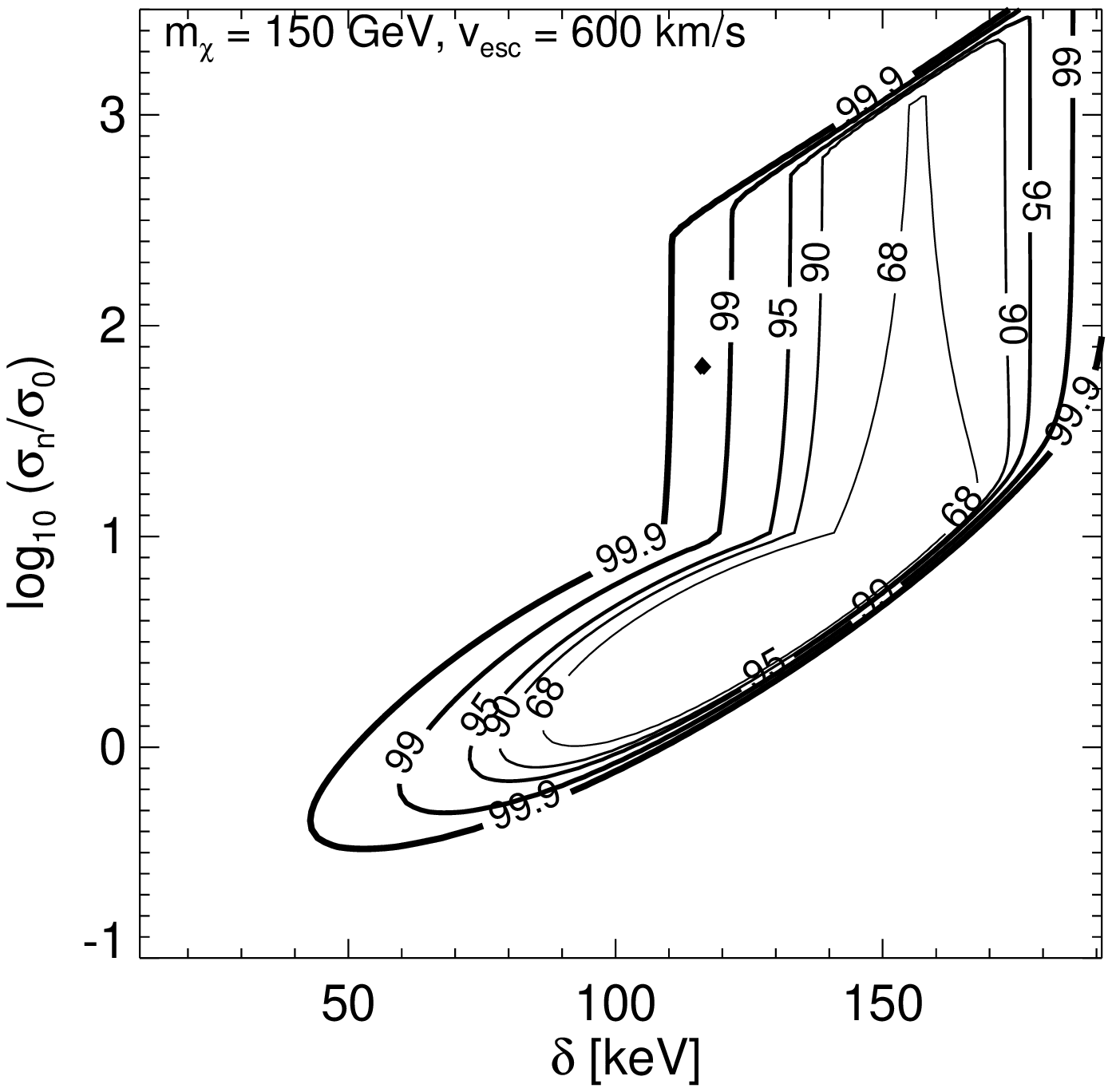}\hskip 0.2in
  \includegraphics[width=.3\textwidth]{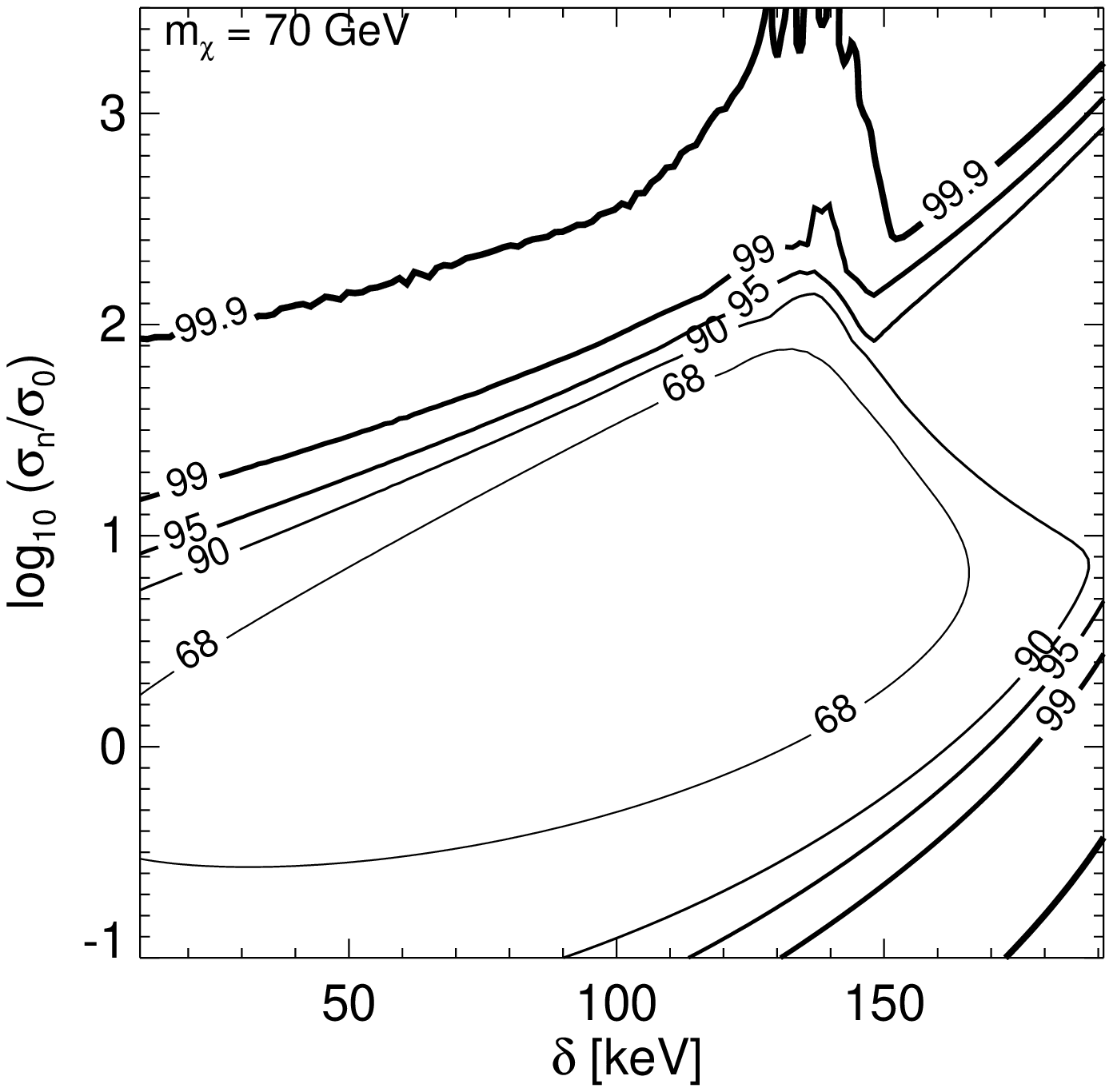}\hskip 0.2in
  \includegraphics[width=.3\textwidth]{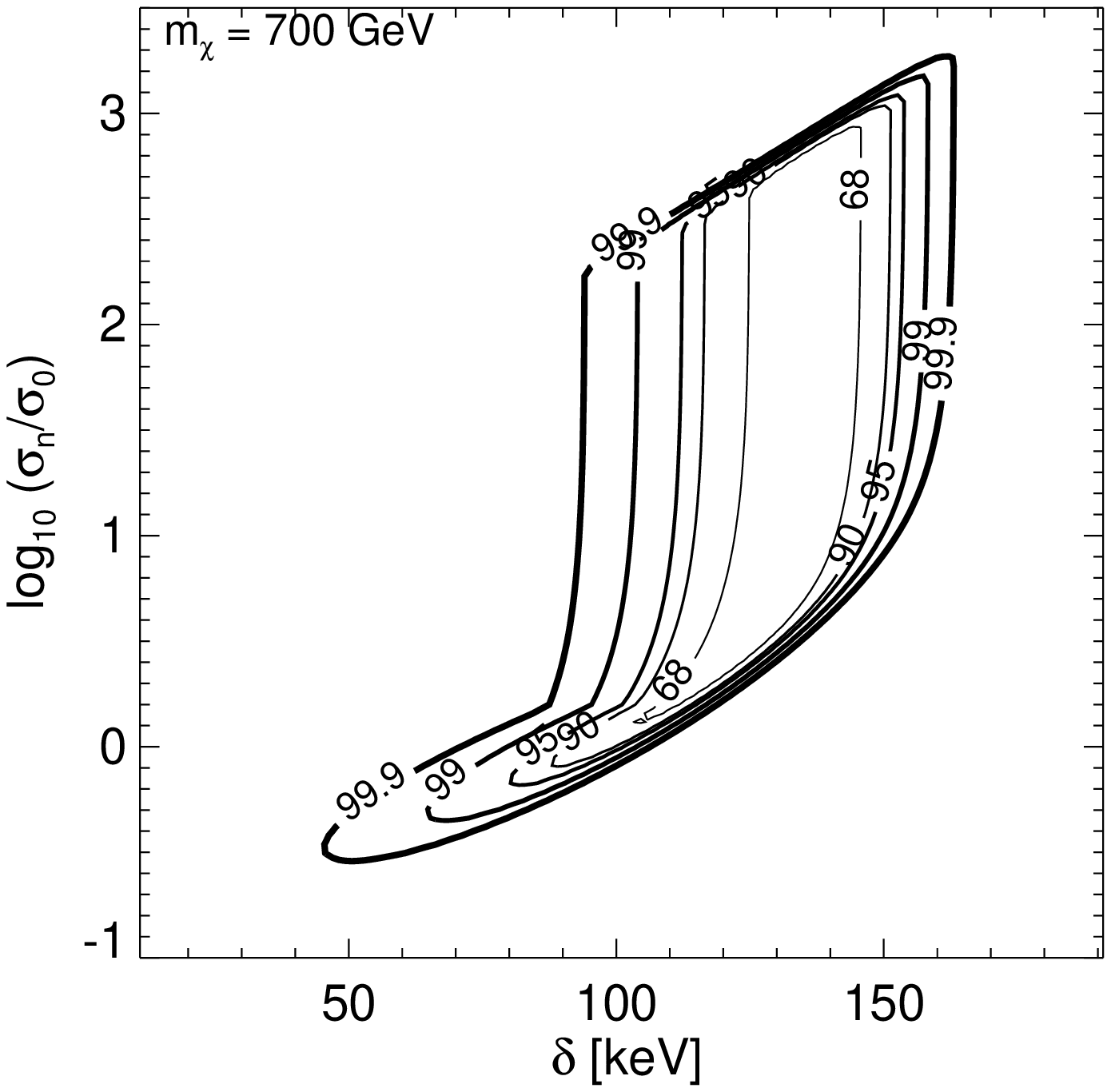}\\
\centering
\caption{Confidence levels for determining $\delta$ and $\sigma$, where $m_\chi$ is unknown, with an exposure of 1000 kg $\cdot$ day.  $\sigma_0 = 10^{-40} cm^2$.}
\label{fig:lnl_deltasigma}
\end{figure*}

\begin{figure*}[htpb]
\centering
\includegraphics[width=.3\textwidth]{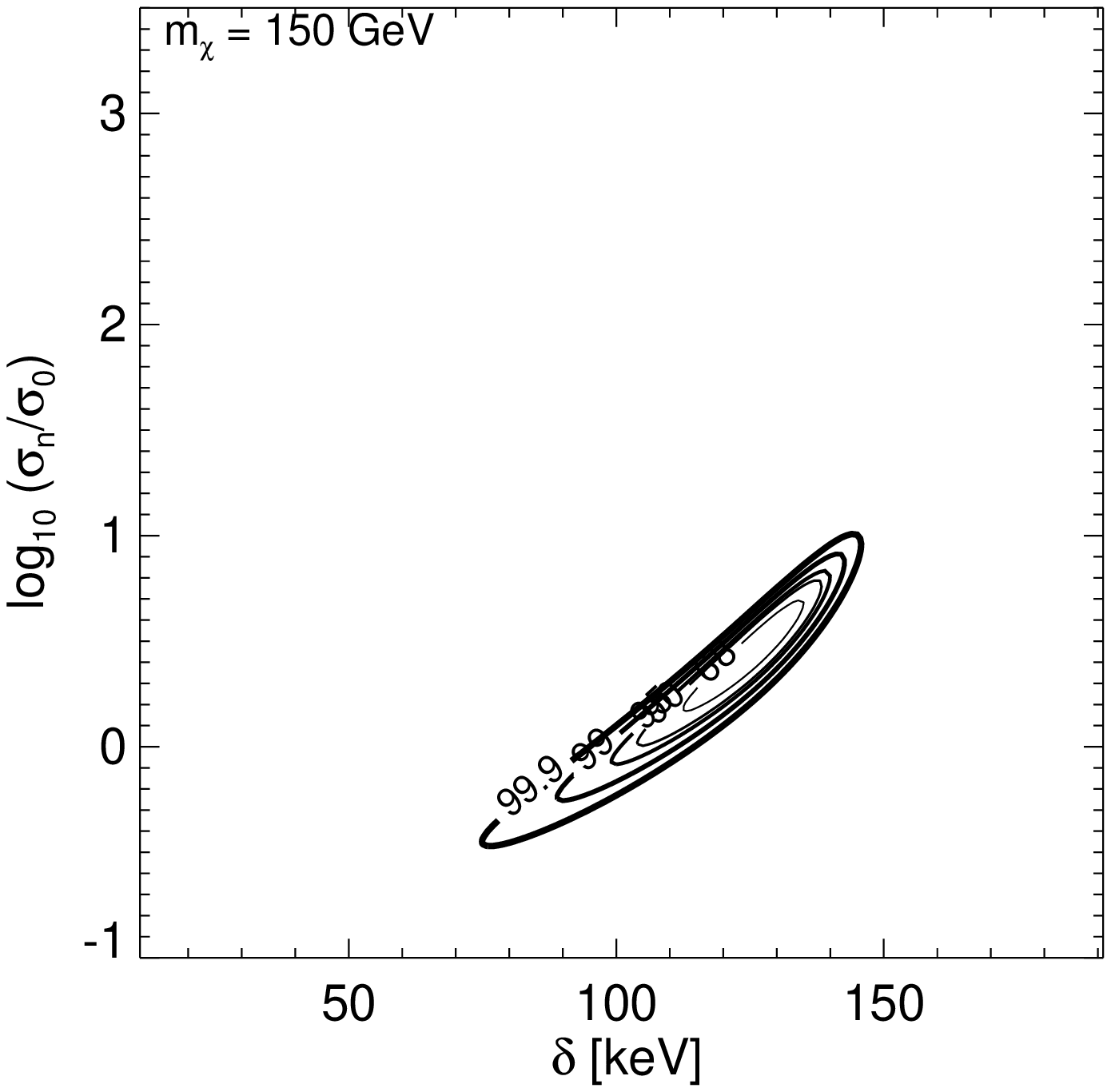}\hskip 0.2in
  \includegraphics[width=.3\textwidth]{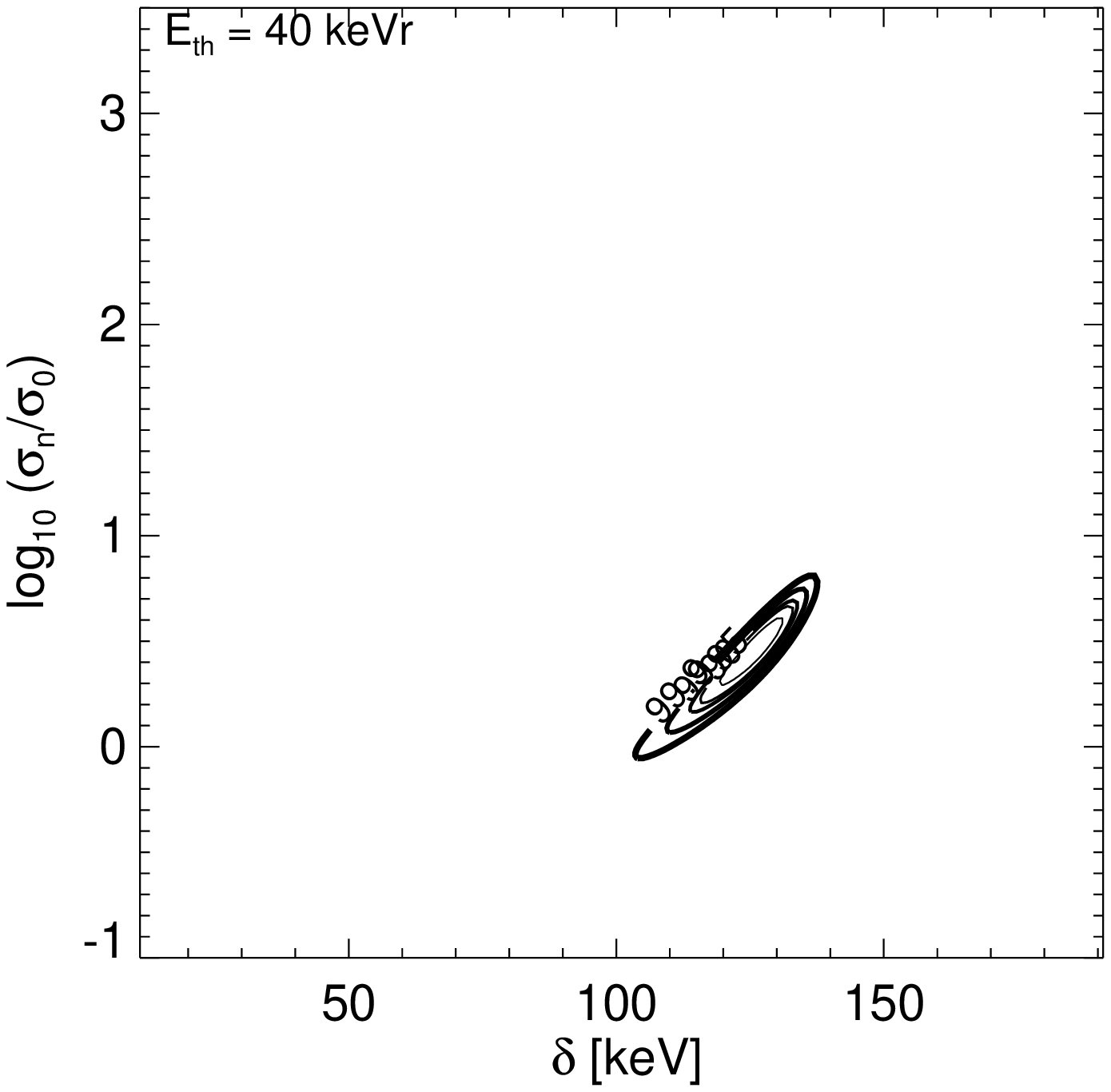}\hskip 0.2in
  \includegraphics[width=.3\textwidth]{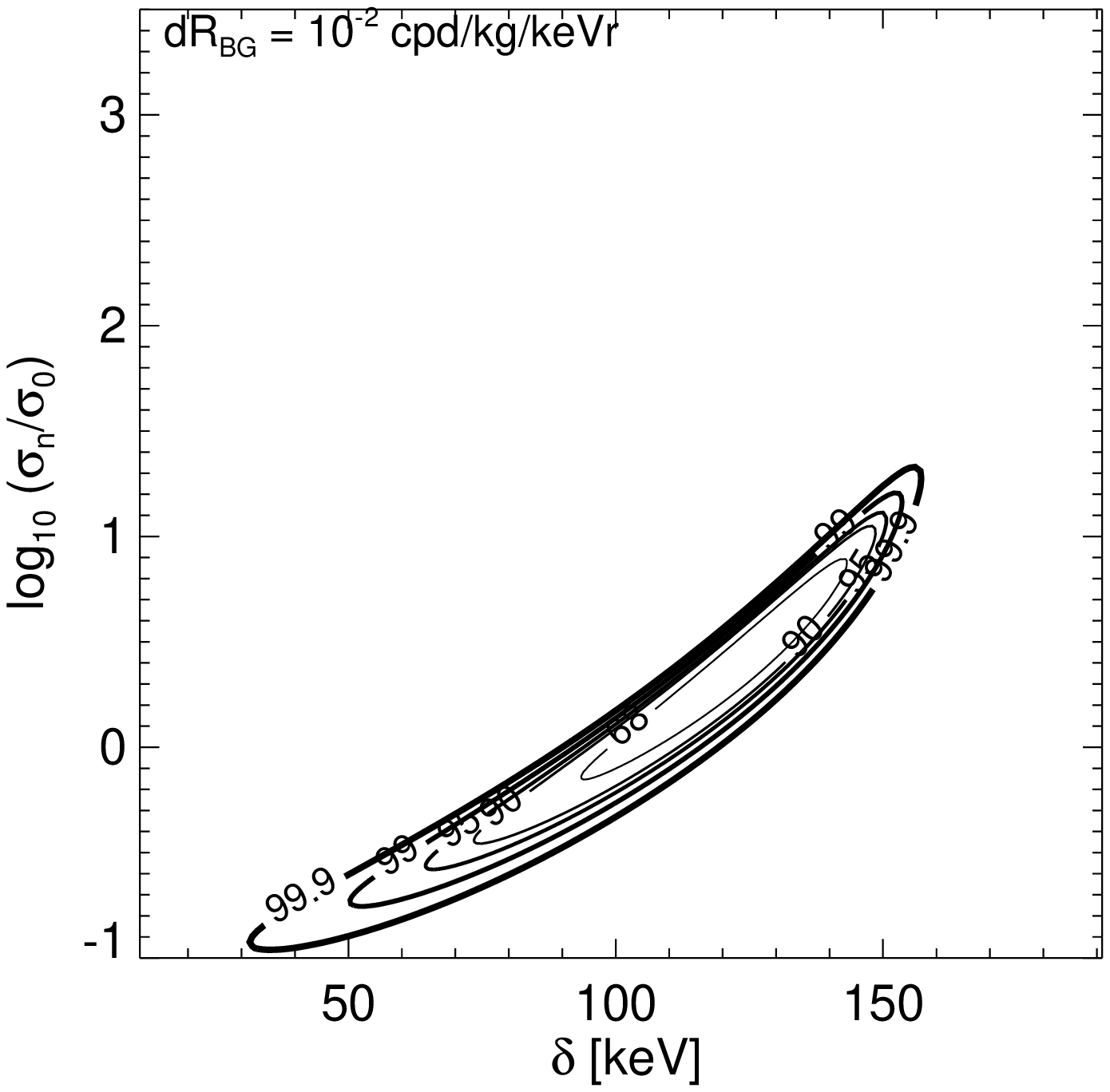}\\
\includegraphics[width=.3\textwidth]{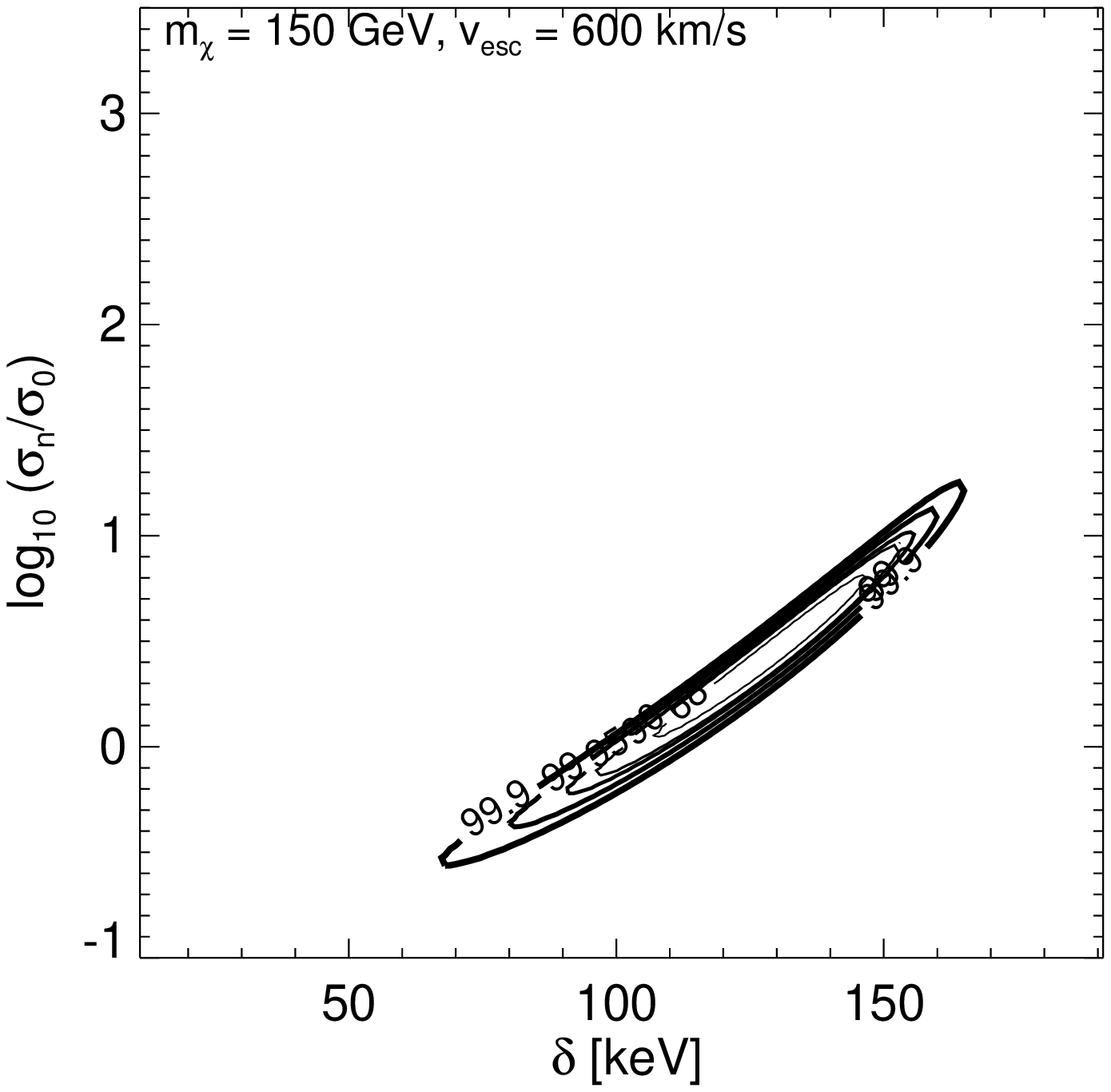}\hskip 0.2in
  \includegraphics[width=.3\textwidth]{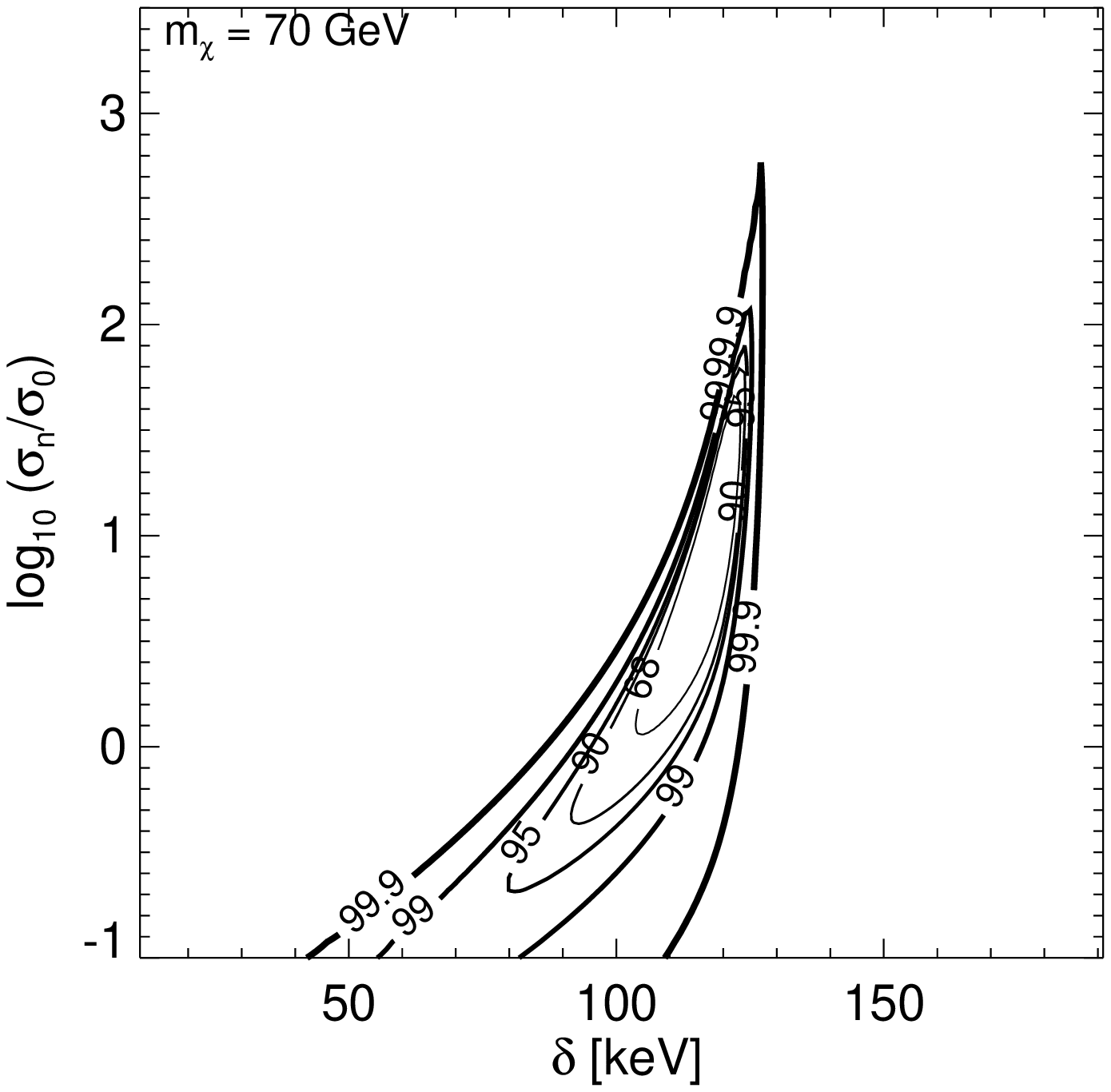}\hskip 0.2in
  \includegraphics[width=.3\textwidth]{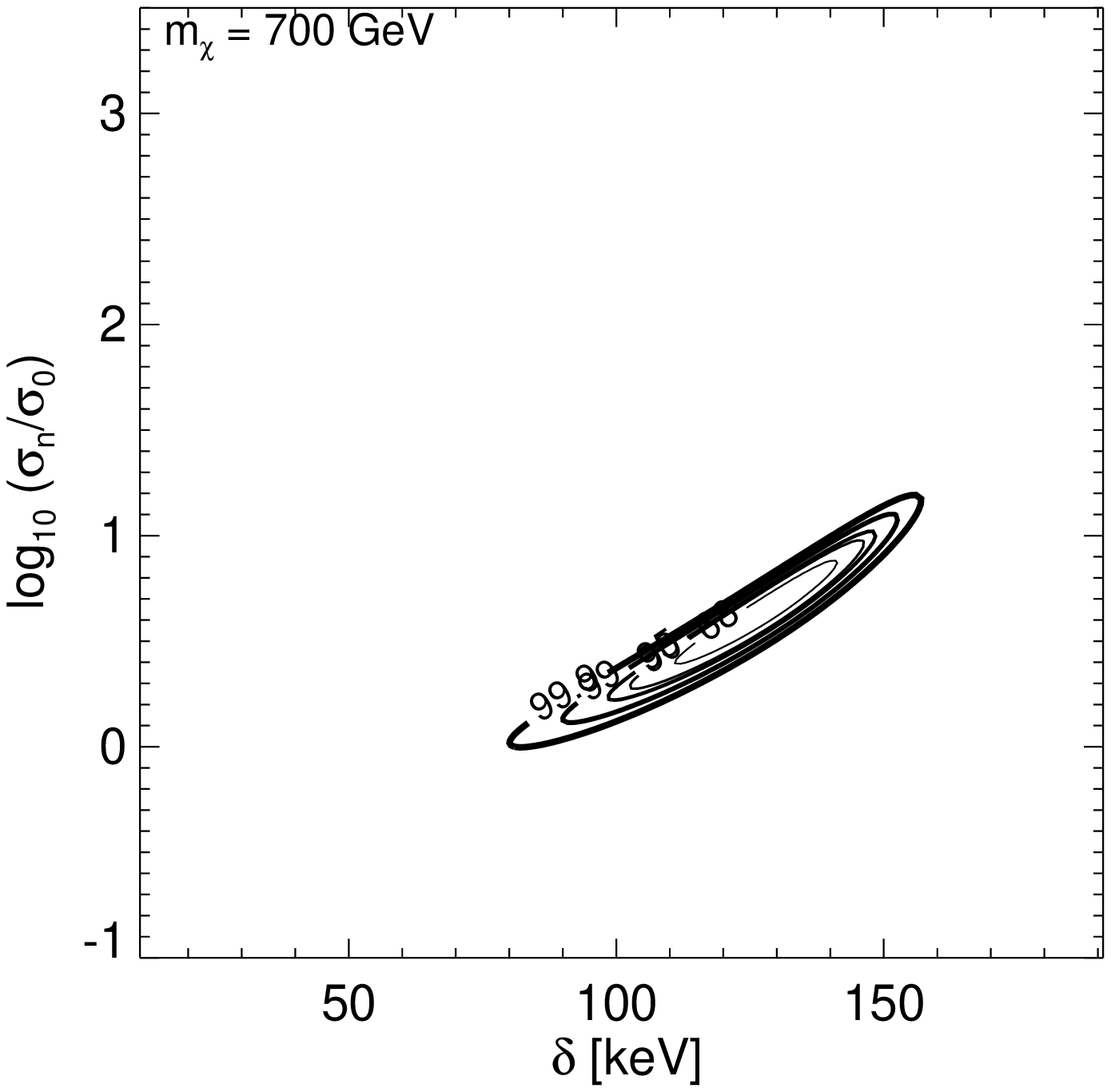}\\
\centering
\caption{Confidence levels for determining $\delta$ and $\sigma_n$, where $m_\chi$ is {\it known}, with an exposure of 1000 kg $\cdot$ day. $\sigma_0 = 10^{-40} cm^2$.}
\label{fig:lnl_fixedMass}
\end{figure*}

\begin{figure*}[htpb]
\centering
\includegraphics[width=.3\textwidth]{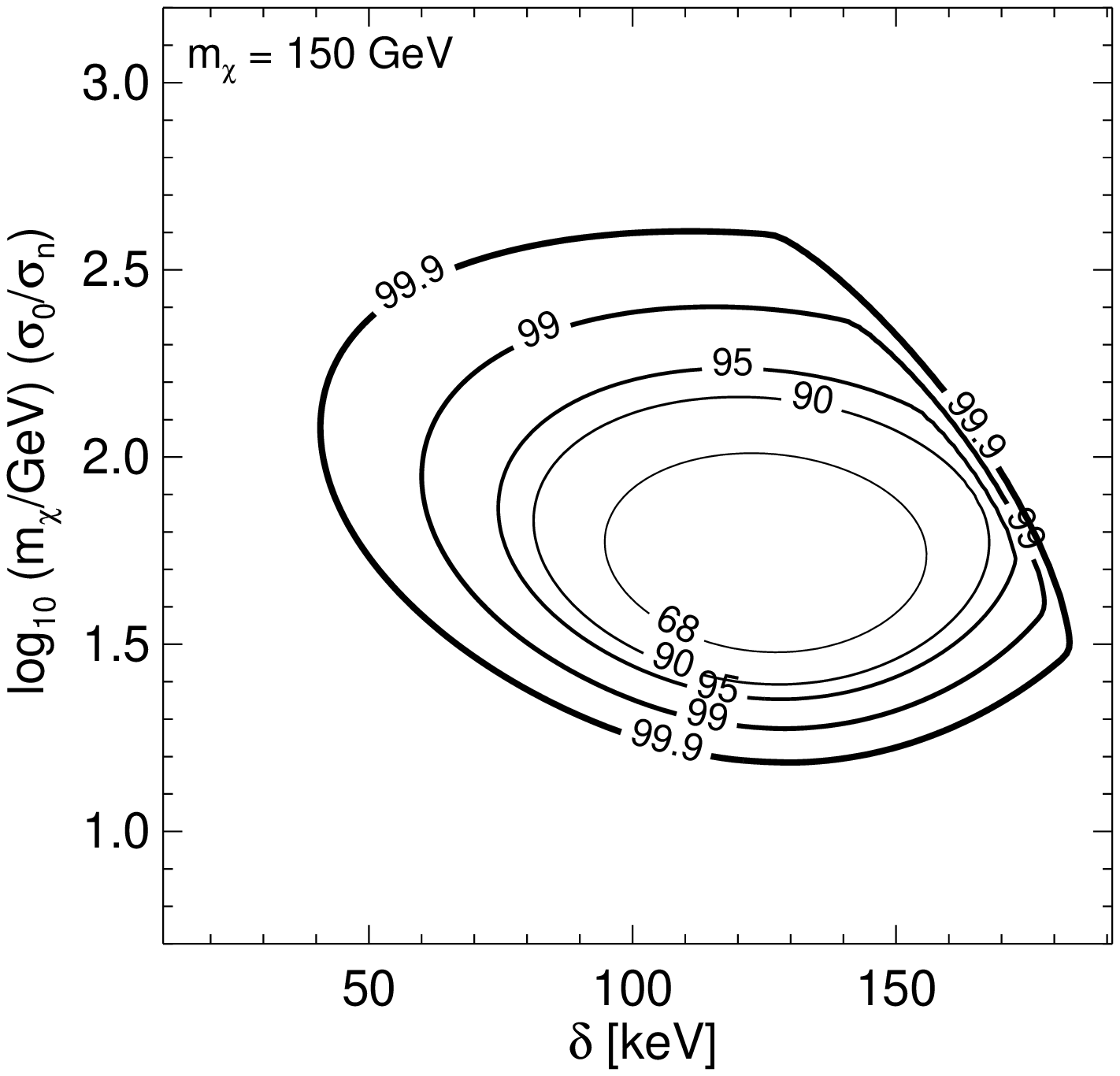}\hskip 0.2in
  \includegraphics[width=.3\textwidth]{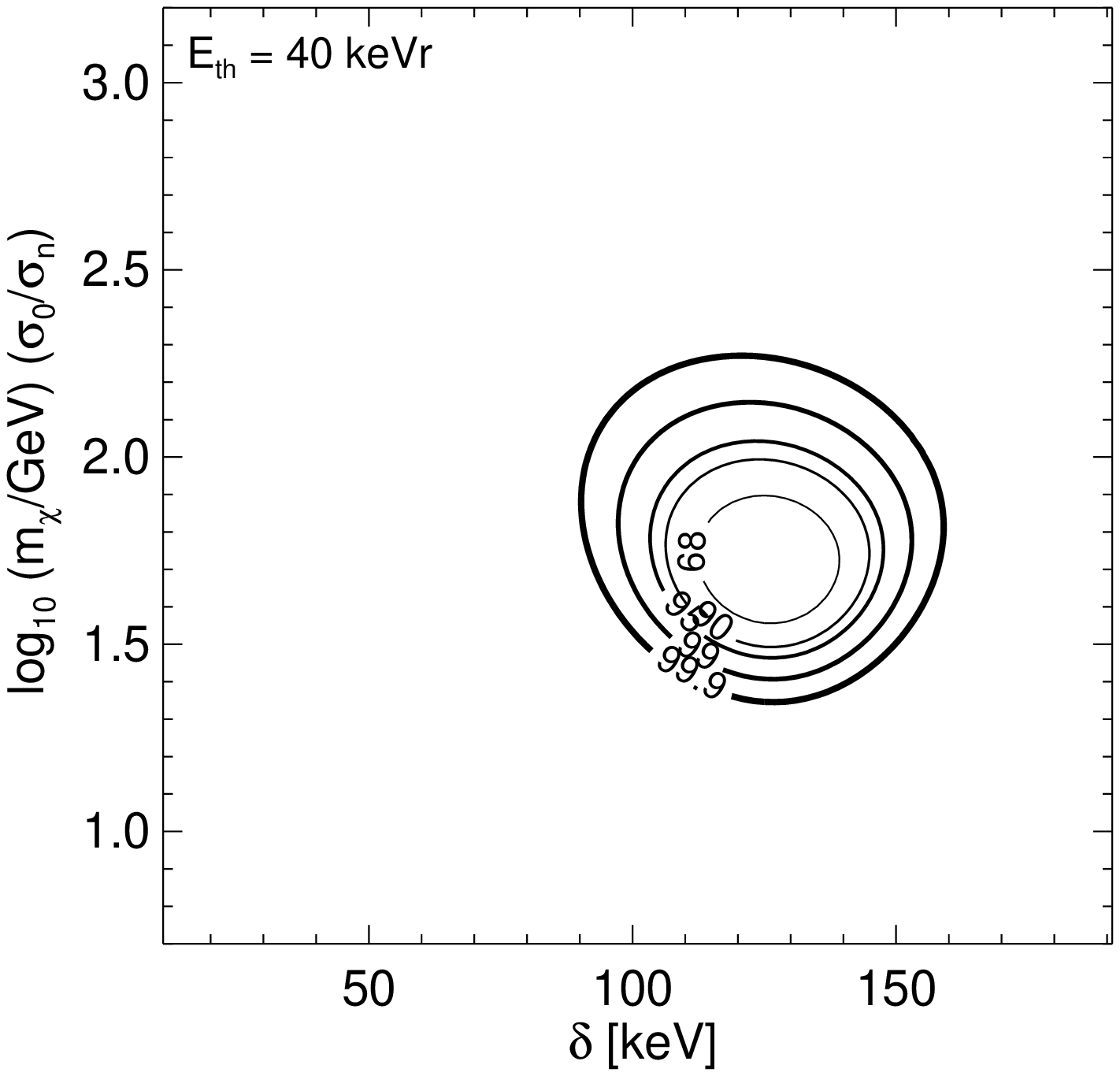}\hskip 0.2in
  \includegraphics[width=.3\textwidth]{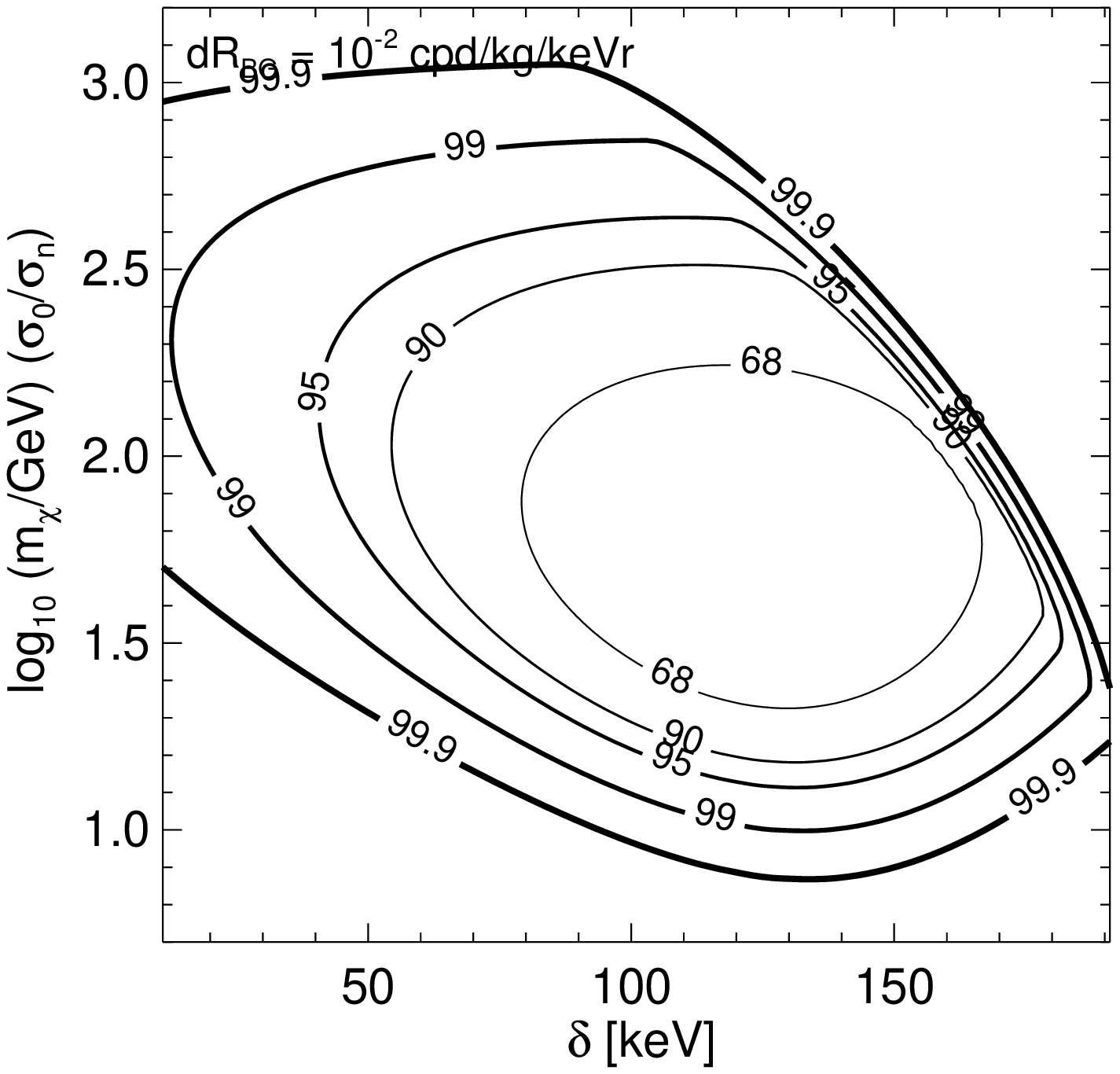}\\
\includegraphics[width=.3\textwidth]{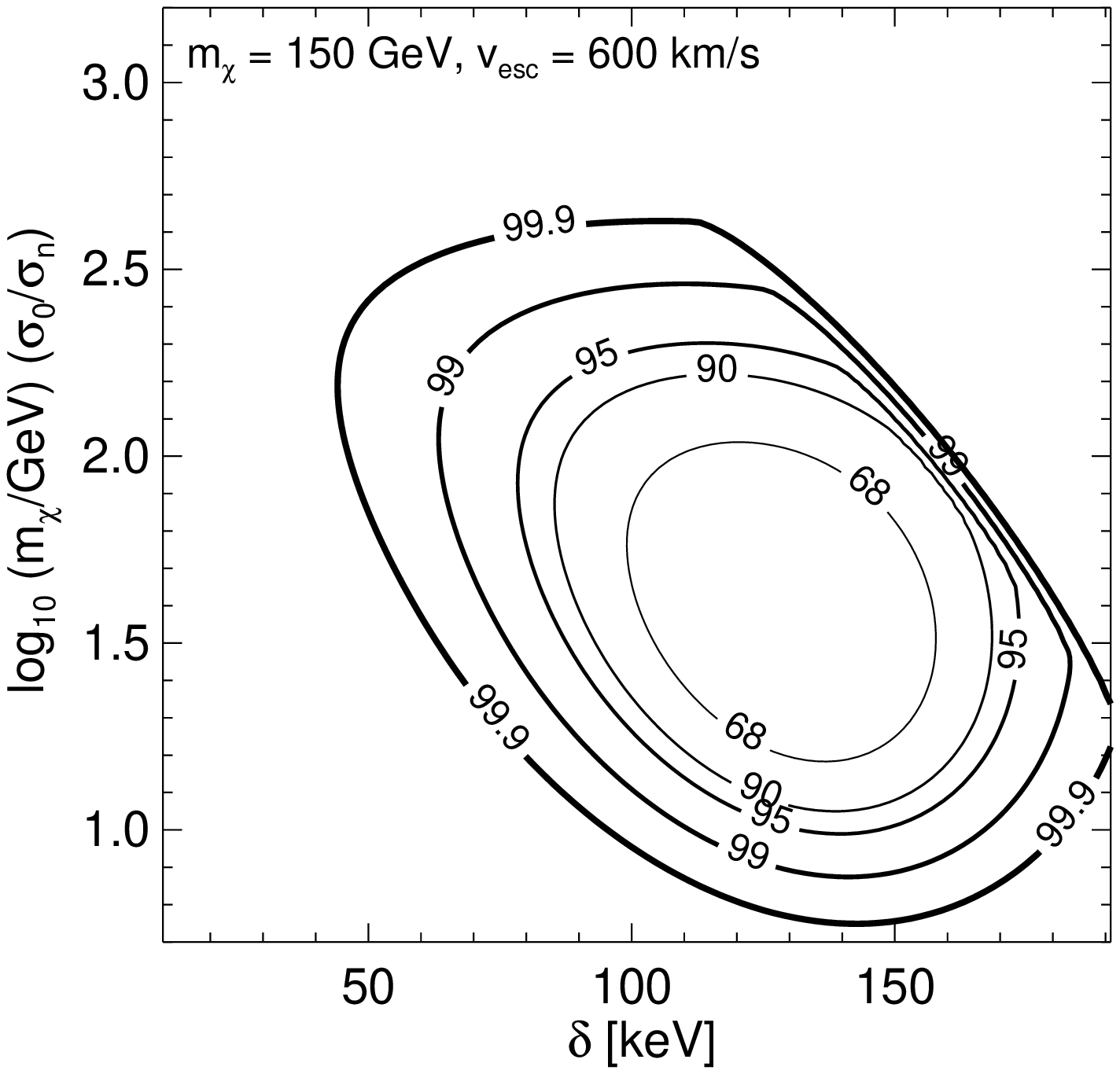}\hskip 0.2in
  \includegraphics[width=.3\textwidth]{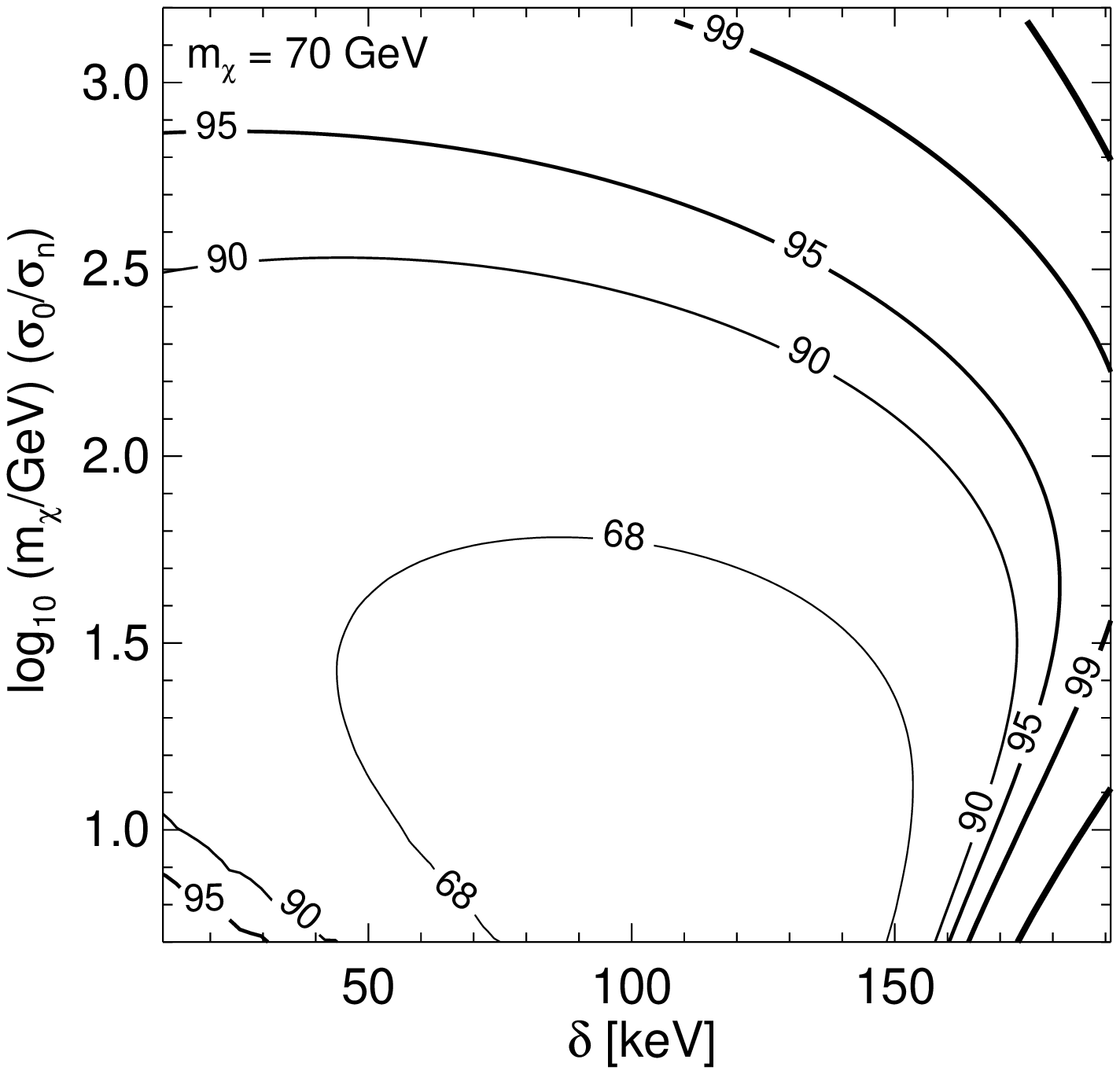}\hskip 0.2in
  \includegraphics[width=.3\textwidth]{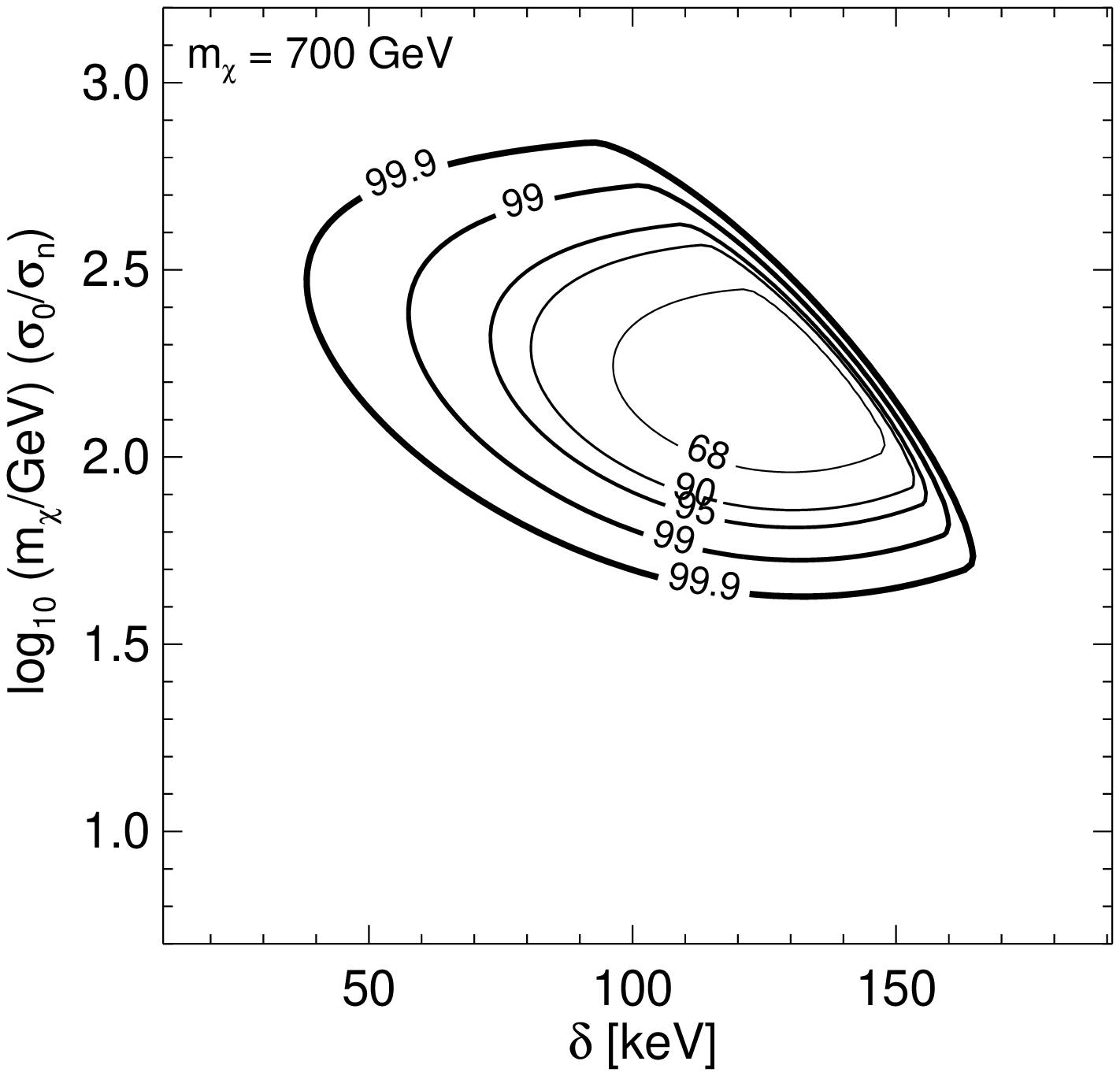}\\
\centering
\caption{Confidence levels for determining $\delta$ and
  $m_\chi/\sigma_n$, where $m_\chi$ is unknown, with an exposure of 1000
  kg $\cdot$ day, taking $\sigma_0 = 10^{-40}\cm^2$.  Over most of the
  parameter space, some value of $m_\chi$ (and therefore $\sigma_n$) can
  be found to produce enough events for the given $\delta$.  However, in the
case of large $\delta$ and large $m_\chi/\sigma_n$, no solution is
possible in some cases.}
\label{fig:lnl_Msig}
\end{figure*}

In Figs.~\ref{fig:lnl_msigma}-\ref{fig:lnl_Msig} we show confidence
levels of (68, 90, 95, 99, and 99.9\%) on the WIMP parameters for an exposure
of 1000 kg $\cdot$ day.
To obtain the probability, or likelihood, at a point in the $m_\chi-\delta$ plane, we 
either: 1) find the likelihood as a function of $\sigma_n$ and maximize with respect 
to $\sigma_n$ or 2) assume $\sigma_n$ is exactly known from some other experiment. We can
do the same also for points in $m_\chi-\sigma_n$ plane and $\sigma_n -\delta$ plane. 
The full log likelihood function lives in the full 3 dimensional parameter space. Here we show
possible slices through that space.

For each possible slice, we have shown several variations on the real WIMP parameters or 
experimental parameters. In the default scenario, we consider the $m_\chi$=150 GeV benchmark
with $E_{th}$=50 \kevr, a background rate of $dR_{BG} = 10^{-3}$~cpd/kg/keVr,
and $v_{esc} = 500\kms$. We consider the
following independent variations:
\begin{itemize}
	\item Lower energy threshold ($E_{th} \to$ 40 \kevr)
	\item Higher background ($dR_{BG} \to 10^{-2}$~cpd/kg/keVr)
	\item Higher escape velocity ($v_{esc} = 600$~km/s)
	\item Lower WIMP mass ($m_\chi \to$ 70 GeV benchmark)
	\item Higher WIMP mass ($m_\chi \to$ 700 GeV benchmark)
\end{itemize}

In each case, as $m_\chi$ and $v_{esc}$ vary, $\sigma_n$ and $\delta$
are adjusted to agree with benchmark fits to DAMA, using the parameters
in Table~\ref{tab:benchmarks}. At masses above 250 GeV, there is increasing tension between the DAMA result and other experiments, notably CDMS. This tension is highly dependent on the high velocity tail of the WIMP velocity distribution, and can be alleviated by considering non-Maxwellian velocity distributions, for instance from the Via Lactea simulation \cite{MarchRussell:2008dy,VLIDM}. Thus, we consider these points, but it should be emphasized that the non-Maxwellian halos generally tend to lead to a {\em larger} signal at DAMA (relative to the other experiments), and thus on a xenon target (because of the similar kinematics), and thus we expect that our use of a Maxwellian distribution is conservative for these points.

At masses much larger than the nucleus mass, the threshold velocity $v_{min}$ 
is independent of mass and the spectrum depends on $m_\chi$ only through 
the local WIMP density $\rho_\chi/m_\chi$. 
In these regions $m_\chi$ and $\sigma_n$ are completely degenerate since only 
the combination $\rho_\chi \sigma_n/m_\chi$ ever appears, as a prefactor determining the overall
rate. This can be clearly seen in Fig.~\ref{fig:lnl_msigma}, which shows confidence
intervals in the $m_\chi -\sigma_n$ plane. Note that because the contours never close,
we have have imposed the (rather conservative) constraint that $m_\chi < 100$ TeV 
based on the unitarity bound \cite{Griest:1989wd} for a thermal relic.

The effects of the $m_\chi -\sigma_n$ degeneracy can also be seen in the $m_\chi -\delta$
plane, shown in Fig.~\ref{fig:lnl_mdelta}. Here high masses are all equally 
likely (given a fixed $\delta$) because $\sigma_n$ can be adjusted accordingly. 

In the  $\delta-\sigma_n$ plane, Fig.~\ref{fig:lnl_deltasigma}, 
there is a sharp discontinuity since low masses are favored
at smaller $\sigma_n$ and very high masses are favored at high $\sigma_n$.
This is because
at low scattering cross section, in order to boost the rates such that it matches the observed
number of events, one can lower $\delta$ or adjust the mass to optimize the number of rates. 
(The scattering rate is maximized when the mass of the WIMP $\sim$ the mass of the nuclei.)
However, at high scattering cross section, one can increase $\delta$ but only increase the mass
to very high masses to reduce the rates. Though lowering the mass drastically also decreases
the rate, the angular shape at very low masses is very distinct (see Fig.~\ref{fig:rates}) and
thus unfavored. 
The cutoff in  Fig.~\ref{fig:lnl_deltasigma} at high $\sigma_n$ is a result of the unitarity
bound on the mass.

These effects can make it difficult to constrain the WIMP mass at low exposures;
however, it is easier to constrain the ratio $m_\chi/\sigma_n$,
which we have shown in Fig.~\ref{fig:lnl_Msig}. 

Finally, we note that in these figures we have assumed the earth velocity is 
unmodulated. For the benchmark where $m_\chi = 70$ GeV, our worst-case scenario,
the effects of the annual modulation in velocity can improve the confidence levels 
significantly if the experiment is done during the summer.

The disadvantage of the likelihood analysis is its model dependence. 
We used the truncated Maxwell-Boltzmann profile,
whereas in reality it is likely there is more structure in the dark matter profile.
However we expect the results to roughly be the same
for many more complicated velocity distributions, and in fact can improve
for inelastic dark matter, as mentioned above. Furthermore, because of the velocity
threshold due to $\delta$, the inelastic scenario is not very sensitive to streams
because most streams are below the threshold velocity. 
Anisotropies in the halo profile do not significantly affect the results here. 
To see the effect of using less simplistic halo models on the \textit{elastic} scattering
spectrum and sensitivity, see \cite{Alenazi:2007sy} and \cite{Copi:2000tv}.

\section{Conclusions}

Motivated by the finding \cite{Chang:2008gd} that inelastic dark
matter (iDM) is compatible with both the DAMA annual modulation signal
at $22-66\kevr$ and limits from other experiments at lower energies,
we have investigated prospects for directional detection in the
context of the iDM model.  We are encouraged by the fact that
ZEPLIN-III has also detected a number of events in the $40-80\kevr$
range \cite{Lebedenko:2008gb}.  This has \emph{not} been claimed as
evidence of WIMP scattering, but makes it impossible to rule out iDM
with such data.  In the near future, LUX
\footnote{http://lux.brown.edu} and XENON100 \cite{Aprile:2009yh} will
have greatly improved sensitivity and lower backgrounds, and will
provide a sharp test of the iDM/DAMA scenario.  If these experiments
also detect an excess of events above background in the appropriate
energy range, a major effort in directional detection will be
justified.

Directional detection with a gaseous detector containing a heavy gas
(e.g. Xe) may not require the huge exposure times implied by the
elastic scattering limits.  For a threshold energy of $E_{th} = 50$
\kevr, we find that exposures of order $\sim 1000$ kg $\cdot$ day in a
directional experiment can convincingly refute or support the claims
of DAMA in the context of the inelastic dark matter model.  At zero
background, roughly 18 events are needed for a clear detection of WIMP
scattering. Even with larger backgrounds, the required exposure is a
few hundred kg~$\cdot$~day, over most of the iDM parameter space that
can explain both DAMA and other direct detection experiments.  With
roughly $1000$ kg $\cdot$ day, it is possible to obtain a measurement
of $\delta > 0$ at high significance and also the parameter $m_\chi/\sigma_n$ to within an
order of magnitude.

Furthermore, if it is possible to roughly determine one of the WIMP parameters, 
for example $\delta \sim 120$ keV, via another experiment, the mass and nucleon
scattering cross section are highly constrained with an exposure of
a few hundred kg $\cdot$ day because of the distinctive shape of 
the energy-angle recoil spectrum.

Significantly lower exposures are needed if the threshold energy is decreased.
As discussed in Section~\ref{sec:experiment}, 
because of the uncertainties in the nuclear recoil energies of the DAMA signal,
it is crucial to reduce the threshold energy as much as possible. 
For low masses, the recoil spectrum is sharply distributed in
energy and angle. However, typical recoil energies are smaller. Thus
with an energy threshold of $E_{th} = 50$ \kevr\ most of the events for $m_\chi =70$ GeV
are not seen. With an energy threshold of 100 \kevr\ and $m_\chi =70$ GeV,
none of the WIMP recoils can be seen. Though the required volume increases
and angular resolution decreases when $E_{th}$ is lowered, we found that a poor 
angular resolution ($\sim 60^\circ$) does not significantly 
affect the results, assuming that 3D reconstruction of the track and 
determining the sense is still possible.


We acknowledge helpful discussions about directional detection with
Peter Fisher, Jocelyn Monroe, and Gabriella Sciolla.  Rick Gaitskell,
Dan McKinsey, and Peter Sorensen provided helpful advice and
much-needed skepticism.  DPF is partially supported by NASA LTSA grant
NAG5-12972.  TL is supported by an NSF Graduate Fellowship.  This work
was partially supported by the Director, Office of Science, of the
U.S.  Department of Energy under Contract No. DE-AC02-05CH11231.  NW
is supported by NSF CAREER grant PHY-0449818 and DOE OJI grant
\#DE-FG02-06E R41417.

\onecolumngrid
\bibliography{crucis}
\bibliographystyle{apsrev}

\end{document}